\documentclass[journal = nalefd,manuscript=article]{achemso}
\setkeys{acs}{articletitle = false}
\usepackage{graphicx,verbatim}
\usepackage{braket}
\usepackage{color}
\usepackage{amssymb}
\usepackage{mciteplus}
\mciteErrorOnUnknownfalse
\usepackage[sort&compress,numbers,super]{natbib}
\usepackage{achemso} 

\usepackage{soul}
\usepackage[normalem]{ulem}
\usepackage[x11names]{xcolor}

\title{Shaping electronic flows with strongly correlated physics}

\author{A. Erpenbeck}
\affiliation{
	Department of Physics, University of Michigan, Ann Arbor, Michigan 48109, USA
	}

\author{E.\ Gull}
\affiliation{
	Department of Physics, University of Michigan, Ann Arbor, Michigan 48109, USA
	}
	
\author{G.\ Cohen}
\affiliation{
    The Raymond and Beverley Sackler Center for Computational Molecular and Materials Science, Tel Aviv University, Tel Aviv 6997801, Israel
    }
\alsoaffiliation{
    School of Chemistry, Tel Aviv University, Tel Aviv 69978, Israel
    }

\date{\today}

\begin{document}

\begin{abstract}
            Nonequilibrium quantum transport is of central importance in nanotechnology.
            Its description requires the understanding of strong electronic correlations, which couple atomic-scale phenomena to the nanoscale.
            So far, research in correlated transport focused predominantly on few-channel transport, precluding the investigation of cross-scale effects.
            Recent theoretical advances enable the solution of  models that capture the interplay between quantum correlations and confinement beyond a few channels.
            This problem is the focus of this study.
			We consider an atomic impurity embedded in a metallic nanosheet spanning two leads, showing that transport is significantly altered by tuning only the phase of a single, local hopping parameter.
			Furthermore---depending on this phase---correlations reshape the electronic flow throughout the sheet, either funneling it through the impurity or scattering it away from a much larger region.
            This demonstrates the potential for quantum correlations to bridge length scales in the design of nanoelectronic devices and sensors.

\end{abstract}

    Transport through quantum systems is a central paradigm in nanoscience.
    Mesoscopic\cite{datta_electronic_1997,kouwenhoven_introduction_1997} and molecular\cite{nitzan_electron_2003,ratner_brief_2013} junctions have provided experimental windows into phenomena ranging from interference\cite{Ballmann_Experimental_2012, Frisenda_Mechanically_2016, Bai_Anti_2019, Greenwald_Highly_2021, Iwakiri_Gate_2022} and dephasing\cite{Hansen_Mesoscopic_2001, Yamamoto_Electrical_2012, Finck_Phase_2014, Duprez_Macroscopic_2019, Poggini_Chemisorption_2021} to highly entangled quantum phases \cite{Roch_Quantum_2008, Custers_Destruction_2012, Hartman_Direct_2018, Paschen_Quantum_2021}.
    The combination of quantum correlation effects and nonequilibrium physics continues to pose a challenge to the theoretical description of transport.
    On one hand, advanced ab-initio techniques are now available for weakly correlated systems \cite{Brandbyge_Density_2002,groth_kwant:_2014}, and are often limited chiefly by experimental uncertainties\cite{Evers_Advances_2020}.
    On the other hand, an accurate description of transport in the presence of strong many-body correlations remains elusive beyond relatively simple models \cite{Evers_Advances_2020,Cohen_Green_2020}.

	The study of strongly correlated transport in bulk metals with magnetic impurities dates back almost a century\cite{de_haas_electrical_1934}.
	At high temperature, electrons scatter off impurity atoms.
	In contrast, at low temperature, a Kondo screening cloud forms around the impurity, resulting in a greatly enhanced scattering cross section and resistance \cite{Hewson_Kondo_1997}.
	The same mechanism is responsible for the opposite effect---suppressed resistance---in mesoscopic and molecular transport: there, electrons from one lead scatter into the other through a thin channel that becomes effectively transparent in the Kondo regime \cite{ng_-site_1988,glazman_resonant_1988,meir_low-temperature_1993,Goldhaber_Kondo_1998,cronenwett_tunable_1998}.
	Although the Kondo effect results from an atomic impurity or small quantum dot, the size of the scattering region can be orders of magnitude larger, and is characterized by the Kondo correlation length $\xi \approx v_f/T_K$; here $v_f$ is the Fermi velocity and $T_K$ is the Kondo temperature\cite{Affleck_Detecting_2001,Affleck_Friedel_2008,Affleck_Kondo_2010,Erpenbeck_Resolving_2021}.
	
	Experiments on quantum dots are generally well described by models where an impurity couples two baths that are not otherwise connected \cite{pustilnik_kondo_2004, DiVentra_electrical_2008, Sohn_Mesoscopic_2013}.
	However, there are also cases where impurities are embedded in a nonequilibrium environment and multiple transport channels are present.
	Perhaps the simplest examples are side-coupled quantum dots\cite{kang_anti-kondo_2001, Aligia_Kondo_2002, Sato_Observation_2005, Feng_Anti_2005,sasaki_fano-kondo_2009,tamura_fanokondo_2010,zitko_fano-kondo_2010, Kiss_Numerical_2011, Huo_Fano_2015, 
    Wang_Unified_2022, Lara_Kondo_2023} and magnetic break junctions\cite{houck_kondo_2005,parks_tuning_2007,calvo_kondo_2009,frisenda_kondo_2015}.
	More complex examples include scanning tunneling microscopy of magnetic atoms\cite{li_kondo_1998,madhavan_tunneling_1998,manoharan_quantum_2000,Knorr_Kondo_2002}, small molecules\cite{Iancu_Manipulating_2006,Iancu_Manipulation_2006,rakhmilevitch_electron-vibration_2014, DaRocha_Curvature_2015} and more recently graphene-like nanostructures\cite{li_electrically_2019,li_single_2019,Tuovinen_Time_2019,li_uncovering_2020,su_atomically_2020,zheng_engineering_2020,Allerdt_Many_2020,Friedrich_Addressing_2022} and molecular chains\cite{wackerlin_role_2022,zhao_quantum_2023}.
	Finally, junctions comprising strongly correlated nanostructures can be approximately mapped onto embedded impurity models \cite{Florens_Nanoscale_2007, Jacob_Kondo_2009, Jacob_Dynamical_2010, Turkowski_Dynamical_2012, Ferrer_Gollum_2014, Schuler_Realistic_2017, Kurth_Nonequilibrium_2019}.
	In all these cases, a controlled theoretical treatment is challenging, because numerical methods able to reliably access the correlated regime of nonequilibrium quantum impurity models are typically either limited in the level of detail in their description of the baths, especially out of equilibrium; or limited in accuracy by the need to go to high perturbation order (see Supplementary Material).
    As a result, theoretical work focuses on aspects of the weakly correlated regime \cite{Solomon_Exploring_2010, Bouatou_Visualizing_2022, Gao_Tunable_2023, Leitherer_Electromigration_2023} or is confined to single- or few-channel correlated transport \cite{Ujsaghy_Theory_2000, Agam_Projecting_2001,Bulka_Fano_2001,Hofstetter_Kondo_2001,torio_kondo_2002,Luo_Fano_2004,DaSilva_Transport_2008,Heidrich_Transport_2009,Dilullo_Molecular_2012}.
	
    Here, we leverage advances in the inchworm Quantum Monte Carlo (iQMC) method \cite{Cohen_Taming_2015, Erpenbeck_Quantum_2023} to overcome these theoretical challenges and simulate a simple model describing a correlated impurity embedded in a metallic nanosheet within a junction, and driven away from equilibrium.
    We explore correlation effects in the overall current and current density, show how microscopic manipulations of the system's structure can manifest in a system-wide reshaping of the flow of electronic currents, and discuss the experimental signatures of strong correlation physics.

    \paragraph{The system:}
    
\begin{figure}
            \centering
            \includegraphics[width=8.6cm]{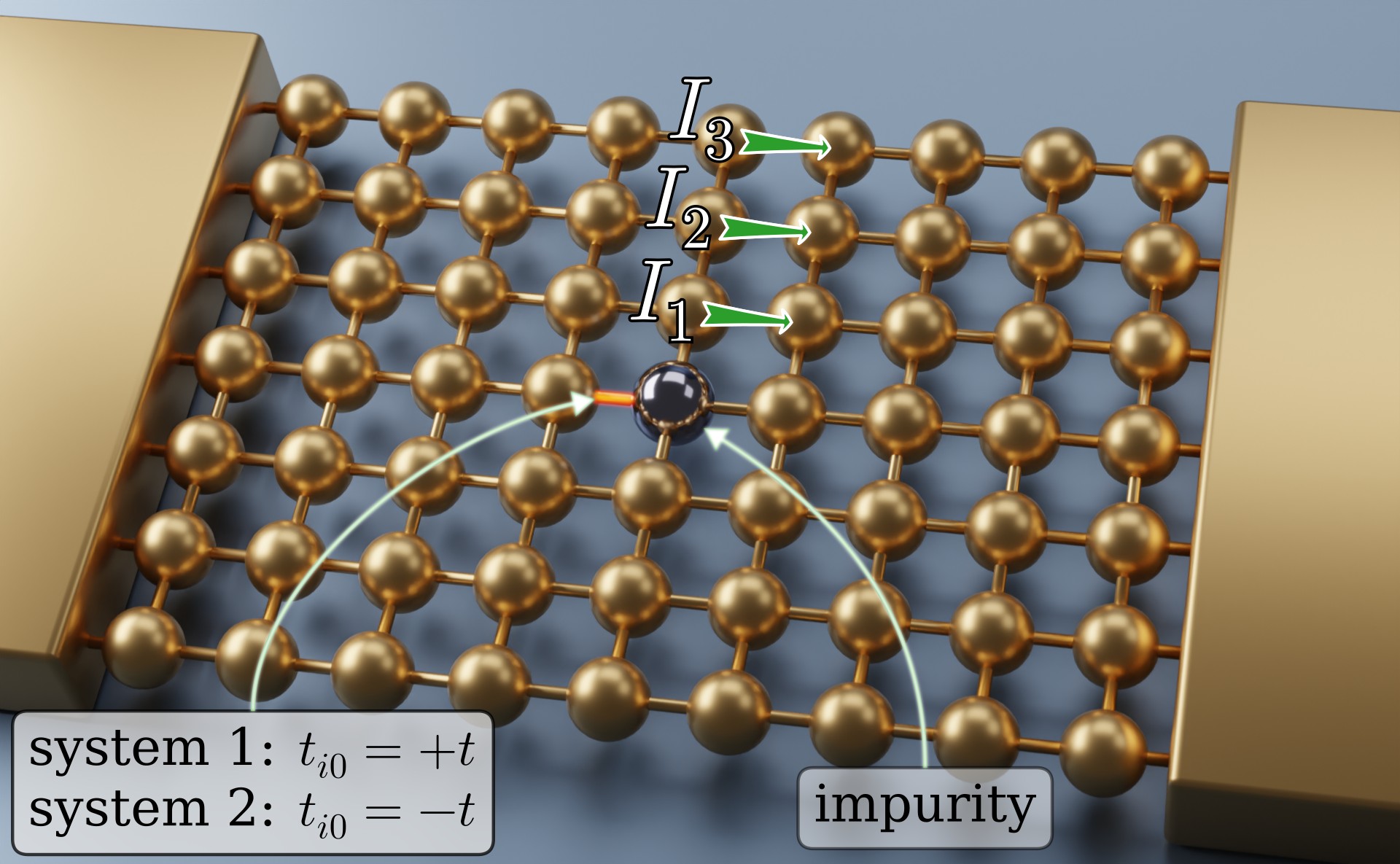}
            \caption{The system:
                     noninteracting lattice with an interacting impurity (dark sphere) at its center. 
                     Other spheres are lattice sites connected by nearest-neighbor hopping.
                     Two macroscopic leads are coupled to the right and left boundaries of the lattice.
                     The highlighted coupling between the impurity and its left neighbor takes one of two possible values $\pm t$.
                     Several bond currents in the cross section of the lattice containing the impurity (green arrows) are discussed below. 
                     }
            \label{fig:system}
        \end{figure}

        Our model comprises a set of sites---or localized orbitals---forming a two-dimensional square lattice.
        To be concrete, we will consider the $9\times7$ case (i.e. a total of 63 spin-degenerate orbitals).
        A single site at the center of the lattice is referred to as the impurity, and will contain the only two-body interaction in the system.
        The left and right boundaries of the lattice are coupled to extended noninteracting electrodes (see Fig.~\ref{fig:system}).
        The Hamiltonian includes terms describing the impurity, the lattice, the leads, and impurity--lattice as well as lattice--lead coupling,
        \begin{eqnarray}
            H &=& H_{\mathrm{imp}} + H_{\mathrm{lat}} + H_{\mathrm{imp-lat}} + H_{\mathrm{leads}} + H_{\mathrm{leads-lat}}  . \label{eq:H_full} 
        \end{eqnarray}
        The impurity Hamiltonian is
        \begin{eqnarray}
            H_{\mathrm{imp}}	&=& 	\sum_{\sigma} \epsilon_0 d_{0\sigma}^\dagger d_{0\sigma} + U d_{0\uparrow}^\dagger d_{0\uparrow} d_{0\downarrow}^\dagger d_{0\downarrow}, \label{eq:H_I} 
        \end{eqnarray}
        with spin index $\sigma\in\lbrace\uparrow,\downarrow\rbrace$, and $0$ the index of the impurity site.
        $\epsilon_0$ is the impurity occupation energy, and $U$ is the Coulomb interaction strength.
        $d_{0\sigma}^{\left( \dagger\right)}$ are electron annihilation (creation) operators at the impurity site.
        The lattice is described by
        \begin{eqnarray}
            H_{\mathrm{lat}} &=& - t \sum_{\sigma}\sum_{\braket{i, j}\atop i,j\neq 0} d^\dagger_{i\sigma} d_{j\sigma} ,
        \end{eqnarray}
        where $\braket{i,j}$ indicates two neighboring lattice sites. 
        $t$ is the hopping strength and $d_{i\sigma}^{\left( \dagger\right)}$ are the annihilation(creation) operators for spin $\sigma$ at site $i$. 
        The lead Hamiltonian assumes the form 
        \begin{eqnarray}
            H_{\mathrm{leads}} &=& \sum_{\sigma}\sum_{k\in\lbrace\mathrm{L/R}\rbrace} \epsilon_{k} c^\dagger_{k\sigma} c_{k\sigma}.
        \end{eqnarray}
        Here, $\epsilon_{k}$ is the energy of state $k$ in the left/right (L/R) lead with annihilation (creation) operator $c_{k\sigma}^{\left( \dagger\right)}$.
        The coupling between the leads and lattice is given by
        \begin{eqnarray}
            H_{\mathrm{leads-lat}} &=& \sum_{\sigma, i}\sum_{k\in\lbrace\mathrm{L/R}\rbrace} V_{ik} c^\dagger_{k\sigma} d_{i\sigma} + \mathrm{h.c.} ,
        \end{eqnarray}
        with coupling strengths $V_{ik}$.
        Only sites at the left at right boundaries couple to the left and right lead.
        The coupling between the impurity and the lattice is 
        \begin{eqnarray}
            H_{\mathrm{imp-lat}} &=& \sum_{\sigma,i\in N_0} -t_{i0} d^\dagger_{i\sigma} d_{0\sigma} + \mathrm{h.c.},
        \end{eqnarray}
        where $N_0$ are the sites adjacent to the impurity, coupled to the impurity with coupling strength $t_{i0}$.
        We consider two cases that differ only by the phase of one local term in the Hamiltonian.
        In ``system 1'', the impurity is coupled to all neighbors with the same strength as the other lattice sites, $t_{i0} = t$ $\forall i \in N_0$. In ``system 2'', the impurity's coupling to its left neighbor (highlighted link in Fig.~\ref{fig:system}) has the same magnitude but opposite sign, and is therefore equal to $-t$ (see lightly shaded box in Fig.~\ref{fig:system}).
        We propose this as a minimal example of an atomically local experimental intervention, like the introduction of a single, charge-neutral interstitial atom between the impurity lattice site and its nearest neighbor.
        Since the Hamiltonian is spin independent, we henceforth drop all spin indices.\\

    \paragraph{Embedding scheme and numerical solution:}
        We calculate the Green's function (GF) of the system using a procedure analogous to the cavity method in dynamical mean field theory \cite{georges_dynamical_1996}.
        The quadratic cavity Hamiltonian, 
        $H_{\mathrm{cavity}} = H_{\mathrm{lattice}} + H_{\mathrm{leads}} + H_{\mathrm{leads-lattice}}$, 
        represents the system without the impurity site $0$.
        The exact GF of $H_{\mathrm{cavity}}$, $G_{\mathrm{cavity}}(\epsilon)$, is given by \cite{Haug_Qauntum_2008}:
        \begin{eqnarray}
            G_{\mathrm{cavity}}^{r/a}(\epsilon) &=& \left[\epsilon - H_{\mathrm{cavity}} - \Sigma_{\mathrm{leads}}^{r/a}(\epsilon)\right]^{-1} , \\ 
            G_{\mathrm{cavity}}^\lessgtr(\epsilon)    &=& G_{\mathrm{cavity}}^r(\epsilon)\Sigma_{\mathrm{leads}}^\lessgtr(\epsilon) G_{\mathrm{cavity}}^a(\epsilon) .
        \end{eqnarray}
        \label{eq:cavity_GF}
    	$\Sigma_{\mathrm{leads}}^{r/a/\lessgtr}$ is the self-energy of the leads, which are modeled in the wide band limit, $\left[\Gamma_\mathrm{L/R}(\epsilon)\right]_{ij}=\Gamma\delta_{ij}$ for the left/right end sites, and zero otherwise. 
    	In this limit, the non-zero elements of the self-energy are  
    	$\Sigma_{\mathrm{leads}}^r(\epsilon) = -i \Gamma$, 
    	and
    	$\Sigma_{\mathrm{leads}}^\lessgtr(\epsilon) = \pm i \left(\Gamma f_\mathrm{L}^\pm(\epsilon) + \Gamma f_\mathrm{R}^\pm(\epsilon)\right)$.
    	The current scales with the coupling strength $\Gamma$, which is the energy unit.
        $f_\mathrm{L/R}^+(\epsilon)$ is the Fermi function of the left/right lead  and $f_\mathrm{L/R}^-(\epsilon) = 1-f_\mathrm{L/R}^+(\epsilon)$. 
        The bias voltage $\Phi = \mu_{\mathrm{L}} - \mu_{\mathrm{R}}$ is realized by shifting the chemical potentials symmetrically, $\mu_{\mathrm{L}} = - \mu_{\mathrm{R}}$.

        Given the cavity GF, we construct an exact hybridization self-energy for the embedded impurity,
        \begin{eqnarray}
            \Delta^\lessgtr(\epsilon) &=& \sum_{i, j \in N_0} t_{i0}t_{j0}^* \left[G^\lessgtr_{\mathrm{cavity}}(\epsilon)\right]_{ij} .
        \end{eqnarray}
        This fully describes the effect of the bath on the impurity GF, and the calculation of $G_{\mathrm{imp}}$ takes the form of a standard quantum impurity problem, albeit with an intrinsically nonequilibrium bath action.
        Different approximations and numerically exact methods could solve this problem \cite{Wilson_renormalization_1975, Grewe_Diagrammatic_1981, Kuramoto_Self_1983, Bickers_Review_1987, Tanimura_Time_1989, Pruschke_Anderson_1989, Keiter_NCA_1990, Anders_Perturbational_1994, Anders_Beyond_1995, Segal_Electron_2000, Haule_Anderson_2001, White_Real_2004, Schollwock_density-matrix_2005, Tanimura_Stochastic_2006, Anders_Steady-State_2008, Bulla_Numerical_2008, Jin_Exact_2008, Grewe_Conserving_2008, Myohanen_Kadanoff_2009, Balzer_Nonequilibrium_2009, Zheng_Complex_2009, Schiro_Real_2009, Segal_Numerically_2010, Schiro_Time_2010, Schollwock_density-matrix_2011, Li_Hierarchical_2012, Hartle_Decoherence_2013, Tuovinen_Time_2014, Hartle_Transport_2015,schwarz_lindblad-driven_2016, Erpenbeck_Extending_2018, Erpenbeck_Hierarchical_2019, Mundinar_Iterative_2019, Allerdt_Numerically_2019,Lode_Colloquium_2020, Tanimura_Numerically_2020, nuseler_efficient_2020,lotem_renormalized_2020, Erpenbeck_Revealing_2021, Erpenbeck_Resolving_2021, purkayastha_periodically_2021, Cirac_Matrix_2021, Nunez_Learning_2022, Erpenbeck_Tensor_2023,cygorek_simulation_2022,ng_real-time_2023,thoenniss_efficient_2023}, but relatively few are capable of producing reliable results in the strongly correlated nonequilibrium regime in which we are interested (see Supplementary Materials). 
        We employ the recently developed steady-state inchworm Quantum Monte Carlo method \cite{Erpenbeck_Quantum_2023}.
        This is a continuous-time quantum Monte Carlo method \cite{Gull_Continuous-Time_2011} based on an expansion in the impurity--bath hybridization \cite{Keiter_Perturbation_1970, Pruschke_Anderson_1989, Werner_Continuous-Time_2006, Werner_Hybridization_2006, Haule_Quantum_2007, Muhlbacher_Real-Time_2008, Gull_Bold-line_2010, Cohen_Greens_2014, Cohen_Greens_2014_1}.
        The inchworm technique\cite{Cohen_Taming_2015, Antipov_Currents_2017, Chen_Inchworm_2017, Chen_Inchworm_2017_2, Cai_Inchworm_2020, Cai_Numerical_2020, Cai_Fast_2022, Boag_Inclusion-Exclusion_2018, Ridley_Numerically_2018, Ridley_Lead_2019, Ridley_Numerically_2019, Eidelstein_Multiorbital_2020, Kim_Pseudoparticle_2022, Li_Interaction-Expansion_2022, Pollock_Reduced_2022, Dong_Quantum_2017, Krivenko_Dynamics_2019, Kleinhenz_Dynamic_2020, Kleinhenz_Kondo_2022} allows for a formulation that can overcome the dynamical sign problem at least in some cases.

        Given $G_{\mathrm{imp}}$, the self-energy of the entire system can be reconstructed,
        \begin{eqnarray}
            \Sigma_{U}^{r/a}(\epsilon) &=& 
                                            \epsilon - \epsilon_0 - \Sigma_{\mathrm{embed}}^{r/a}(\epsilon) - 1/G_{\mathrm{imp}^{r/a}(\epsilon)} ,
                                            \\
            \Sigma_{U}^\lessgtr(\epsilon) &=& 
                                           G_{\mathrm{imp}}^\lessgtr(\epsilon) / G_{\mathrm{imp}}^r(\epsilon)G_{\mathrm{imp}}^a(\epsilon) - \Sigma_{\mathrm{embed}}^\lessgtr(\epsilon) .
        \end{eqnarray}
        Since the Coulomb interaction only acts at the impurity site $0$, the full lattice GF is then exactly given by
        \begin{eqnarray}
            G^r(\epsilon) &=& \left[\epsilon - H - \Sigma_{\mathrm{leads}}^r(\epsilon) - \Sigma_U^r(\epsilon) \right]^{-1} , \\ 
            G^\lessgtr(\epsilon)    &=& G^r(\epsilon) \left[ \Sigma_{\mathrm{leads}}^\lessgtr(\epsilon) - \Sigma_U^\lessgtr(\epsilon) \right] G^a(\epsilon) ,
        \end{eqnarray}
        where $\left[\Sigma_U^{r\lessgtr}(\epsilon)\right]_{ij} = \delta_{i0}\delta_{j0} \Sigma_U^{r\lessgtr}(\epsilon)$.
        \\

    \paragraph{Observables of interest:}
        We study the electronic populations of the lattice sites, 
        $n_i = \int G^<_{ii}(\epsilon) \ \frac{d\epsilon}{2\pi}$,
        and the bond currents between two adjacent lattice sites $i$ and $j$,
        $I_{ij}  = \int H_{ij} G^<_{ij}(\epsilon) - H_{ji} G^<_{ji} \ \frac{d\epsilon}{\pi}$ \cite{Cresti_Keldysh_2003}.
        The total current $I$ across the system is calculated using the Meir--Wingreen formula \cite{Meir_Landauer_1992},
        $\int \mathrm{Tr}\left\lbrace \Sigma_\mathrm{L}^<(\epsilon) G^>(\epsilon) - \Sigma_\mathrm{L}^>(\epsilon) G^<(\epsilon)  \right\rbrace \frac{d\epsilon}{\pi}$.

\paragraph{Transport in the presence of an impurity:}

        \begin{figure*}[htb!]
            \raggedright \hspace*{0.65cm} System 1, $\Phi=0.33\Gamma$ \hspace*{1.07cm} System 1, $\Phi=1.33\Gamma$ \hspace*{1.07cm} System 1, $\Phi=2.66\Gamma$
            \vspace{-0.25cm} \\
            \centering
            \includegraphics[width=0.3\textwidth]{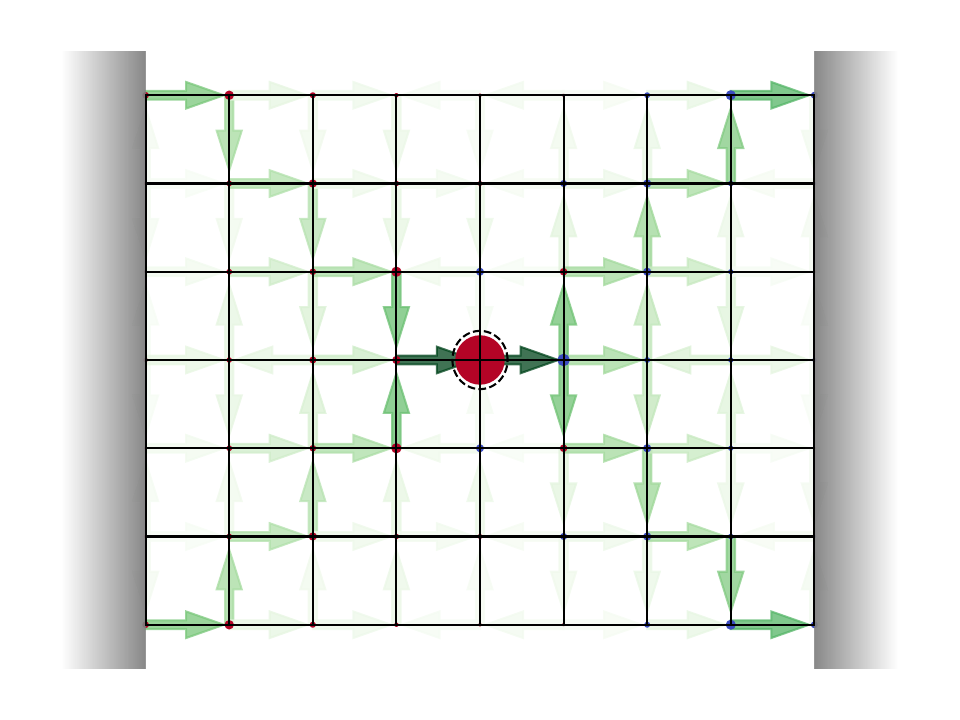}
            \includegraphics[width=0.3\textwidth]{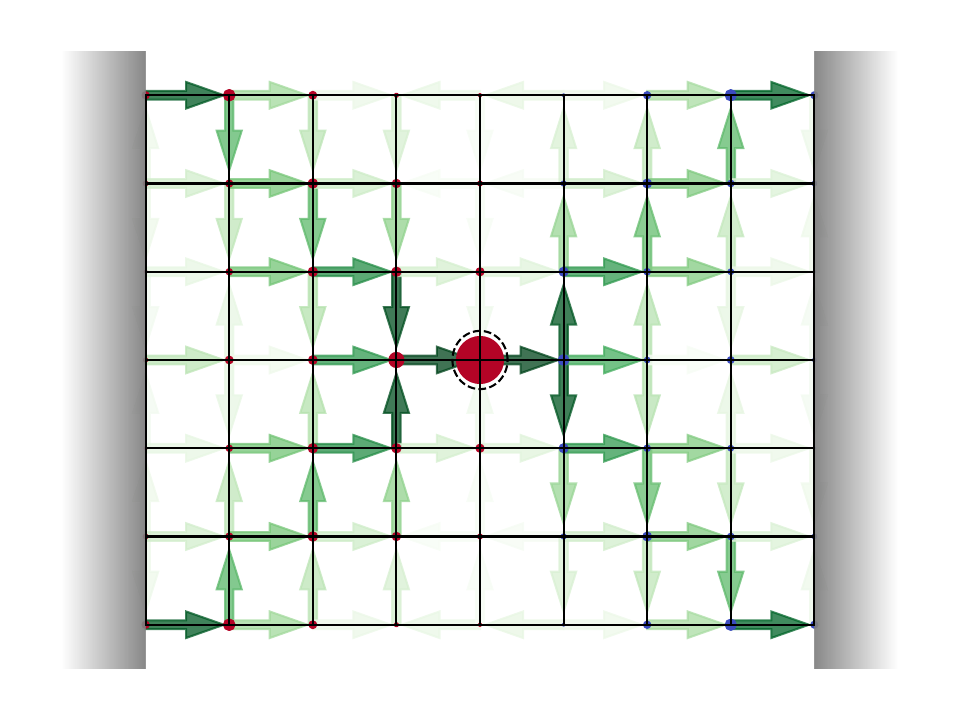}
            \includegraphics[width=0.3\textwidth]{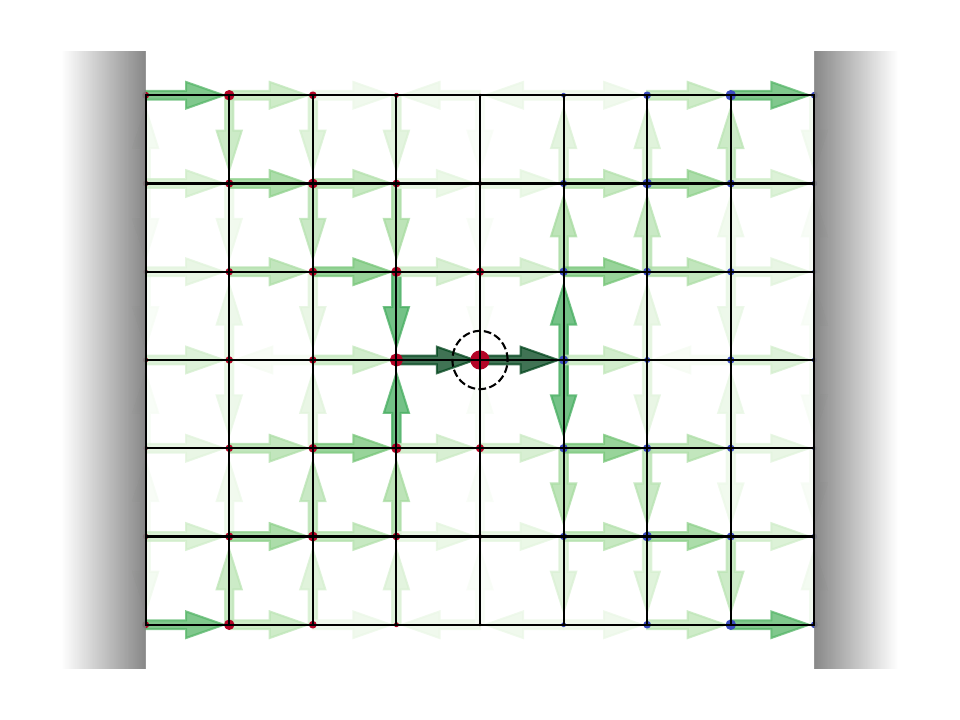}
            \hspace{-0.5cm}
            \includegraphics[width=0.083\textwidth]{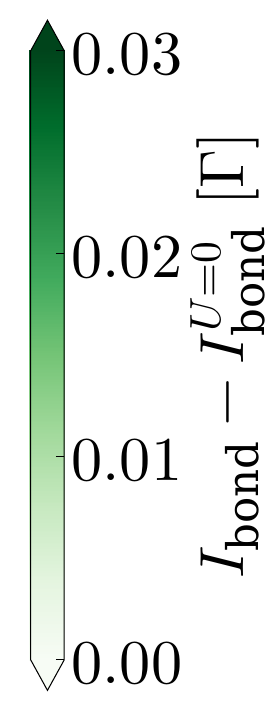}\\
            \raggedright \hspace*{0.65cm} System 2, $\Phi=0.33\Gamma$ \hspace*{1.07cm} System 2, $\Phi=1.33\Gamma$ \hspace*{1.07cm} System 2, $\Phi=2.66\Gamma$
            \vspace{-0.25cm} \\
            \centering
            \includegraphics[width=0.3\textwidth]{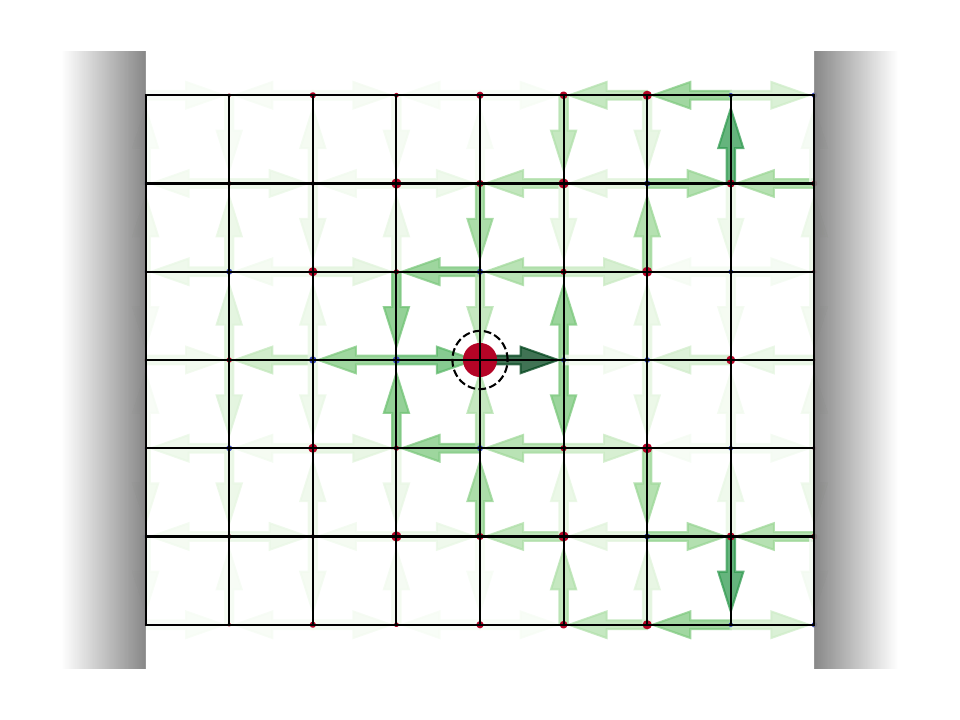}
            \includegraphics[width=0.3\textwidth]{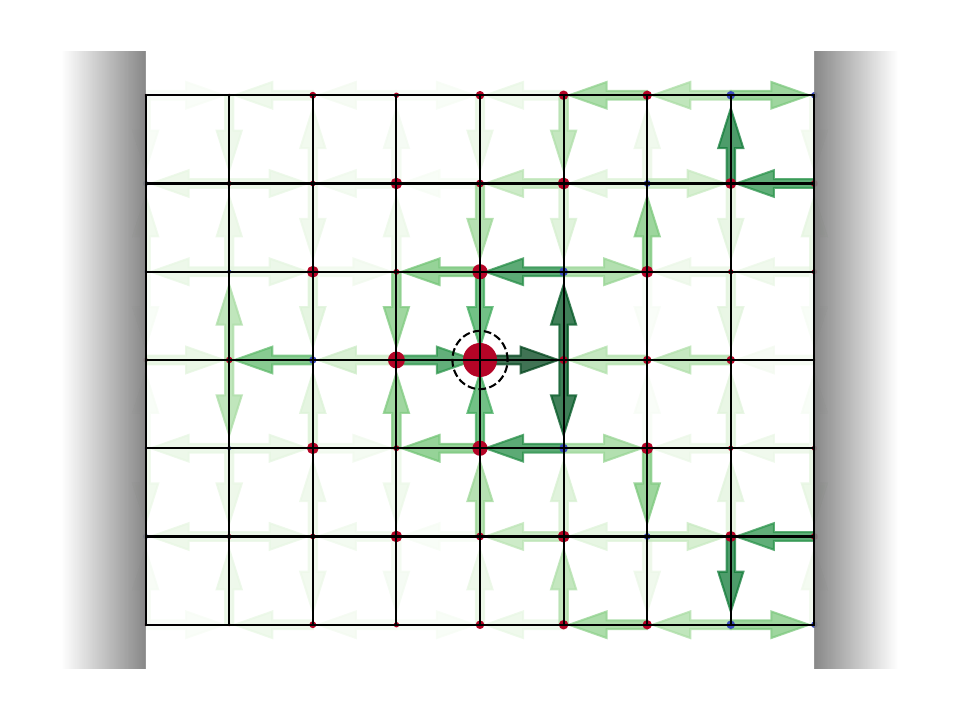}
            \includegraphics[width=0.3\textwidth]{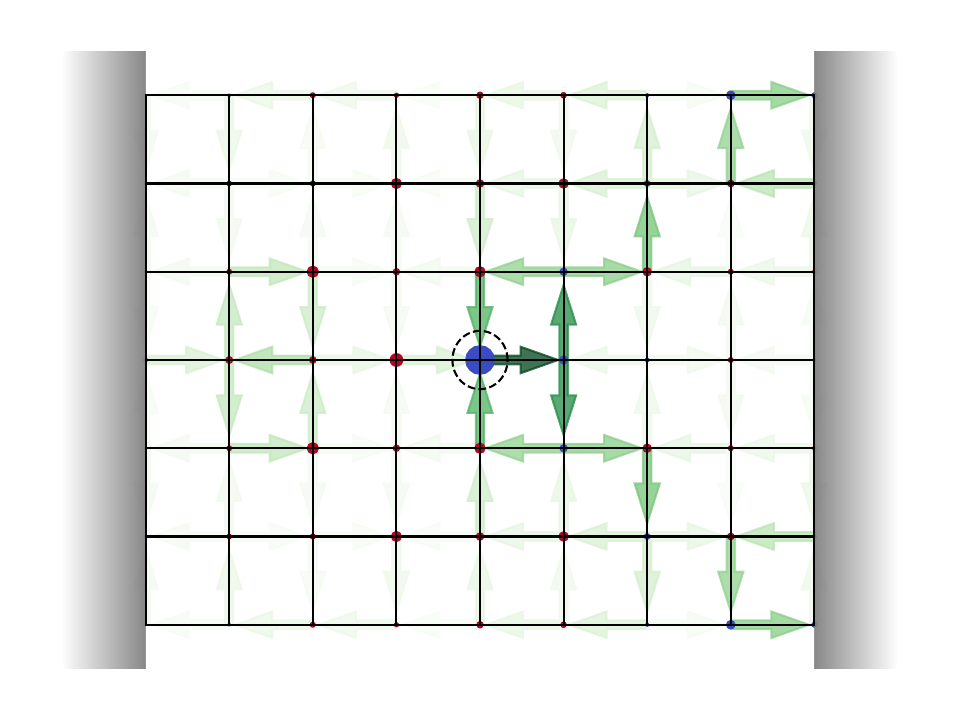}
            \hspace{-0.5cm}
            \includegraphics[width=0.084\textwidth]{{colorbar}.pdf}\\
            \caption{
                    Difference of current and charge distribution with respect to the noninteracting system for three representative bias voltages (left to right) for system 1 (top) and system 2 (bottom) at temperature $T=0.167\Gamma$, $t=\Gamma/3$, and $U=2\Gamma$. 
                    Green arrows: difference between the interacting bond current, $I_\mathrm{bond}$, and the bond current in the noninteracting case, $I_\mathrm{bond}^{U=0}$. 
                    As the overall current flows from left to right ($\mu_\mathrm{L} = \Phi/2 = -\mu_\mathrm{R}$), arrows pointing from left to right indicate an increased current, arrows pointing from right to left indicate a reduced current. 
                    Red/blue circles: less/more charge accumulated as compared to the system without interaction, with the size of the circle indicating the amount of charge. 
                    Dashed circle: position of the impurity.
                    }
            \label{fig:transport_map}
        \end{figure*}

    We show how the Kondo mechanism, by modifying atomic-scale features, can control transport throughout the lattice. Specifically, adjusting the phase of the hopping parameter between the impurity and its adjacent site distinguishes system 1 from system 2. We expect that this $\pi$-phase shift leads to opposing interference effects between the Kondo channel and the normal channels through the lattice, constructive in system 1 and destructive in system 2. Notably, changes in hopping phase/magnitude and shifts in energy levels/capacitance at different lattice points are also viable options (e.g. from proximity to an STM tip). The choice depends on the experimental setup, and our theoretical framework allows for a variety of scenarios.

        To understand the influence of the impurity on the current, we consider how transport properties change at low temperatures (here $T=\Gamma/6$), when the interaction strength is taken from the noninteracting limit, $U=0$, to $U=2\Gamma$.
        We set $\epsilon_0 = -\Gamma$.
        The two cases are shown separately in the Supplementary Materials. Fig.~\ref{fig:transport_map} shows the difference in bond current and charge between the interacting and noninteracting case.
        Top panels are for system 1,  bottom panels for system 2.
        The bias voltage increases from left to right.
        
        In system 1 (top panels), the interaction enhances transport and channels the current flow through the impurity, resulting in an overall increase in conductivity.
        The channeling is visible in the ``funnel'' of arrows apparent in all panels, which reveals that turning on the interaction displaces current from the entire cross-section of the lattice towards the highly conductive impurity site.
        Contrary to this, in system 2, the interaction suppresses the current parallel to it, and creates a wake-like structure to the right of the impurity.
        These phenomena gradually diminish with increasing bias voltage, but survive well past the linear regime.
        The contrasting current flow patterns in systems 1 and 2 demonstrate that flow at the nanoscale can be shaped by controlling the phase of a single, atomic-scale hopping element.
        The magnitude of the shaping effect is sizable in the correlated regime: bond currents passing through the impurity are enhanced by more than 100\%, while currents parallel to the impurity can be suppressed by more than 10\%.
        For the charge distribution, the largest changes occur at the impurity site (red circles in Fig.~\ref{fig:transport_map} and are related to the on-site energy at the impurity.
        \\

    \paragraph{Transport and the Kondo effect:}
        To link current shaping and the Kondo effect, we examine temperature-dependent currents in systems 1 and 2. As a rough indicator for the onset of Kondo physics, the Kondo temperature based on the Bethe ansatz \cite{Hewson_Kondo_1997} is
        $T_K \approx U \sqrt{\Delta/2U} e^{-\pi U/8 \Delta + \pi\Delta/2U} \approx 1.25\Gamma$  with 
        $\Delta = \pi\rho_0|t|^2$ and $\rho_0\approx3.2/\Gamma$ being the average spectral density from all neighboring sites.
        
        \begin{figure}[htb!]
            \centering
            \vspace*{-0.3cm}
            \hspace*{-9cm}
            a)\\             
            \includegraphics{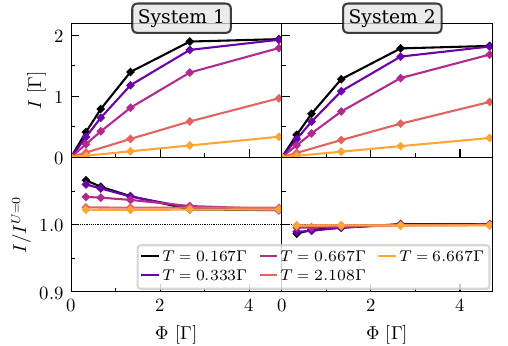}\\
            \centering
            \vspace*{-0.3cm}
            \hspace*{-9cm}
            b)\\       
            \includegraphics{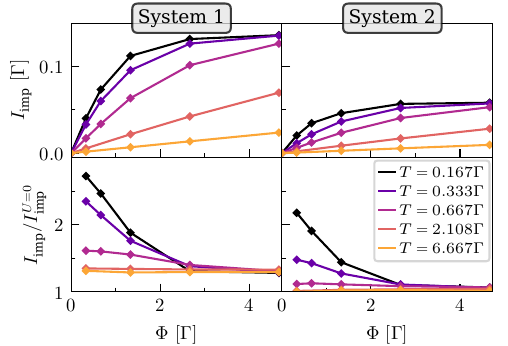}\\
            \centering 
            \vspace*{-0.3cm}
            \hspace*{-9cm}
            c)\\       
            \includegraphics{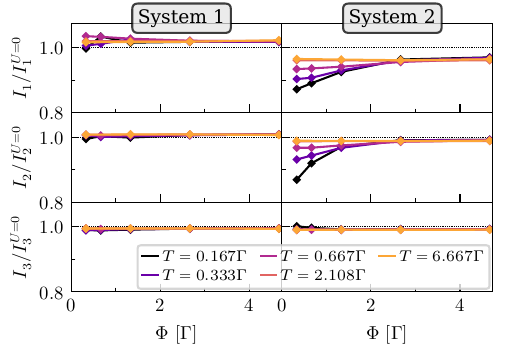}
            \caption{
                Currents and current ratios for system 1 (left) and system 2 (right) at different temperatures.
                a: current--voltage characteristics of the entire system (top), ratio of this current and its noninteracting counterpart (bottom).
                b: current flowing through the impurity (top), ratio of this current and its noninteracting counterpart (bottom).
                c: ratio of the currents $I_1$, $I_2$, and $I_3$ flowing parallel to the impurity with respect to their counterpart for $U=0$.
                The bonds are as indicated in Fig.~\ref{fig:system}.
                }
            \label{fig:Kondo}
        \end{figure}
        Fig.~\ref{fig:Kondo} presents results for the interacting total current and the bond currents at $U=2\Gamma$, and the ratio between these interacting currents and the corresponding noninteracting currents at $U=0$.
        We'll refer to the latter as current ratios (CRs).
        System 1 and 2 are plotted in the left and right panels at a sequence of representative temperatures below and above the Kondo temperature.
        
        Fig.~\ref{fig:Kondo}a shows the total current (top) and total CR (bottom).
        The current increases with bias voltage in both systems, eventually plateauing at its maximal value (only visible at low temperatures here).
        System 1 exhibits a slightly better overall conductivity than system 2.
        The CR reveals that the interaction enhances the total current in system 1 by a small, temperature-and-voltage-independent factor of ~2\%, and by an additional ~5\% at low temperatures and voltages.
        While the small constant term in the enhancement is a mean field effect, the larger temperature and voltage dependent term appears only in the Kondo regime and can be attributed to Kondo physics.
        In system 2, the CR is ~1 at high temperature and voltage, indicating that there is almost no mean field effect.
        A smaller suppression of ~1\% is visible in the Kondo regime.

        Fig.~\ref{fig:Kondo}b shows the bond current flowing into the impurity from the left.
        In system 1, this current is approximately two to three times larger than in system 2.
        The CR (bottom panels of Fig.~\ref{fig:Kondo}b) shows that both systems exhibit a more than twofold enhancement of current through the impurity in the Kondo regime.
        The enhancement is stronger for system 1, where the total current was also enhanced; but remains in system 2, where it was suppressed.
        The enhancement is consistent with the increased conductivity observed for single-channel quantum dots in the Kondo regime \cite{Goldhaber_Kondo_1998, Borzenets_Observation_2020}.

        Fig.~\ref{fig:Kondo} shows the CRs for currents flowing parallel to the impurity (represented by $I_1$, $I_2$, and $I_3$ as given in Fig.~\ref{fig:system}). 
        In system 1, these CRs remain largely temperature and voltage-independent, aside from a minor increase near the impurity site's nearest neighbor, primarily due to mean field behavior.
        The enhancement of the total current is therefore mainly due to the opening of the Kondo channel.
        In system 2, the currents parallel to the impurity are suppressed by more than $\sim$10\%.
        This suppression consists of a small mean field reduction of $\sim$2\% appearing only at the nearest neighbor, and of a larger, more spatially widespread suppression in the Kondo regime.
        This suppression in system 2 suggests scattering off a Kondo cloud formed around the impurity.
        The negative tunneling element introduces a negative phase shift in this scattering, resulting in antiresonant transport.
        More support for this interpretation can be found by looking at the local spectral function, which can be accessed within linear response in transport setups based on STMs (see Supplementary Materials).\\

    \paragraph{Identifying correlation effects at the atomic level:}
        
        Below, we present a method for identifying and characterizing correlation effects in the current, solely using interacting currents and a noninteracting reference system. This approach effectively discerns system differences, such as between system 1 and system 2, without prior knowledge of impurity or hopping parameter modifications. 
        This addresses experimentalists, where the interacting current is experimentally available, but the detailed structure of the system is unknown. 
        Our noninteracting reference system, labeled ``ref'' captures the system's overall structure while lacking impurity-specific details. Specifically, our reference system uses Eq.~(\ref{eq:H_full}) with parameters $\epsilon_0=U=0$ and $t_{i0}=t\ \forall i\in N_i$.

        Consider the total interacting current $I(\Phi, T)$ at bias $\Phi$ and temperature $T$. 
        Beyond the correlated regime, this current can be accurately described by a mean field model, suggesting that current in the reference system could have a similar dependence on bias voltage and temperature in this regime.
        To facilitate a comparison between the interacting and noninteracting systems, we define $\tilde I(\Phi, T) = I(\Phi, T) / I(\Phi, T_{\mathrm{max}})$, accounting for conductance differences between the interacting and reference system.
        Here, $T_{\mathrm{max}}$ represents a high temperature beyond the correlated regime.
        We expect that $\tilde I(\Phi, T)/\tilde I^{\mathrm{ref}}(\Phi, T) \simeq 1$ outside of the correlated regime.
        Any deviation of this ratio from unity is indicative of either correlation effects, or a severe mismatch between the reference and physical system.
        The latter is less likely, because the reference system is noninteracting, and therefore unlikely to feature a dramatic dependence on temperature.
        
        \begin{figure}[tb!]
            \centering
            \includegraphics{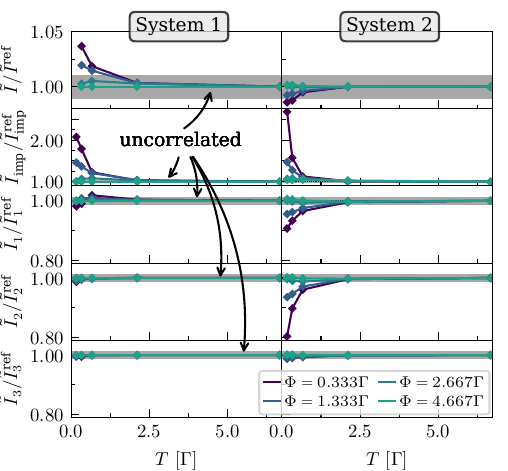}
            \caption{
                    Temperature dependence of the different currents of the interacting system $\tilde I$, divided by their noninteracting counterparts of the reference system $\tilde I^{\mathrm{ref}}$.
                    Left panels: system 1. Right panels: system 2.
                    Top two panels display to total current and the current trough the impurity, respectively.
                    Bottom three panels show the currents parallel to the impurity, $I_1$, $I_2$, and $I_3$, as indicated in Fig.~\ref{fig:system}.
                    The gray bar highlights the regime where the currents essentially behave like noninteracting currents.
                    }
            \label{fig:temperature}
        \end{figure}
        
        The top panels of Fig.~\ref{fig:temperature} show the ratios $\tilde I/\tilde I^{\mathrm{ref}}$ for system 1 and 2 as a function of temperature for representative bias voltages. 
        At low bias voltages and temperatures, the data significantly deviates from unity (as indicated by the shaded gray area), serving as an indicator of correlation effects.
        While the reference system possesses no prior knowledge of the distinctions between system 1 and 2, the results accurately predict an increased current due to correlations in system 1, and a mild suppression for system 2.

        We now extend our analysis to the individual bond currents. 
        The second row of panels in Fig.~\ref{fig:temperature} shows $\tilde I_{\mathrm{imp}}$ pertaining the current flowing through the impurity.
        Here, $\tilde I_{\mathrm{imp}}/\tilde I_{\mathrm{imp}}^{\mathrm{ref}}$ significantly exceeds unity at low bias voltages and temperatures.
        This captures the fact that the Kondo effect enhances transport through the impurity site in both system 1 and system 2.
        The lower three rows of panels in Fig.~\ref{fig:temperature} expand our analysis to the currents flowing parallel to the impurity.
        The data correctly reflects that the current parallel to the impurity for system 1 remains unaffected by correlations, while system 2 experiences a suppression of current due to the Kondo effect.

        We emphasize that this strategy for identifying correlation effects in the current is remarkably robust against the choice of the noninteracting reference system.
        We have repeated the same analysis with several different noninteracting reference systems, yielding qualitatively similar results (see Supplementary Material).

\paragraph{Conclusion:}

	To conclude, we demonstrated how an interacting impurity embedded in a nanoscale system, in conjunction with temperature and an applied bias voltage, can effectively control current flow from atomic to mesoscopic length scales.
	Leveraging recent advancements in the iQMC method, we presented a versatile methodology capable of describing interacting impurities in complex environments and applicable to a wide variety of scenarios.
    Applying this approach to a two-dimensional metallic nanosheet hosting an impurity in its center and connecting two macroscopic leads, 
    we reconstructed the electronic flow through the system, showcasing that the current can be significantly modified by temperature and by adjusting the coupling between individual atoms.
    This allowed us to derive a scheme capable of identifying correlations and probing the phase of individual bonds in the system based on the interacting current, all without requiring prior knowledge of the atomic details.
    
    Overall, our study has demonstrated the ability to utilize correlations in order to guide the current on an atomic scale across a mesoscopic system, effectively bridging different length scales. 
    The methodology employed here holds significant promise for investigating the intricate dynamics of current flow, a crucial aspect in the design of nanoelectronic devices and sensors. 
    Future extensions of this work may explore systems with multiple impurities, potentially influencing one another and amplifying the observed effects, thus providing even greater control over current flow across various length scales.

\begin{acknowledgement}
    A.E.~was funded by the Deutsche Forschungsgemeinschaft (DFG, German Research Foundation) -- 453644843.
    E.G.~was supported by the U.S. Department of Energy, Office of Science, Office of Advanced Scientific Computing Research and Office of Basic Energy Sciences, Scientific Discovery through Advanced Computing (SciDAC) program under Award Number DE‐SC0022088.
    This research used resources of the National Energy Research Scientific Computing Center, a DOE Office of Science User Facility supported by the Office of Science of the U.S. Department of Energy
    under Contract No. DE-AC02-05CH11231 using NERSC award BES-ERCAP0021805.
    G.C.~acknowledges support by the Israel Science Foundation (Grants No.~2902/21 and 218/19) and by the PAZY foundation (Grant No.~308/19).
\end{acknowledgement}

\bibliography{Bib}

\providecommand{\latin}[1]{#1}
\makeatletter
\providecommand{\doi}
  {\begingroup\let\do\@makeother\dospecials
  \catcode`\{=1 \catcode`\}=2 \doi@aux}
\providecommand{\doi@aux}[1]{\endgroup\texttt{#1}}
\makeatother
\providecommand*\mcitethebibliography{\thebibliography}
\csname @ifundefined\endcsname{endmcitethebibliography}
  {\let\endmcitethebibliography\endthebibliography}{}
\begin{mcitethebibliography}{172}
\providecommand*\natexlab[1]{#1}
\providecommand*\mciteSetBstSublistMode[1]{}
\providecommand*\mciteSetBstMaxWidthForm[2]{}
\providecommand*\mciteBstWouldAddEndPuncttrue
  {\def\EndOfBibitem{\unskip.}}
\providecommand*\mciteBstWouldAddEndPunctfalse
  {\let\EndOfBibitem\relax}
\providecommand*\mciteSetBstMidEndSepPunct[3]{}
\providecommand*\mciteSetBstSublistLabelBeginEnd[3]{}
\providecommand*\EndOfBibitem{}
\mciteSetBstSublistMode{f}
\mciteSetBstMaxWidthForm{subitem}{(\alph{mcitesubitemcount})}
\mciteSetBstSublistLabelBeginEnd
  {\mcitemaxwidthsubitemform\space}
  {\relax}
  {\relax}

\bibitem[Datta(1997)]{datta_electronic_1997}
Datta,~S. \emph{Electronic {{Transport}} in {{Mesoscopic Systems}}}; {Cambridge
  UniversityPress}, 1997\relax
\mciteBstWouldAddEndPuncttrue
\mciteSetBstMidEndSepPunct{\mcitedefaultmidpunct}
{\mcitedefaultendpunct}{\mcitedefaultseppunct}\relax
\EndOfBibitem
\bibitem[Kouwenhoven \latin{et~al.}(1997)Kouwenhoven, Sch{\"o}n, and
  Sohn]{kouwenhoven_introduction_1997}
Kouwenhoven,~L.~P.; Sch{\"o}n,~G.; Sohn,~L.~L. In \emph{Mesoscopic {{Electron
  Transport}}}; Sohn,~L.~L., Kouwenhoven,~L.~P., Sch{\"o}n,~G., Eds.; {{NATO
  ASI Series}}; {Springer Netherlands}: {Dordrecht}, 1997; pp 1--44\relax
\mciteBstWouldAddEndPuncttrue
\mciteSetBstMidEndSepPunct{\mcitedefaultmidpunct}
{\mcitedefaultendpunct}{\mcitedefaultseppunct}\relax
\EndOfBibitem
\bibitem[Nitzan and Ratner(2003)Nitzan, and Ratner]{nitzan_electron_2003}
Nitzan,~A.; Ratner,~M.~A. \emph{Science} \textbf{2003}, \emph{300},
  1384--1389\relax
\mciteBstWouldAddEndPuncttrue
\mciteSetBstMidEndSepPunct{\mcitedefaultmidpunct}
{\mcitedefaultendpunct}{\mcitedefaultseppunct}\relax
\EndOfBibitem
\bibitem[Ratner(2013)]{ratner_brief_2013}
Ratner,~M. \emph{Nat. Nanotechnol.} \textbf{2013}, \emph{8}, 378--381\relax
\mciteBstWouldAddEndPuncttrue
\mciteSetBstMidEndSepPunct{\mcitedefaultmidpunct}
{\mcitedefaultendpunct}{\mcitedefaultseppunct}\relax
\EndOfBibitem
\bibitem[Ballmann \latin{et~al.}(2012)Ballmann, H\"artle, Coto, Elbing, Mayor,
  Bryce, Thoss, and Weber]{Ballmann_Experimental_2012}
Ballmann,~S.; H\"artle,~R.; Coto,~P.~B.; Elbing,~M.; Mayor,~M.; Bryce,~M.~R.;
  Thoss,~M.; Weber,~H.~B. \emph{Phys. Rev. Lett.} \textbf{2012}, \emph{109},
  056801\relax
\mciteBstWouldAddEndPuncttrue
\mciteSetBstMidEndSepPunct{\mcitedefaultmidpunct}
{\mcitedefaultendpunct}{\mcitedefaultseppunct}\relax
\EndOfBibitem
\bibitem[Frisenda \latin{et~al.}(2016)Frisenda, Janssen, Grozema, Van Der~Zant,
  and Renaud]{Frisenda_Mechanically_2016}
Frisenda,~R.; Janssen,~V.~A.; Grozema,~F.~C.; Van Der~Zant,~H.~S.; Renaud,~N.
  \emph{Nat. Chem.} \textbf{2016}, \emph{8}, 1099--1104\relax
\mciteBstWouldAddEndPuncttrue
\mciteSetBstMidEndSepPunct{\mcitedefaultmidpunct}
{\mcitedefaultendpunct}{\mcitedefaultseppunct}\relax
\EndOfBibitem
\bibitem[Bai \latin{et~al.}(2019)Bai, Daaoub, Sangtarash, Li, Tang, Zou,
  Sadeghi, Liu, Huang, Tan, \latin{et~al.} others]{Bai_Anti_2019}
Bai,~J.; Daaoub,~A.; Sangtarash,~S.; Li,~X.; Tang,~Y.; Zou,~Q.; Sadeghi,~H.;
  Liu,~S.; Huang,~X.; Tan,~Z., \latin{et~al.}  \emph{Nat. Mater.}
  \textbf{2019}, \emph{18}, 364--369\relax
\mciteBstWouldAddEndPuncttrue
\mciteSetBstMidEndSepPunct{\mcitedefaultmidpunct}
{\mcitedefaultendpunct}{\mcitedefaultseppunct}\relax
\EndOfBibitem
\bibitem[Greenwald \latin{et~al.}(2021)Greenwald, Cameron, Findlay, Fu,
  Gunasekaran, Skabara, and Venkataraman]{Greenwald_Highly_2021}
Greenwald,~J.~E.; Cameron,~J.; Findlay,~N.~J.; Fu,~T.; Gunasekaran,~S.;
  Skabara,~P.~J.; Venkataraman,~L. \emph{Nat. Nanotechnol.} \textbf{2021},
  \emph{16}, 313--317\relax
\mciteBstWouldAddEndPuncttrue
\mciteSetBstMidEndSepPunct{\mcitedefaultmidpunct}
{\mcitedefaultendpunct}{\mcitedefaultseppunct}\relax
\EndOfBibitem
\bibitem[Iwakiri \latin{et~al.}(2022)Iwakiri, de~Vries, Portol{\'e}s, Zheng,
  Taniguchi, Watanabe, Ihn, and Ensslin]{Iwakiri_Gate_2022}
Iwakiri,~S.; de~Vries,~F.~K.; Portol{\'e}s,~E.; Zheng,~G.; Taniguchi,~T.;
  Watanabe,~K.; Ihn,~T.; Ensslin,~K. \emph{Nano Lett.} \textbf{2022},
  \emph{22}, 6292--6297\relax
\mciteBstWouldAddEndPuncttrue
\mciteSetBstMidEndSepPunct{\mcitedefaultmidpunct}
{\mcitedefaultendpunct}{\mcitedefaultseppunct}\relax
\EndOfBibitem
\bibitem[Hansen \latin{et~al.}(2001)Hansen, Kristensen, Pedersen, S\o{}rensen,
  and Lindelof]{Hansen_Mesoscopic_2001}
Hansen,~A.~E.; Kristensen,~A.; Pedersen,~S.; S\o{}rensen,~C.~B.;
  Lindelof,~P.~E. \emph{Phys. Rev. B} \textbf{2001}, \emph{64}, 045327\relax
\mciteBstWouldAddEndPuncttrue
\mciteSetBstMidEndSepPunct{\mcitedefaultmidpunct}
{\mcitedefaultendpunct}{\mcitedefaultseppunct}\relax
\EndOfBibitem
\bibitem[Yaamamoto \latin{et~al.}(2012)Yaamamoto, Takada, B{\"a}uerle,
  Watanabe, Wieck, and Tarucha]{Yamamoto_Electrical_2012}
Yaamamoto,~M.; Takada,~S.; B{\"a}uerle,~C.; Watanabe,~K.; Wieck,~A.~D.;
  Tarucha,~S. \emph{Nat. Nanotechnol.} \textbf{2012}, \emph{7}, 247--251\relax
\mciteBstWouldAddEndPuncttrue
\mciteSetBstMidEndSepPunct{\mcitedefaultmidpunct}
{\mcitedefaultendpunct}{\mcitedefaultseppunct}\relax
\EndOfBibitem
\bibitem[Finck \latin{et~al.}(2014)Finck, Kurter, Hor, and
  Van~Harlingen]{Finck_Phase_2014}
Finck,~A. D.~K.; Kurter,~C.; Hor,~Y.~S.; Van~Harlingen,~D.~J. \emph{Phys. Rev.
  X} \textbf{2014}, \emph{4}, 041022\relax
\mciteBstWouldAddEndPuncttrue
\mciteSetBstMidEndSepPunct{\mcitedefaultmidpunct}
{\mcitedefaultendpunct}{\mcitedefaultseppunct}\relax
\EndOfBibitem
\bibitem[Duprez \latin{et~al.}(2019)Duprez, Sivre, Anthore, Aassime, Cavanna,
  Ouerghi, Gennser, and Pierre]{Duprez_Macroscopic_2019}
Duprez,~H.; Sivre,~E.; Anthore,~A.; Aassime,~A.; Cavanna,~A.; Ouerghi,~A.;
  Gennser,~U.; Pierre,~F. \emph{Phys. Rev. X} \textbf{2019}, \emph{9},
  021030\relax
\mciteBstWouldAddEndPuncttrue
\mciteSetBstMidEndSepPunct{\mcitedefaultmidpunct}
{\mcitedefaultendpunct}{\mcitedefaultseppunct}\relax
\EndOfBibitem
\bibitem[Poggini \latin{et~al.}(2021)Poggini, Lunghi, Collauto, Barbon,
  Armelao, Magnani, Caneschi, Totti, Sorace, and
  Mannini]{Poggini_Chemisorption_2021}
Poggini,~L.; Lunghi,~A.; Collauto,~A.; Barbon,~A.; Armelao,~L.; Magnani,~A.;
  Caneschi,~A.; Totti,~F.; Sorace,~L.; Mannini,~M. \emph{Nanoscale}
  \textbf{2021}, \emph{13}, 7613--7621\relax
\mciteBstWouldAddEndPuncttrue
\mciteSetBstMidEndSepPunct{\mcitedefaultmidpunct}
{\mcitedefaultendpunct}{\mcitedefaultseppunct}\relax
\EndOfBibitem
\bibitem[Roch \latin{et~al.}(2008)Roch, Florens, Bouchiat, Wernsdorfer, and
  Balestro]{Roch_Quantum_2008}
Roch,~N.; Florens,~S.; Bouchiat,~V.; Wernsdorfer,~W.; Balestro,~F.
  \emph{Nature} \textbf{2008}, \emph{453}, 633--637\relax
\mciteBstWouldAddEndPuncttrue
\mciteSetBstMidEndSepPunct{\mcitedefaultmidpunct}
{\mcitedefaultendpunct}{\mcitedefaultseppunct}\relax
\EndOfBibitem
\bibitem[Custers \latin{et~al.}(2012)Custers, Lorenzer, M{\"u}ller, Prokofiev,
  Sidorenko, Winkler, Strydom, Shimura, Sakakibara, Yu, \latin{et~al.}
  others]{Custers_Destruction_2012}
Custers,~J.; Lorenzer,~K.; M{\"u}ller,~M.; Prokofiev,~A.; Sidorenko,~A.;
  Winkler,~H.; Strydom,~A.; Shimura,~Y.; Sakakibara,~T.; Yu,~R., \latin{et~al.}
   \emph{Nat. Mater.} \textbf{2012}, \emph{11}, 189--194\relax
\mciteBstWouldAddEndPuncttrue
\mciteSetBstMidEndSepPunct{\mcitedefaultmidpunct}
{\mcitedefaultendpunct}{\mcitedefaultseppunct}\relax
\EndOfBibitem
\bibitem[Hartman \latin{et~al.}(2018)Hartman, Olsen, L{\"u}scher, Samani,
  Fallahi, Gardner, Manfra, and Folk]{Hartman_Direct_2018}
Hartman,~N.; Olsen,~C.; L{\"u}scher,~S.; Samani,~M.; Fallahi,~S.;
  Gardner,~G.~C.; Manfra,~M.; Folk,~J. \emph{Nat. Phys.} \textbf{2018},
  \emph{14}, 1083--1086\relax
\mciteBstWouldAddEndPuncttrue
\mciteSetBstMidEndSepPunct{\mcitedefaultmidpunct}
{\mcitedefaultendpunct}{\mcitedefaultseppunct}\relax
\EndOfBibitem
\bibitem[Paschen and Si(2021)Paschen, and Si]{Paschen_Quantum_2021}
Paschen,~S.; Si,~Q. \emph{Nat. Rev. Phys.} \textbf{2021}, \emph{3}, 9--26\relax
\mciteBstWouldAddEndPuncttrue
\mciteSetBstMidEndSepPunct{\mcitedefaultmidpunct}
{\mcitedefaultendpunct}{\mcitedefaultseppunct}\relax
\EndOfBibitem
\bibitem[Brandbyge \latin{et~al.}(2002)Brandbyge, Mozos, Ordej\'on, Taylor, and
  Stokbro]{Brandbyge_Density_2002}
Brandbyge,~M.; Mozos,~J.-L.; Ordej\'on,~P.; Taylor,~J.; Stokbro,~K. \emph{Phys.
  Rev. B} \textbf{2002}, \emph{65}, 165401\relax
\mciteBstWouldAddEndPuncttrue
\mciteSetBstMidEndSepPunct{\mcitedefaultmidpunct}
{\mcitedefaultendpunct}{\mcitedefaultseppunct}\relax
\EndOfBibitem
\bibitem[Groth \latin{et~al.}(2014)Groth, Wimmer, Akhmerov, and
  Waintal]{groth_kwant:_2014}
Groth,~C.~W.; Wimmer,~M.; Akhmerov,~A.~R.; Waintal,~X. \emph{New J. Phys.}
  \textbf{2014}, \emph{16}, 063065\relax
\mciteBstWouldAddEndPuncttrue
\mciteSetBstMidEndSepPunct{\mcitedefaultmidpunct}
{\mcitedefaultendpunct}{\mcitedefaultseppunct}\relax
\EndOfBibitem
\bibitem[Evers \latin{et~al.}(2020)Evers, Koryt{\'a}r, Tewari, and {van
  Ruitenbeek}]{Evers_Advances_2020}
Evers,~F.; Koryt{\'a}r,~R.; Tewari,~S.; {van Ruitenbeek},~J.~M. \emph{Rev. Mod.
  Phys.} \textbf{2020}, \emph{92}, 035001\relax
\mciteBstWouldAddEndPuncttrue
\mciteSetBstMidEndSepPunct{\mcitedefaultmidpunct}
{\mcitedefaultendpunct}{\mcitedefaultseppunct}\relax
\EndOfBibitem
\bibitem[Cohen and Galperin(2020)Cohen, and Galperin]{Cohen_Green_2020}
Cohen,~G.; Galperin,~M. \emph{J. Chem. Phys.} \textbf{2020}, \emph{152},
  090901\relax
\mciteBstWouldAddEndPuncttrue
\mciteSetBstMidEndSepPunct{\mcitedefaultmidpunct}
{\mcitedefaultendpunct}{\mcitedefaultseppunct}\relax
\EndOfBibitem
\bibitem[{de Haas} \latin{et~al.}(1934){de Haas}, {de Boer}, and {van d{\"e}n
  Berg}]{de_haas_electrical_1934}
{de Haas},~W.~J.; {de Boer},~J.; {van d{\"e}n Berg},~G.~J. \emph{Physica}
  \textbf{1934}, \emph{1}, 1115--1124\relax
\mciteBstWouldAddEndPuncttrue
\mciteSetBstMidEndSepPunct{\mcitedefaultmidpunct}
{\mcitedefaultendpunct}{\mcitedefaultseppunct}\relax
\EndOfBibitem
\bibitem[Hewson and Edwards(1997)Hewson, and Edwards]{Hewson_Kondo_1997}
Hewson,~A.; Edwards,~D. \emph{The Kondo Problem to Heavy Fermions}; Cambridge
  Studies in Magnetism; 1997\relax
\mciteBstWouldAddEndPuncttrue
\mciteSetBstMidEndSepPunct{\mcitedefaultmidpunct}
{\mcitedefaultendpunct}{\mcitedefaultseppunct}\relax
\EndOfBibitem
\bibitem[Ng and Lee(1988)Ng, and Lee]{ng_-site_1988}
Ng,~T.~K.; Lee,~P.~A. \emph{Phy. Rev. Lett.} \textbf{1988}, \emph{61},
  1768--1771\relax
\mciteBstWouldAddEndPuncttrue
\mciteSetBstMidEndSepPunct{\mcitedefaultmidpunct}
{\mcitedefaultendpunct}{\mcitedefaultseppunct}\relax
\EndOfBibitem
\bibitem[Glazman and Raikh(1988)Glazman, and Raikh]{glazman_resonant_1988}
Glazman,~L.~I.; Raikh,~M.~{\'E}. \emph{Sov. J. Exp. Theor. Phys.}
  \textbf{1988}, \emph{47}, 452\relax
\mciteBstWouldAddEndPuncttrue
\mciteSetBstMidEndSepPunct{\mcitedefaultmidpunct}
{\mcitedefaultendpunct}{\mcitedefaultseppunct}\relax
\EndOfBibitem
\bibitem[Meir \latin{et~al.}(1993)Meir, Wingreen, and
  Lee]{meir_low-temperature_1993}
Meir,~Y.; Wingreen,~N.~S.; Lee,~P.~A. \emph{Phys. Rev. Lett.} \textbf{1993},
  \emph{70}, 2601\relax
\mciteBstWouldAddEndPuncttrue
\mciteSetBstMidEndSepPunct{\mcitedefaultmidpunct}
{\mcitedefaultendpunct}{\mcitedefaultseppunct}\relax
\EndOfBibitem
\bibitem[Goldhaber-Gordon \latin{et~al.}(1998)Goldhaber-Gordon, Shtrikman,
  Mahalu, Abusch-Magder, Meirav, and Kastner]{Goldhaber_Kondo_1998}
Goldhaber-Gordon,~D.; Shtrikman,~H.; Mahalu,~D.; Abusch-Magder,~D.; Meirav,~U.;
  Kastner,~M. \emph{Nature} \textbf{1998}, \emph{391}, 156--159\relax
\mciteBstWouldAddEndPuncttrue
\mciteSetBstMidEndSepPunct{\mcitedefaultmidpunct}
{\mcitedefaultendpunct}{\mcitedefaultseppunct}\relax
\EndOfBibitem
\bibitem[Cronenwett \latin{et~al.}(1998)Cronenwett, Oosterkamp, and
  Kouwenhoven]{cronenwett_tunable_1998}
Cronenwett,~S.~M.; Oosterkamp,~T.~H.; Kouwenhoven,~L.~P. \emph{Science}
  \textbf{1998}, \emph{281}, 540--544\relax
\mciteBstWouldAddEndPuncttrue
\mciteSetBstMidEndSepPunct{\mcitedefaultmidpunct}
{\mcitedefaultendpunct}{\mcitedefaultseppunct}\relax
\EndOfBibitem
\bibitem[Affleck and Simon(2001)Affleck, and Simon]{Affleck_Detecting_2001}
Affleck,~I.; Simon,~P. \emph{Phys. Rev. Lett.} \textbf{2001}, \emph{86},
  2854--2857\relax
\mciteBstWouldAddEndPuncttrue
\mciteSetBstMidEndSepPunct{\mcitedefaultmidpunct}
{\mcitedefaultendpunct}{\mcitedefaultseppunct}\relax
\EndOfBibitem
\bibitem[Affleck \latin{et~al.}(2008)Affleck, Borda, and
  Saleur]{Affleck_Friedel_2008}
Affleck,~I.; Borda,~L.; Saleur,~H. \emph{Phys. Rev. B} \textbf{2008},
  \emph{77}, 180404\relax
\mciteBstWouldAddEndPuncttrue
\mciteSetBstMidEndSepPunct{\mcitedefaultmidpunct}
{\mcitedefaultendpunct}{\mcitedefaultseppunct}\relax
\EndOfBibitem
\bibitem[Affleck(2010)]{Affleck_Kondo_2010}
Affleck,~I. \emph{Perspectives of {{Mesoscopic Physics}}}; World Scientific,
  2010; pp 1--44\relax
\mciteBstWouldAddEndPuncttrue
\mciteSetBstMidEndSepPunct{\mcitedefaultmidpunct}
{\mcitedefaultendpunct}{\mcitedefaultseppunct}\relax
\EndOfBibitem
\bibitem[Erpenbeck and Cohen(2021)Erpenbeck, and
  Cohen]{Erpenbeck_Resolving_2021}
Erpenbeck,~A.; Cohen,~G. \emph{SciPost Phys.} \textbf{2021}, \emph{10},
  142\relax
\mciteBstWouldAddEndPuncttrue
\mciteSetBstMidEndSepPunct{\mcitedefaultmidpunct}
{\mcitedefaultendpunct}{\mcitedefaultseppunct}\relax
\EndOfBibitem
\bibitem[Pustilnik and Glazman(2004)Pustilnik, and
  Glazman]{pustilnik_kondo_2004}
Pustilnik,~M.; Glazman,~L. \emph{J. Phys. Condens. Matter} \textbf{2004},
  \emph{16}, R513\relax
\mciteBstWouldAddEndPuncttrue
\mciteSetBstMidEndSepPunct{\mcitedefaultmidpunct}
{\mcitedefaultendpunct}{\mcitedefaultseppunct}\relax
\EndOfBibitem
\bibitem[Di~Ventra(2008)]{DiVentra_electrical_2008}
Di~Ventra,~M. \emph{Electrical Transport in Nanoscale Systems}; Cambridge
  University Press, 2008\relax
\mciteBstWouldAddEndPuncttrue
\mciteSetBstMidEndSepPunct{\mcitedefaultmidpunct}
{\mcitedefaultendpunct}{\mcitedefaultseppunct}\relax
\EndOfBibitem
\bibitem[Sohn \latin{et~al.}(2013)Sohn, Kouwenhoven, and
  Sch{\"o}n]{Sohn_Mesoscopic_2013}
Sohn,~L.; Kouwenhoven,~L.; Sch{\"o}n,~G. \emph{Mesoscopic Electron Transport};
  NATO Science Series E:; Springer Netherlands, 2013\relax
\mciteBstWouldAddEndPuncttrue
\mciteSetBstMidEndSepPunct{\mcitedefaultmidpunct}
{\mcitedefaultendpunct}{\mcitedefaultseppunct}\relax
\EndOfBibitem
\bibitem[Kang \latin{et~al.}(2001)Kang, Cho, Kim, and
  Shin]{kang_anti-kondo_2001}
Kang,~K.; Cho,~S.~Y.; Kim,~J.-J.; Shin,~S.-C. \emph{Phys. Rev. B}
  \textbf{2001}, \emph{63}, 113304\relax
\mciteBstWouldAddEndPuncttrue
\mciteSetBstMidEndSepPunct{\mcitedefaultmidpunct}
{\mcitedefaultendpunct}{\mcitedefaultseppunct}\relax
\EndOfBibitem
\bibitem[Aligia and Proetto(2002)Aligia, and Proetto]{Aligia_Kondo_2002}
Aligia,~A.~A.; Proetto,~C.~R. \emph{Phys. Rev. B} \textbf{2002}, \emph{65},
  165305\relax
\mciteBstWouldAddEndPuncttrue
\mciteSetBstMidEndSepPunct{\mcitedefaultmidpunct}
{\mcitedefaultendpunct}{\mcitedefaultseppunct}\relax
\EndOfBibitem
\bibitem[Sato \latin{et~al.}(2005)Sato, Aikawa, Kobayashi, Katsumoto, and
  Iye]{Sato_Observation_2005}
Sato,~M.; Aikawa,~H.; Kobayashi,~K.; Katsumoto,~S.; Iye,~Y. \emph{Phys. Rev.
  Lett.} \textbf{2005}, \emph{95}, 066801\relax
\mciteBstWouldAddEndPuncttrue
\mciteSetBstMidEndSepPunct{\mcitedefaultmidpunct}
{\mcitedefaultendpunct}{\mcitedefaultseppunct}\relax
\EndOfBibitem
\bibitem[Feng \latin{et~al.}(2005)Feng, Jiang, Zhong, and
  Jiang]{Feng_Anti_2005}
Feng,~J.-F.; Jiang,~X.-F.; Zhong,~J.-L.; Jiang,~S.-S. \emph{Physica B Condens.
  Matter} \textbf{2005}, \emph{365}, 20--26\relax
\mciteBstWouldAddEndPuncttrue
\mciteSetBstMidEndSepPunct{\mcitedefaultmidpunct}
{\mcitedefaultendpunct}{\mcitedefaultseppunct}\relax
\EndOfBibitem
\bibitem[Sasaki \latin{et~al.}(2009)Sasaki, Tamura, Akazaki, and
  Fujisawa]{sasaki_fano-kondo_2009}
Sasaki,~S.; Tamura,~H.; Akazaki,~T.; Fujisawa,~T. \emph{Phys. Rev. Lett.}
  \textbf{2009}, \emph{103}, 266806\relax
\mciteBstWouldAddEndPuncttrue
\mciteSetBstMidEndSepPunct{\mcitedefaultmidpunct}
{\mcitedefaultendpunct}{\mcitedefaultseppunct}\relax
\EndOfBibitem
\bibitem[Tamura and Sasaki(2010)Tamura, and Sasaki]{tamura_fanokondo_2010}
Tamura,~H.; Sasaki,~S. \emph{Phys. E: Low-Dimens.} \textbf{2010}, \emph{42},
  864--867\relax
\mciteBstWouldAddEndPuncttrue
\mciteSetBstMidEndSepPunct{\mcitedefaultmidpunct}
{\mcitedefaultendpunct}{\mcitedefaultseppunct}\relax
\EndOfBibitem
\bibitem[{\v Z}itko(2010)]{zitko_fano-kondo_2010}
{\v Z}itko,~R. \emph{Phys. Rev. B} \textbf{2010}, \emph{81}, 115316\relax
\mciteBstWouldAddEndPuncttrue
\mciteSetBstMidEndSepPunct{\mcitedefaultmidpunct}
{\mcitedefaultendpunct}{\mcitedefaultseppunct}\relax
\EndOfBibitem
\bibitem[Kiss \latin{et~al.}(2011)Kiss, Kuramoto, and
  Hoshino]{Kiss_Numerical_2011}
Kiss,~A.; Kuramoto,~Y.; Hoshino,~S. \emph{Phys. Rev. B} \textbf{2011},
  \emph{84}, 174402\relax
\mciteBstWouldAddEndPuncttrue
\mciteSetBstMidEndSepPunct{\mcitedefaultmidpunct}
{\mcitedefaultendpunct}{\mcitedefaultseppunct}\relax
\EndOfBibitem
\bibitem[Huo(2015)]{Huo_Fano_2015}
Huo,~D.-M. \emph{Z. Naturforsch.} \textbf{2015}, \emph{70}, 961--967\relax
\mciteBstWouldAddEndPuncttrue
\mciteSetBstMidEndSepPunct{\mcitedefaultmidpunct}
{\mcitedefaultendpunct}{\mcitedefaultseppunct}\relax
\EndOfBibitem
\bibitem[Wang \latin{et~al.}(2022)Wang, Zhou, Yan, Li, Nan, Zhang, Ma, Wang,
  Ma, Luo, and Xiong]{Wang_Unified_2022}
Wang,~J.-N.; Zhou,~W.-H.; Yan,~Y.-X.; Li,~W.; Nan,~N.; Zhang,~J.; Ma,~Y.-N.;
  Wang,~P.-C.; Ma,~X.-R.; Luo,~S.-J.; Xiong,~Y.-C. \emph{Phys. Rev. B}
  \textbf{2022}, \emph{106}, 035428\relax
\mciteBstWouldAddEndPuncttrue
\mciteSetBstMidEndSepPunct{\mcitedefaultmidpunct}
{\mcitedefaultendpunct}{\mcitedefaultseppunct}\relax
\EndOfBibitem
\bibitem[Lara \latin{et~al.}(2023)Lara, Ramos-Andrade, Zambrano, and
  Orellana]{Lara_Kondo_2023}
Lara,~G.; Ramos-Andrade,~J.; Zambrano,~D.; Orellana,~P. \emph{Phys. E:
  Low-Dimens.} \textbf{2023}, \emph{152}, 115743\relax
\mciteBstWouldAddEndPuncttrue
\mciteSetBstMidEndSepPunct{\mcitedefaultmidpunct}
{\mcitedefaultendpunct}{\mcitedefaultseppunct}\relax
\EndOfBibitem
\bibitem[Houck \latin{et~al.}(2005)Houck, Labaziewicz, Chan, Folk, and
  Chuang]{houck_kondo_2005}
Houck,~A.~A.; Labaziewicz,~J.; Chan,~E.~K.; Folk,~J.~A.; Chuang,~I.~L.
  \emph{Nano Lett.} \textbf{2005}, \emph{5}, 1685--1688\relax
\mciteBstWouldAddEndPuncttrue
\mciteSetBstMidEndSepPunct{\mcitedefaultmidpunct}
{\mcitedefaultendpunct}{\mcitedefaultseppunct}\relax
\EndOfBibitem
\bibitem[Parks \latin{et~al.}(2007)Parks, Champagne, Hutchison,
  {Flores-Torres}, Abru{\~n}a, and Ralph]{parks_tuning_2007}
Parks,~J.~J.; Champagne,~A.~R.; Hutchison,~G.~R.; {Flores-Torres},~S.;
  Abru{\~n}a,~H.~D.; Ralph,~D.~C. \emph{Phys. Rev. Lett.} \textbf{2007},
  \emph{99}, 026601\relax
\mciteBstWouldAddEndPuncttrue
\mciteSetBstMidEndSepPunct{\mcitedefaultmidpunct}
{\mcitedefaultendpunct}{\mcitedefaultseppunct}\relax
\EndOfBibitem
\bibitem[Calvo \latin{et~al.}(2009)Calvo, {Fern{\'a}ndez-Rossier}, Palacios,
  Jacob, Natelson, and Untiedt]{calvo_kondo_2009}
Calvo,~M.~R.; {Fern{\'a}ndez-Rossier},~J.; Palacios,~J.~J.; Jacob,~D.;
  Natelson,~D.; Untiedt,~C. \emph{Nature} \textbf{2009}, \emph{458},
  1150--1153\relax
\mciteBstWouldAddEndPuncttrue
\mciteSetBstMidEndSepPunct{\mcitedefaultmidpunct}
{\mcitedefaultendpunct}{\mcitedefaultseppunct}\relax
\EndOfBibitem
\bibitem[Frisenda \latin{et~al.}(2015)Frisenda, Gaudenzi, Franco,
  {Mas-Torrent}, Rovira, Veciana, Alcon, Bromley, Burzur{\'i}, and {van der
  Zant}]{frisenda_kondo_2015}
Frisenda,~R.; Gaudenzi,~R.; Franco,~C.; {Mas-Torrent},~M.; Rovira,~C.;
  Veciana,~J.; Alcon,~I.; Bromley,~S.~T.; Burzur{\'i},~E.; {van der Zant},~H.
  S.~J. \emph{Nano Lett.} \textbf{2015}, \emph{15}, 3109--3114\relax
\mciteBstWouldAddEndPuncttrue
\mciteSetBstMidEndSepPunct{\mcitedefaultmidpunct}
{\mcitedefaultendpunct}{\mcitedefaultseppunct}\relax
\EndOfBibitem
\bibitem[Li \latin{et~al.}(1998)Li, Schneider, Berndt, and
  Delley]{li_kondo_1998}
Li,~J.; Schneider,~W.-D.; Berndt,~R.; Delley,~B. \emph{Phys. Rev. Lett.}
  \textbf{1998}, \emph{80}, 2893--2896\relax
\mciteBstWouldAddEndPuncttrue
\mciteSetBstMidEndSepPunct{\mcitedefaultmidpunct}
{\mcitedefaultendpunct}{\mcitedefaultseppunct}\relax
\EndOfBibitem
\bibitem[Madhavan \latin{et~al.}(1998)Madhavan, Chen, Jamneala, Crommie, and
  Wingreen]{madhavan_tunneling_1998}
Madhavan,~V.; Chen,~W.; Jamneala,~T.; Crommie,~M.~F.; Wingreen,~N.~S.
  \emph{Science} \textbf{1998}, \emph{280}, 567--569\relax
\mciteBstWouldAddEndPuncttrue
\mciteSetBstMidEndSepPunct{\mcitedefaultmidpunct}
{\mcitedefaultendpunct}{\mcitedefaultseppunct}\relax
\EndOfBibitem
\bibitem[Manoharan \latin{et~al.}(2000)Manoharan, Lutz, and
  Eigler]{manoharan_quantum_2000}
Manoharan,~H.~C.; Lutz,~C.~P.; Eigler,~D.~M. \emph{Nature} \textbf{2000},
  \emph{403}, 512--515\relax
\mciteBstWouldAddEndPuncttrue
\mciteSetBstMidEndSepPunct{\mcitedefaultmidpunct}
{\mcitedefaultendpunct}{\mcitedefaultseppunct}\relax
\EndOfBibitem
\bibitem[Knorr \latin{et~al.}(2002)Knorr, Schneider, Diekh{\"o}ner, Wahl, and
  Kern]{Knorr_Kondo_2002}
Knorr,~N.; Schneider,~M.~A.; Diekh{\"o}ner,~L.; Wahl,~P.; Kern,~K. \emph{Phys.
  Rev. Lett.} \textbf{2002}, \emph{88}, 096804\relax
\mciteBstWouldAddEndPuncttrue
\mciteSetBstMidEndSepPunct{\mcitedefaultmidpunct}
{\mcitedefaultendpunct}{\mcitedefaultseppunct}\relax
\EndOfBibitem
\bibitem[Iancu \latin{et~al.}(2006)Iancu, Deshpande, and
  Hla]{Iancu_Manipulating_2006}
Iancu,~V.; Deshpande,~A.; Hla,~S.-W. \emph{Nano Lett.} \textbf{2006}, \emph{6},
  820--823\relax
\mciteBstWouldAddEndPuncttrue
\mciteSetBstMidEndSepPunct{\mcitedefaultmidpunct}
{\mcitedefaultendpunct}{\mcitedefaultseppunct}\relax
\EndOfBibitem
\bibitem[Iancu \latin{et~al.}(2006)Iancu, Deshpande, and
  Hla]{Iancu_Manipulation_2006}
Iancu,~V.; Deshpande,~A.; Hla,~S.-W. \emph{Phys. Rev. Lett.} \textbf{2006},
  \emph{97}, 266603\relax
\mciteBstWouldAddEndPuncttrue
\mciteSetBstMidEndSepPunct{\mcitedefaultmidpunct}
{\mcitedefaultendpunct}{\mcitedefaultseppunct}\relax
\EndOfBibitem
\bibitem[Rakhmilevitch \latin{et~al.}(2014)Rakhmilevitch, Koryt{\'a}r, Bagrets,
  Evers, and Tal]{rakhmilevitch_electron-vibration_2014}
Rakhmilevitch,~D.; Koryt{\'a}r,~R.; Bagrets,~A.; Evers,~F.; Tal,~O. \emph{Phys.
  Rev. Lett.} \textbf{2014}, \emph{113}, 236603\relax
\mciteBstWouldAddEndPuncttrue
\mciteSetBstMidEndSepPunct{\mcitedefaultmidpunct}
{\mcitedefaultendpunct}{\mcitedefaultseppunct}\relax
\EndOfBibitem
\bibitem[da~Rocha \latin{et~al.}(2015)da~Rocha, Tuovinen, van Leeuwen, and
  Koskinen]{DaRocha_Curvature_2015}
da~Rocha,~C.~G.; Tuovinen,~R.; van Leeuwen,~R.; Koskinen,~P. \emph{Nanoscale}
  \textbf{2015}, \emph{7}, 8627--8635\relax
\mciteBstWouldAddEndPuncttrue
\mciteSetBstMidEndSepPunct{\mcitedefaultmidpunct}
{\mcitedefaultendpunct}{\mcitedefaultseppunct}\relax
\EndOfBibitem
\bibitem[Li \latin{et~al.}(2019)Li, Friedrich, Merino, {de Oteyza}, Pe{\~n}a,
  Jacob, and Pascual]{li_electrically_2019}
Li,~J.; Friedrich,~N.; Merino,~N.; {de Oteyza},~D.~G.; Pe{\~n}a,~D.; Jacob,~D.;
  Pascual,~J.~I. \emph{Nano Lett.} \textbf{2019}, \emph{19}, 3288--3294\relax
\mciteBstWouldAddEndPuncttrue
\mciteSetBstMidEndSepPunct{\mcitedefaultmidpunct}
{\mcitedefaultendpunct}{\mcitedefaultseppunct}\relax
\EndOfBibitem
\bibitem[Li \latin{et~al.}(2019)Li, Sanz, Corso, Choi, Pe{\~n}a, Frederiksen,
  and Pascual]{li_single_2019}
Li,~J.; Sanz,~S.; Corso,~M.; Choi,~D.~J.; Pe{\~n}a,~D.; Frederiksen,~T.;
  Pascual,~J.~I. \emph{Nat. Commun.} \textbf{2019}, \emph{10}, 200\relax
\mciteBstWouldAddEndPuncttrue
\mciteSetBstMidEndSepPunct{\mcitedefaultmidpunct}
{\mcitedefaultendpunct}{\mcitedefaultseppunct}\relax
\EndOfBibitem
\bibitem[Tuovinen \latin{et~al.}(2019)Tuovinen, Sentef, da~Rocha, and
  Ferreira]{Tuovinen_Time_2019}
Tuovinen,~R.; Sentef,~M.~A.; da~Rocha,~C.~G.; Ferreira,~M.~S. \emph{Nanoscale}
  \textbf{2019}, \emph{11}, 12296--12304\relax
\mciteBstWouldAddEndPuncttrue
\mciteSetBstMidEndSepPunct{\mcitedefaultmidpunct}
{\mcitedefaultendpunct}{\mcitedefaultseppunct}\relax
\EndOfBibitem
\bibitem[Li \latin{et~al.}(2020)Li, Sanz, {Castro-Esteban}, {Vilas-Varela},
  Friedrich, Frederiksen, Pe{\~n}a, and Pascual]{li_uncovering_2020}
Li,~J.; Sanz,~S.; {Castro-Esteban},~J.; {Vilas-Varela},~M.; Friedrich,~N.;
  Frederiksen,~T.; Pe{\~n}a,~D.; Pascual,~J.~I. \emph{Phys. Rev. Lett.}
  \textbf{2020}, \emph{124}, 177201\relax
\mciteBstWouldAddEndPuncttrue
\mciteSetBstMidEndSepPunct{\mcitedefaultmidpunct}
{\mcitedefaultendpunct}{\mcitedefaultseppunct}\relax
\EndOfBibitem
\bibitem[Su \latin{et~al.}(2020)Su, Li, Du, Tao, Wang, and
  Yu]{su_atomically_2020}
Su,~X.; Li,~C.; Du,~Q.; Tao,~K.; Wang,~S.; Yu,~P. \emph{Nano Lett.}
  \textbf{2020}, \emph{20}, 6859--6864\relax
\mciteBstWouldAddEndPuncttrue
\mciteSetBstMidEndSepPunct{\mcitedefaultmidpunct}
{\mcitedefaultendpunct}{\mcitedefaultseppunct}\relax
\EndOfBibitem
\bibitem[Zheng \latin{et~al.}(2020)Zheng, Li, Zhao, Beyer, Wang, Xu, Yue, Chen,
  Guan, Li, Zheng, Liu, Luo, Feng, Wang, and Jia]{zheng_engineering_2020}
Zheng,~Y. \latin{et~al.}  \emph{Phys. Rev. Lett.} \textbf{2020}, \emph{124},
  147206\relax
\mciteBstWouldAddEndPuncttrue
\mciteSetBstMidEndSepPunct{\mcitedefaultmidpunct}
{\mcitedefaultendpunct}{\mcitedefaultseppunct}\relax
\EndOfBibitem
\bibitem[Allerdt \latin{et~al.}(2020)Allerdt, Hafiz, Barbiellini, Bansil, and
  Feiguin]{Allerdt_Many_2020}
Allerdt,~A.; Hafiz,~H.; Barbiellini,~B.; Bansil,~A.; Feiguin,~A.~E. \emph{Appl.
  Sci.} \textbf{2020}, \emph{10}, 2542\relax
\mciteBstWouldAddEndPuncttrue
\mciteSetBstMidEndSepPunct{\mcitedefaultmidpunct}
{\mcitedefaultendpunct}{\mcitedefaultseppunct}\relax
\EndOfBibitem
\bibitem[Friedrich \latin{et~al.}(2022)Friedrich, Mench{\'o}n, Pozo, Hieulle,
  Vegliante, Li, S{\'a}nchez-Portal, Pe\~{n}a, Garcia-Lekue, and
  Pascual]{Friedrich_Addressing_2022}
Friedrich,~N.; Mench{\'o}n,~R.~E.; Pozo,~I.; Hieulle,~J.; Vegliante,~A.;
  Li,~J.; S{\'a}nchez-Portal,~D.; Pe\~{n}a,~D.; Garcia-Lekue,~A.;
  Pascual,~J.~I. \emph{ACS Nano} \textbf{2022}, \emph{16}, 14819--14826\relax
\mciteBstWouldAddEndPuncttrue
\mciteSetBstMidEndSepPunct{\mcitedefaultmidpunct}
{\mcitedefaultendpunct}{\mcitedefaultseppunct}\relax
\EndOfBibitem
\bibitem[W{\"a}ckerlin \latin{et~al.}(2022)W{\"a}ckerlin, Cahl{\'i}k,
  Goikoetxea, Stetsovych, Medvedeva, Redondo, {\v S}vec, Delley, Ondr{\'a}{\v
  c}ek, Pinar, {Blanco-Rey}, Koloren{\v c}, Arnau, and
  Jel{\'i}nek]{wackerlin_role_2022}
W{\"a}ckerlin,~C.; Cahl{\'i}k,~A.; Goikoetxea,~J.; Stetsovych,~O.;
  Medvedeva,~D.; Redondo,~J.; {\v S}vec,~M.; Delley,~B.; Ondr{\'a}{\v c}ek,~M.;
  Pinar,~A.; {Blanco-Rey},~M.; Koloren{\v c},~J.; Arnau,~A.; Jel{\'i}nek,~P.
  \emph{ACS Nano} \textbf{2022}, \emph{16}, 16402--16413\relax
\mciteBstWouldAddEndPuncttrue
\mciteSetBstMidEndSepPunct{\mcitedefaultmidpunct}
{\mcitedefaultendpunct}{\mcitedefaultseppunct}\relax
\EndOfBibitem
\bibitem[Zhao \latin{et~al.}(2023)Zhao, Jiang, Li, Liu, Zhu, Pizzochero,
  Kaxiras, Guan, Li, Zheng, Liu, Jia, Qin, Zhuang, and Wang]{zhao_quantum_2023}
Zhao,~Y.; Jiang,~K.; Li,~C.; Liu,~Y.; Zhu,~G.; Pizzochero,~M.; Kaxiras,~E.;
  Guan,~D.; Li,~Y.; Zheng,~H.; Liu,~C.; Jia,~J.; Qin,~M.; Zhuang,~X.; Wang,~S.
  \emph{Nat. Chem.} \textbf{2023}, \emph{15}, 53--60\relax
\mciteBstWouldAddEndPuncttrue
\mciteSetBstMidEndSepPunct{\mcitedefaultmidpunct}
{\mcitedefaultendpunct}{\mcitedefaultseppunct}\relax
\EndOfBibitem
\bibitem[Florens(2007)]{Florens_Nanoscale_2007}
Florens,~S. \emph{Phys. Rev. Lett.} \textbf{2007}, \emph{99}, 046402\relax
\mciteBstWouldAddEndPuncttrue
\mciteSetBstMidEndSepPunct{\mcitedefaultmidpunct}
{\mcitedefaultendpunct}{\mcitedefaultseppunct}\relax
\EndOfBibitem
\bibitem[Jacob \latin{et~al.}(2009)Jacob, Haule, and Kotliar]{Jacob_Kondo_2009}
Jacob,~D.; Haule,~K.; Kotliar,~G. \emph{Phys. Rev. Lett.} \textbf{2009},
  \emph{103}, 016803\relax
\mciteBstWouldAddEndPuncttrue
\mciteSetBstMidEndSepPunct{\mcitedefaultmidpunct}
{\mcitedefaultendpunct}{\mcitedefaultseppunct}\relax
\EndOfBibitem
\bibitem[Jacob \latin{et~al.}(2010)Jacob, Haule, and
  Kotliar]{Jacob_Dynamical_2010}
Jacob,~D.; Haule,~K.; Kotliar,~G. \emph{Phys. Rev. B} \textbf{2010}, \emph{82},
  195115\relax
\mciteBstWouldAddEndPuncttrue
\mciteSetBstMidEndSepPunct{\mcitedefaultmidpunct}
{\mcitedefaultendpunct}{\mcitedefaultseppunct}\relax
\EndOfBibitem
\bibitem[Turkowski \latin{et~al.}(2012)Turkowski, Kabir, Nayyar, and
  Rahman]{Turkowski_Dynamical_2012}
Turkowski,~V.; Kabir,~A.; Nayyar,~N.; Rahman,~T.~S. \emph{J. Chem. Phys.}
  \textbf{2012}, \emph{136}, 114108\relax
\mciteBstWouldAddEndPuncttrue
\mciteSetBstMidEndSepPunct{\mcitedefaultmidpunct}
{\mcitedefaultendpunct}{\mcitedefaultseppunct}\relax
\EndOfBibitem
\bibitem[Ferrer \latin{et~al.}(2014)Ferrer, Lambert, {Garc{\'i}a-Su{\'a}rez},
  Manrique, Visontai, Oroszlany, {Rodr{\'i}guez-Ferrad{\'a}s}, Grace, Bailey,
  Gillemot, Sadeghi, and Algharagholy]{Ferrer_Gollum_2014}
Ferrer,~J.; Lambert,~C.~J.; {Garc{\'i}a-Su{\'a}rez},~V.~M.; Manrique,~D.~Z.;
  Visontai,~D.; Oroszlany,~L.; {Rodr{\'i}guez-Ferrad{\'a}s},~R.; Grace,~I.;
  Bailey,~S. W.~D.; Gillemot,~K.; Sadeghi,~H.; Algharagholy,~L.~A. \emph{New J.
  Phys.} \textbf{2014}, \emph{16}, 093029\relax
\mciteBstWouldAddEndPuncttrue
\mciteSetBstMidEndSepPunct{\mcitedefaultmidpunct}
{\mcitedefaultendpunct}{\mcitedefaultseppunct}\relax
\EndOfBibitem
\bibitem[Sch{\"u}ler \latin{et~al.}(2017)Sch{\"u}ler, Barthel, Wehling,
  Karolak, Valli, and Sangiovanni]{Schuler_Realistic_2017}
Sch{\"u}ler,~M.; Barthel,~S.; Wehling,~T.; Karolak,~M.; Valli,~A.;
  Sangiovanni,~G. \emph{Eur. Phys. J. Spec. Top.} \textbf{2017}, \emph{226},
  2615--2640\relax
\mciteBstWouldAddEndPuncttrue
\mciteSetBstMidEndSepPunct{\mcitedefaultmidpunct}
{\mcitedefaultendpunct}{\mcitedefaultseppunct}\relax
\EndOfBibitem
\bibitem[Kurth \latin{et~al.}(2019)Kurth, Jacob, Sobrino, and
  Stefanucci]{Kurth_Nonequilibrium_2019}
Kurth,~S.; Jacob,~D.; Sobrino,~N.; Stefanucci,~G. \emph{Phys. Rev. B}
  \textbf{2019}, \emph{100}, 085114\relax
\mciteBstWouldAddEndPuncttrue
\mciteSetBstMidEndSepPunct{\mcitedefaultmidpunct}
{\mcitedefaultendpunct}{\mcitedefaultseppunct}\relax
\EndOfBibitem
\bibitem[Solomon \latin{et~al.}(2010)Solomon, Herrmann, Hansen, Mujica, and
  Ratner]{Solomon_Exploring_2010}
Solomon,~G.~C.; Herrmann,~C.; Hansen,~T.; Mujica,~V.; Ratner,~M.~A. \emph{Nat.
  Chem.} \textbf{2010}, \emph{2}, 223--228\relax
\mciteBstWouldAddEndPuncttrue
\mciteSetBstMidEndSepPunct{\mcitedefaultmidpunct}
{\mcitedefaultendpunct}{\mcitedefaultseppunct}\relax
\EndOfBibitem
\bibitem[Bouatou \latin{et~al.}(2022)Bouatou, Chacon, Lorentzen, Ngo, Girard,
  Repain, Bellec, Rousset, Brandbyge, Dappe, and
  Lagoute]{Bouatou_Visualizing_2022}
Bouatou,~M.; Chacon,~C.; Lorentzen,~A.~B.; Ngo,~H.~T.; Girard,~Y.; Repain,~V.;
  Bellec,~A.; Rousset,~S.; Brandbyge,~M.; Dappe,~Y.~J.; Lagoute,~J. \emph{Adv.
  Funct. Mater.} \textbf{2022}, \emph{32}, 2208048\relax
\mciteBstWouldAddEndPuncttrue
\mciteSetBstMidEndSepPunct{\mcitedefaultmidpunct}
{\mcitedefaultendpunct}{\mcitedefaultseppunct}\relax
\EndOfBibitem
\bibitem[Gao \latin{et~al.}(2023)Gao, Mench{\'o}n, Garcia-Lekue, and
  Brandbyge]{Gao_Tunable_2023}
Gao,~F.; Mench{\'o}n,~R.~E.; Garcia-Lekue,~A.; Brandbyge,~M. \emph{Commun.
  Phys.} \textbf{2023}, \emph{6}, 115\relax
\mciteBstWouldAddEndPuncttrue
\mciteSetBstMidEndSepPunct{\mcitedefaultmidpunct}
{\mcitedefaultendpunct}{\mcitedefaultseppunct}\relax
\EndOfBibitem
\bibitem[Leitherer \latin{et~al.}(2023)Leitherer, Brandbyge, and
  Solomon]{Leitherer_Electromigration_2023}
Leitherer,~S.; Brandbyge,~M.; Solomon,~G.~C. \emph{ChemRxiv} \textbf{2023},
  \relax
\mciteBstWouldAddEndPunctfalse
\mciteSetBstMidEndSepPunct{\mcitedefaultmidpunct}
{}{\mcitedefaultseppunct}\relax
\EndOfBibitem
\bibitem[{\'U}js{\'a}ghy \latin{et~al.}(2000){\'U}js{\'a}ghy, Kroha, Szunyogh,
  and Zawadowski]{Ujsaghy_Theory_2000}
{\'U}js{\'a}ghy,~O.; Kroha,~J.; Szunyogh,~L.; Zawadowski,~A. \emph{Phys. Rev.
  Lett.} \textbf{2000}, \emph{85}, 2557--2560\relax
\mciteBstWouldAddEndPuncttrue
\mciteSetBstMidEndSepPunct{\mcitedefaultmidpunct}
{\mcitedefaultendpunct}{\mcitedefaultseppunct}\relax
\EndOfBibitem
\bibitem[Agam and Schiller(2001)Agam, and Schiller]{Agam_Projecting_2001}
Agam,~O.; Schiller,~A. \emph{Phys. Rev. Lett.} \textbf{2001}, \emph{86},
  484--487\relax
\mciteBstWouldAddEndPuncttrue
\mciteSetBstMidEndSepPunct{\mcitedefaultmidpunct}
{\mcitedefaultendpunct}{\mcitedefaultseppunct}\relax
\EndOfBibitem
\bibitem[Bu{\l}ka and Stefa{\'n}ski(2001)Bu{\l}ka, and
  Stefa{\'n}ski]{Bulka_Fano_2001}
Bu{\l}ka,~B.~R.; Stefa{\'n}ski,~P. \emph{Phys. Rev. Lett.} \textbf{2001},
  \emph{86}, 5128--5131\relax
\mciteBstWouldAddEndPuncttrue
\mciteSetBstMidEndSepPunct{\mcitedefaultmidpunct}
{\mcitedefaultendpunct}{\mcitedefaultseppunct}\relax
\EndOfBibitem
\bibitem[Hofstetter \latin{et~al.}(2001)Hofstetter, K{\"o}nig, and
  Schoeller]{Hofstetter_Kondo_2001}
Hofstetter,~W.; K{\"o}nig,~J.; Schoeller,~H. \emph{Phys. Rev. Lett.}
  \textbf{2001}, \emph{87}, 156803\relax
\mciteBstWouldAddEndPuncttrue
\mciteSetBstMidEndSepPunct{\mcitedefaultmidpunct}
{\mcitedefaultendpunct}{\mcitedefaultseppunct}\relax
\EndOfBibitem
\bibitem[Torio \latin{et~al.}(2002)Torio, Hallberg, Ceccatto, and
  Proetto]{torio_kondo_2002}
Torio,~M.~E.; Hallberg,~K.; Ceccatto,~A.~H.; Proetto,~C.~R. \emph{Phys. Rev. B}
  \textbf{2002}, \emph{65}, 085302\relax
\mciteBstWouldAddEndPuncttrue
\mciteSetBstMidEndSepPunct{\mcitedefaultmidpunct}
{\mcitedefaultendpunct}{\mcitedefaultseppunct}\relax
\EndOfBibitem
\bibitem[Luo \latin{et~al.}(2004)Luo, Xiang, Wang, Su, and Yu]{Luo_Fano_2004}
Luo,~H.~G.; Xiang,~T.; Wang,~X.~Q.; Su,~Z.~B.; Yu,~L. \emph{Phys. Rev. Lett.}
  \textbf{2004}, \emph{92}, 256602\relax
\mciteBstWouldAddEndPuncttrue
\mciteSetBstMidEndSepPunct{\mcitedefaultmidpunct}
{\mcitedefaultendpunct}{\mcitedefaultseppunct}\relax
\EndOfBibitem
\bibitem[Dias~da Silva \latin{et~al.}(2008)Dias~da Silva, Heidrich-Meisner,
  Feiguin, B\"usser, Martins, Anda, and Dagotto]{DaSilva_Transport_2008}
Dias~da Silva,~L. G. G.~V.; Heidrich-Meisner,~F.; Feiguin,~A.~E.;
  B\"usser,~C.~A.; Martins,~G.~B.; Anda,~E.~V.; Dagotto,~E. \emph{Phys. Rev. B}
  \textbf{2008}, \emph{78}, 195317\relax
\mciteBstWouldAddEndPuncttrue
\mciteSetBstMidEndSepPunct{\mcitedefaultmidpunct}
{\mcitedefaultendpunct}{\mcitedefaultseppunct}\relax
\EndOfBibitem
\bibitem[Heidrich-Meisner \latin{et~al.}(2009)Heidrich-Meisner, Martins,
  B{\"u}sser, Al-Hassanieh, Feiguin, Chiappe, Anda, and
  Dagotto]{Heidrich_Transport_2009}
Heidrich-Meisner,~F.; Martins,~G.; B{\"u}sser,~C.; Al-Hassanieh,~K.~A.;
  Feiguin,~A.; Chiappe,~G.; Anda,~E.; Dagotto,~E. \emph{Eur. Phys. J. B}
  \textbf{2009}, \emph{67}, 527--542\relax
\mciteBstWouldAddEndPuncttrue
\mciteSetBstMidEndSepPunct{\mcitedefaultmidpunct}
{\mcitedefaultendpunct}{\mcitedefaultseppunct}\relax
\EndOfBibitem
\bibitem[DiLullo \latin{et~al.}(2012)DiLullo, Chang, Baadji, Clark,
  Kl{\"o}ckner, Prosenc, Sanvito, Wiesendanger, Hoffmann, and
  Hla]{Dilullo_Molecular_2012}
DiLullo,~A.; Chang,~S.-H.; Baadji,~N.; Clark,~K.; Kl{\"o}ckner,~J.-P.;
  Prosenc,~M.-H.; Sanvito,~S.; Wiesendanger,~R.; Hoffmann,~G.; Hla,~S.-W.
  \emph{Nano Lett.} \textbf{2012}, \emph{12}, 3174--3179\relax
\mciteBstWouldAddEndPuncttrue
\mciteSetBstMidEndSepPunct{\mcitedefaultmidpunct}
{\mcitedefaultendpunct}{\mcitedefaultseppunct}\relax
\EndOfBibitem
\bibitem[Cohen \latin{et~al.}(2015)Cohen, Gull, Reichman, and
  Millis]{Cohen_Taming_2015}
Cohen,~G.; Gull,~E.; Reichman,~D.~R.; Millis,~A.~J. \emph{Phys. Rev. Lett.}
  \textbf{2015}, \emph{115}, 266802\relax
\mciteBstWouldAddEndPuncttrue
\mciteSetBstMidEndSepPunct{\mcitedefaultmidpunct}
{\mcitedefaultendpunct}{\mcitedefaultseppunct}\relax
\EndOfBibitem
\bibitem[Erpenbeck \latin{et~al.}(2023)Erpenbeck, Gull, and
  Cohen]{Erpenbeck_Quantum_2023}
Erpenbeck,~A.; Gull,~E.; Cohen,~G. \emph{Phys. Rev. Lett.} \textbf{2023},
  \emph{130}, 186301\relax
\mciteBstWouldAddEndPuncttrue
\mciteSetBstMidEndSepPunct{\mcitedefaultmidpunct}
{\mcitedefaultendpunct}{\mcitedefaultseppunct}\relax
\EndOfBibitem
\bibitem[Georges \latin{et~al.}(1996)Georges, Kotliar, Krauth, and
  Rozenberg]{georges_dynamical_1996}
Georges,~A.; Kotliar,~G.; Krauth,~W.; Rozenberg,~M.~J. \emph{Rev. Mod. Phys.}
  \textbf{1996}, \emph{68}, 13--125\relax
\mciteBstWouldAddEndPuncttrue
\mciteSetBstMidEndSepPunct{\mcitedefaultmidpunct}
{\mcitedefaultendpunct}{\mcitedefaultseppunct}\relax
\EndOfBibitem
\bibitem[Haug and Jauho(2008)Haug, and Jauho]{Haug_Qauntum_2008}
Haug,~H. J.~W.; Jauho,~A.-P. \emph{Quantum Kinetics in Transport and Optics of
  Semiconductors}; Springer Series in Solid-State Sciences: Berlin, Heidelberg,
  2008\relax
\mciteBstWouldAddEndPuncttrue
\mciteSetBstMidEndSepPunct{\mcitedefaultmidpunct}
{\mcitedefaultendpunct}{\mcitedefaultseppunct}\relax
\EndOfBibitem
\bibitem[Wilson(1975)]{Wilson_renormalization_1975}
Wilson,~K.~G. \emph{Rev. Mod. Phys.} \textbf{1975}, \emph{47}, 773--840\relax
\mciteBstWouldAddEndPuncttrue
\mciteSetBstMidEndSepPunct{\mcitedefaultmidpunct}
{\mcitedefaultendpunct}{\mcitedefaultseppunct}\relax
\EndOfBibitem
\bibitem[Grewe and Keiter(1981)Grewe, and Keiter]{Grewe_Diagrammatic_1981}
Grewe,~N.; Keiter,~H. \emph{Phys. Rev. B} \textbf{1981}, \emph{24},
  4420--4444\relax
\mciteBstWouldAddEndPuncttrue
\mciteSetBstMidEndSepPunct{\mcitedefaultmidpunct}
{\mcitedefaultendpunct}{\mcitedefaultseppunct}\relax
\EndOfBibitem
\bibitem[Kuramoto(1983)]{Kuramoto_Self_1983}
Kuramoto,~Y. \emph{Z. Phys. B Con. Mat.} \textbf{1983}, \emph{53}, 37--52\relax
\mciteBstWouldAddEndPuncttrue
\mciteSetBstMidEndSepPunct{\mcitedefaultmidpunct}
{\mcitedefaultendpunct}{\mcitedefaultseppunct}\relax
\EndOfBibitem
\bibitem[Bickers(1987)]{Bickers_Review_1987}
Bickers,~N.~E. \emph{Rev. Mod. Phys.} \textbf{1987}, \emph{59}, 845--939\relax
\mciteBstWouldAddEndPuncttrue
\mciteSetBstMidEndSepPunct{\mcitedefaultmidpunct}
{\mcitedefaultendpunct}{\mcitedefaultseppunct}\relax
\EndOfBibitem
\bibitem[Tanimura and Kubo(1989)Tanimura, and Kubo]{Tanimura_Time_1989}
Tanimura,~Y.; Kubo,~R. \emph{J. Phys. Soc. Jpn} \textbf{1989}, \emph{58},
  101--114\relax
\mciteBstWouldAddEndPuncttrue
\mciteSetBstMidEndSepPunct{\mcitedefaultmidpunct}
{\mcitedefaultendpunct}{\mcitedefaultseppunct}\relax
\EndOfBibitem
\bibitem[Pruschke and Grewe(1989)Pruschke, and Grewe]{Pruschke_Anderson_1989}
Pruschke,~T.; Grewe,~N. \emph{Z. Phys. B} \textbf{1989}, \emph{74},
  439--449\relax
\mciteBstWouldAddEndPuncttrue
\mciteSetBstMidEndSepPunct{\mcitedefaultmidpunct}
{\mcitedefaultendpunct}{\mcitedefaultseppunct}\relax
\EndOfBibitem
\bibitem[Keiter and Qin(1990)Keiter, and Qin]{Keiter_NCA_1990}
Keiter,~H.; Qin,~Q. \emph{Phys. B: Condens. Matter} \textbf{1990}, \emph{163},
  594--596\relax
\mciteBstWouldAddEndPuncttrue
\mciteSetBstMidEndSepPunct{\mcitedefaultmidpunct}
{\mcitedefaultendpunct}{\mcitedefaultseppunct}\relax
\EndOfBibitem
\bibitem[Anders and Grewe(1994)Anders, and Grewe]{Anders_Perturbational_1994}
Anders,~F.~B.; Grewe,~N. \emph{Europhys. Lett.} \textbf{1994}, \emph{26},
  551\relax
\mciteBstWouldAddEndPuncttrue
\mciteSetBstMidEndSepPunct{\mcitedefaultmidpunct}
{\mcitedefaultendpunct}{\mcitedefaultseppunct}\relax
\EndOfBibitem
\bibitem[Anders(1995)]{Anders_Beyond_1995}
Anders,~F.~B. \emph{Phys. B: Condens. Matter} \textbf{1995}, \emph{206-207},
  177--179\relax
\mciteBstWouldAddEndPuncttrue
\mciteSetBstMidEndSepPunct{\mcitedefaultmidpunct}
{\mcitedefaultendpunct}{\mcitedefaultseppunct}\relax
\EndOfBibitem
\bibitem[Segal \latin{et~al.}(2000)Segal, Nitzan, Davis, Wasielewski, and
  Ratner]{Segal_Electron_2000}
Segal,~D.; Nitzan,~A.; Davis,~W.~B.; Wasielewski,~M.~R.; Ratner,~M.~A. \emph{J.
  Phys. Chem. B} \textbf{2000}, \emph{104}, 3817--3829\relax
\mciteBstWouldAddEndPuncttrue
\mciteSetBstMidEndSepPunct{\mcitedefaultmidpunct}
{\mcitedefaultendpunct}{\mcitedefaultseppunct}\relax
\EndOfBibitem
\bibitem[Haule \latin{et~al.}(2001)Haule, Kirchner, Kroha, and
  W\"olfle]{Haule_Anderson_2001}
Haule,~K.; Kirchner,~S.; Kroha,~J.; W\"olfle,~P. \emph{Phys. Rev. B}
  \textbf{2001}, \emph{64}, 155111\relax
\mciteBstWouldAddEndPuncttrue
\mciteSetBstMidEndSepPunct{\mcitedefaultmidpunct}
{\mcitedefaultendpunct}{\mcitedefaultseppunct}\relax
\EndOfBibitem
\bibitem[White and Feiguin(2004)White, and Feiguin]{White_Real_2004}
White,~S.~R.; Feiguin,~A.~E. \emph{Phys. Rev. Lett.} \textbf{2004}, \emph{93},
  076401\relax
\mciteBstWouldAddEndPuncttrue
\mciteSetBstMidEndSepPunct{\mcitedefaultmidpunct}
{\mcitedefaultendpunct}{\mcitedefaultseppunct}\relax
\EndOfBibitem
\bibitem[Schollw\"ock(2005)]{Schollwock_density-matrix_2005}
Schollw\"ock,~U. \emph{Rev. Mod. Phys.} \textbf{2005}, \emph{77},
  259--315\relax
\mciteBstWouldAddEndPuncttrue
\mciteSetBstMidEndSepPunct{\mcitedefaultmidpunct}
{\mcitedefaultendpunct}{\mcitedefaultseppunct}\relax
\EndOfBibitem
\bibitem[Tanimura(2006)]{Tanimura_Stochastic_2006}
Tanimura,~Y. \emph{J. Phys. Soc. Jpn} \textbf{2006}, \emph{75}, 082001\relax
\mciteBstWouldAddEndPuncttrue
\mciteSetBstMidEndSepPunct{\mcitedefaultmidpunct}
{\mcitedefaultendpunct}{\mcitedefaultseppunct}\relax
\EndOfBibitem
\bibitem[Anders(2008)]{Anders_Steady-State_2008}
Anders,~F.~B. \emph{Phys. Rev. Lett.} \textbf{2008}, \emph{101}, 066804\relax
\mciteBstWouldAddEndPuncttrue
\mciteSetBstMidEndSepPunct{\mcitedefaultmidpunct}
{\mcitedefaultendpunct}{\mcitedefaultseppunct}\relax
\EndOfBibitem
\bibitem[Bulla \latin{et~al.}(2008)Bulla, Costi, and
  Pruschke]{Bulla_Numerical_2008}
Bulla,~R.; Costi,~T.~A.; Pruschke,~T. \emph{Rev. Mod. Phys.} \textbf{2008},
  \emph{80}, 395--450\relax
\mciteBstWouldAddEndPuncttrue
\mciteSetBstMidEndSepPunct{\mcitedefaultmidpunct}
{\mcitedefaultendpunct}{\mcitedefaultseppunct}\relax
\EndOfBibitem
\bibitem[Jin \latin{et~al.}(2008)Jin, Zheng, and Yan]{Jin_Exact_2008}
Jin,~J.; Zheng,~X.; Yan,~Y. \emph{J. Chem. Phys.} \textbf{2008}, \emph{128},
  234703\relax
\mciteBstWouldAddEndPuncttrue
\mciteSetBstMidEndSepPunct{\mcitedefaultmidpunct}
{\mcitedefaultendpunct}{\mcitedefaultseppunct}\relax
\EndOfBibitem
\bibitem[Grewe \latin{et~al.}(2008)Grewe, Schmitt, Jabben, and
  Anders]{Grewe_Conserving_2008}
Grewe,~N.; Schmitt,~S.; Jabben,~T.; Anders,~F.~B. \emph{J. Phys.: Condens.
  Matter} \textbf{2008}, \emph{20}, 365217\relax
\mciteBstWouldAddEndPuncttrue
\mciteSetBstMidEndSepPunct{\mcitedefaultmidpunct}
{\mcitedefaultendpunct}{\mcitedefaultseppunct}\relax
\EndOfBibitem
\bibitem[My\"oh\"anen \latin{et~al.}(2009)My\"oh\"anen, Stan, Stefanucci, and
  van Leeuwen]{Myohanen_Kadanoff_2009}
My\"oh\"anen,~P.; Stan,~A.; Stefanucci,~G.; van Leeuwen,~R. \emph{Phys. Rev. B}
  \textbf{2009}, \emph{80}, 115107\relax
\mciteBstWouldAddEndPuncttrue
\mciteSetBstMidEndSepPunct{\mcitedefaultmidpunct}
{\mcitedefaultendpunct}{\mcitedefaultseppunct}\relax
\EndOfBibitem
\bibitem[Balzer \latin{et~al.}(2009)Balzer, Bonitz, van Leeuwen, Stan, and
  Dahlen]{Balzer_Nonequilibrium_2009}
Balzer,~K.; Bonitz,~M.; van Leeuwen,~R.; Stan,~A.; Dahlen,~N.~E. \emph{Phys.
  Rev. B} \textbf{2009}, \emph{79}, 245306\relax
\mciteBstWouldAddEndPuncttrue
\mciteSetBstMidEndSepPunct{\mcitedefaultmidpunct}
{\mcitedefaultendpunct}{\mcitedefaultseppunct}\relax
\EndOfBibitem
\bibitem[Zheng \latin{et~al.}(2009)Zheng, Luo, Jin, and
  Yan]{Zheng_Complex_2009}
Zheng,~X.; Luo,~J.; Jin,~J.; Yan,~Y. \emph{J. Chem. Phys.} \textbf{2009},
  \emph{130}, 124508\relax
\mciteBstWouldAddEndPuncttrue
\mciteSetBstMidEndSepPunct{\mcitedefaultmidpunct}
{\mcitedefaultendpunct}{\mcitedefaultseppunct}\relax
\EndOfBibitem
\bibitem[Schir\'o and Fabrizio(2009)Schir\'o, and Fabrizio]{Schiro_Real_2009}
Schir\'o,~M.; Fabrizio,~M. \emph{Phys. Rev. B} \textbf{2009}, \emph{79},
  153302\relax
\mciteBstWouldAddEndPuncttrue
\mciteSetBstMidEndSepPunct{\mcitedefaultmidpunct}
{\mcitedefaultendpunct}{\mcitedefaultseppunct}\relax
\EndOfBibitem
\bibitem[Segal \latin{et~al.}(2010)Segal, Millis, and
  Reichman]{Segal_Numerically_2010}
Segal,~D.; Millis,~A.~J.; Reichman,~D.~R. \emph{Phys. Rev. B} \textbf{2010},
  \emph{82}, 205323\relax
\mciteBstWouldAddEndPuncttrue
\mciteSetBstMidEndSepPunct{\mcitedefaultmidpunct}
{\mcitedefaultendpunct}{\mcitedefaultseppunct}\relax
\EndOfBibitem
\bibitem[Schir\'o and Fabrizio(2010)Schir\'o, and Fabrizio]{Schiro_Time_2010}
Schir\'o,~M.; Fabrizio,~M. \emph{Phys. Rev. Lett.} \textbf{2010}, \emph{105},
  076401\relax
\mciteBstWouldAddEndPuncttrue
\mciteSetBstMidEndSepPunct{\mcitedefaultmidpunct}
{\mcitedefaultendpunct}{\mcitedefaultseppunct}\relax
\EndOfBibitem
\bibitem[Schollw\"ock(2011)]{Schollwock_density-matrix_2011}
Schollw\"ock,~U. \emph{Ann. Phys.} \textbf{2011}, \emph{326}, 96--192, January
  2011 Special Issue\relax
\mciteBstWouldAddEndPuncttrue
\mciteSetBstMidEndSepPunct{\mcitedefaultmidpunct}
{\mcitedefaultendpunct}{\mcitedefaultseppunct}\relax
\EndOfBibitem
\bibitem[Li \latin{et~al.}(2012)Li, Tong, Zheng, Hou, Wei, Hu, and
  Yan]{Li_Hierarchical_2012}
Li,~Z.; Tong,~N.; Zheng,~X.; Hou,~D.; Wei,~J.; Hu,~J.; Yan,~Y. \emph{Phys. Rev.
  Lett.} \textbf{2012}, \emph{109}, 266403\relax
\mciteBstWouldAddEndPuncttrue
\mciteSetBstMidEndSepPunct{\mcitedefaultmidpunct}
{\mcitedefaultendpunct}{\mcitedefaultseppunct}\relax
\EndOfBibitem
\bibitem[H\"artle \latin{et~al.}(2013)H\"artle, Cohen, Reichman, and
  Millis]{Hartle_Decoherence_2013}
H\"artle,~R.; Cohen,~G.; Reichman,~D.~R.; Millis,~A.~J. \emph{Phys. Rev. B}
  \textbf{2013}, \emph{88}, 235426\relax
\mciteBstWouldAddEndPuncttrue
\mciteSetBstMidEndSepPunct{\mcitedefaultmidpunct}
{\mcitedefaultendpunct}{\mcitedefaultseppunct}\relax
\EndOfBibitem
\bibitem[Tuovinen \latin{et~al.}(2014)Tuovinen, Perfetto, Stefanucci, and van
  Leeuwen]{Tuovinen_Time_2014}
Tuovinen,~R.; Perfetto,~E.; Stefanucci,~G.; van Leeuwen,~R. \emph{Phys. Rev. B}
  \textbf{2014}, \emph{89}, 085131\relax
\mciteBstWouldAddEndPuncttrue
\mciteSetBstMidEndSepPunct{\mcitedefaultmidpunct}
{\mcitedefaultendpunct}{\mcitedefaultseppunct}\relax
\EndOfBibitem
\bibitem[H\"artle \latin{et~al.}(2015)H\"artle, Cohen, Reichman, and
  Millis]{Hartle_Transport_2015}
H\"artle,~R.; Cohen,~G.; Reichman,~D.~R.; Millis,~A.~J. \emph{Phys. Rev. B}
  \textbf{2015}, \emph{92}, 085430\relax
\mciteBstWouldAddEndPuncttrue
\mciteSetBstMidEndSepPunct{\mcitedefaultmidpunct}
{\mcitedefaultendpunct}{\mcitedefaultseppunct}\relax
\EndOfBibitem
\bibitem[Schwarz \latin{et~al.}(2016)Schwarz, Goldstein, Dorda, Arrigoni,
  Weichselbaum, and {von Delft}]{schwarz_lindblad-driven_2016}
Schwarz,~F.; Goldstein,~M.; Dorda,~A.; Arrigoni,~E.; Weichselbaum,~A.; {von
  Delft},~J. \emph{Phys. Rev. B} \textbf{2016}, \emph{94}, 155142\relax
\mciteBstWouldAddEndPuncttrue
\mciteSetBstMidEndSepPunct{\mcitedefaultmidpunct}
{\mcitedefaultendpunct}{\mcitedefaultseppunct}\relax
\EndOfBibitem
\bibitem[Erpenbeck \latin{et~al.}(2018)Erpenbeck, Hertlein, Schinabeck, and
  Thoss]{Erpenbeck_Extending_2018}
Erpenbeck,~A.; Hertlein,~C.; Schinabeck,~C.; Thoss,~M. \emph{J. Chem. Phys.}
  \textbf{2018}, \emph{149}\relax
\mciteBstWouldAddEndPuncttrue
\mciteSetBstMidEndSepPunct{\mcitedefaultmidpunct}
{\mcitedefaultendpunct}{\mcitedefaultseppunct}\relax
\EndOfBibitem
\bibitem[Erpenbeck and Thoss(2019)Erpenbeck, and
  Thoss]{Erpenbeck_Hierarchical_2019}
Erpenbeck,~A.; Thoss,~M. \emph{J. Chem. Phys.} \textbf{2019}, \emph{151},
  191101\relax
\mciteBstWouldAddEndPuncttrue
\mciteSetBstMidEndSepPunct{\mcitedefaultmidpunct}
{\mcitedefaultendpunct}{\mcitedefaultseppunct}\relax
\EndOfBibitem
\bibitem[Mundinar \latin{et~al.}(2019)Mundinar, Stegmann, K\"onig, and
  Weiss]{Mundinar_Iterative_2019}
Mundinar,~S.; Stegmann,~P.; K\"onig,~J.; Weiss,~S. \emph{Phys. Rev. B}
  \textbf{2019}, \emph{99}, 195457\relax
\mciteBstWouldAddEndPuncttrue
\mciteSetBstMidEndSepPunct{\mcitedefaultmidpunct}
{\mcitedefaultendpunct}{\mcitedefaultseppunct}\relax
\EndOfBibitem
\bibitem[Allerdt and Feiguin(2019)Allerdt, and
  Feiguin]{Allerdt_Numerically_2019}
Allerdt,~A.; Feiguin,~A.~E. \emph{Front. Phys.} \textbf{2019}, \emph{7},
  67\relax
\mciteBstWouldAddEndPuncttrue
\mciteSetBstMidEndSepPunct{\mcitedefaultmidpunct}
{\mcitedefaultendpunct}{\mcitedefaultseppunct}\relax
\EndOfBibitem
\bibitem[Lode \latin{et~al.}(2020)Lode, L\'ev\^eque, Madsen, Streltsov, and
  Alon]{Lode_Colloquium_2020}
Lode,~A. U.~J.; L\'ev\^eque,~C.; Madsen,~L.~B.; Streltsov,~A.~I.; Alon,~O.~E.
  \emph{Rev. Mod. Phys.} \textbf{2020}, \emph{92}, 011001\relax
\mciteBstWouldAddEndPuncttrue
\mciteSetBstMidEndSepPunct{\mcitedefaultmidpunct}
{\mcitedefaultendpunct}{\mcitedefaultseppunct}\relax
\EndOfBibitem
\bibitem[Tanimura(2020)]{Tanimura_Numerically_2020}
Tanimura,~Y. \emph{J. Chem. Phys.} \textbf{2020}, \emph{153}, 020901\relax
\mciteBstWouldAddEndPuncttrue
\mciteSetBstMidEndSepPunct{\mcitedefaultmidpunct}
{\mcitedefaultendpunct}{\mcitedefaultseppunct}\relax
\EndOfBibitem
\bibitem[N{\"u}{\ss}eler \latin{et~al.}(2020)N{\"u}{\ss}eler, Dhand, Huelga,
  and Plenio]{nuseler_efficient_2020}
N{\"u}{\ss}eler,~A.; Dhand,~I.; Huelga,~S.~F.; Plenio,~M.~B. \emph{Phys. Rev.
  B} \textbf{2020}, \emph{101}, 155134\relax
\mciteBstWouldAddEndPuncttrue
\mciteSetBstMidEndSepPunct{\mcitedefaultmidpunct}
{\mcitedefaultendpunct}{\mcitedefaultseppunct}\relax
\EndOfBibitem
\bibitem[Lotem \latin{et~al.}(2020)Lotem, Weichselbaum, {von Delft}, and
  Goldstein]{lotem_renormalized_2020}
Lotem,~M.; Weichselbaum,~A.; {von Delft},~J.; Goldstein,~M. \emph{Phys. Rev.
  Res.} \textbf{2020}, \emph{2}, 043052\relax
\mciteBstWouldAddEndPuncttrue
\mciteSetBstMidEndSepPunct{\mcitedefaultmidpunct}
{\mcitedefaultendpunct}{\mcitedefaultseppunct}\relax
\EndOfBibitem
\bibitem[Erpenbeck \latin{et~al.}(2021)Erpenbeck, Gull, and
  Cohen]{Erpenbeck_Revealing_2021}
Erpenbeck,~A.; Gull,~E.; Cohen,~G. \emph{Phys. Rev. B} \textbf{2021},
  \emph{103}, 125431\relax
\mciteBstWouldAddEndPuncttrue
\mciteSetBstMidEndSepPunct{\mcitedefaultmidpunct}
{\mcitedefaultendpunct}{\mcitedefaultseppunct}\relax
\EndOfBibitem
\bibitem[Purkayastha \latin{et~al.}(2021)Purkayastha, Guarnieri, Campbell,
  Prior, and Goold]{purkayastha_periodically_2021}
Purkayastha,~A.; Guarnieri,~G.; Campbell,~S.; Prior,~J.; Goold,~J. \emph{Phys.
  Rev. B} \textbf{2021}, \emph{104}, 045417\relax
\mciteBstWouldAddEndPuncttrue
\mciteSetBstMidEndSepPunct{\mcitedefaultmidpunct}
{\mcitedefaultendpunct}{\mcitedefaultseppunct}\relax
\EndOfBibitem
\bibitem[Cirac \latin{et~al.}(2021)Cirac, P\'erez-Garc\'{\i}a, Schuch, and
  Verstraete]{Cirac_Matrix_2021}
Cirac,~J.~I.; P\'erez-Garc\'{\i}a,~D.; Schuch,~N.; Verstraete,~F. \emph{Rev.
  Mod. Phys.} \textbf{2021}, \emph{93}, 045003\relax
\mciteBstWouldAddEndPuncttrue
\mciteSetBstMidEndSepPunct{\mcitedefaultmidpunct}
{\mcitedefaultendpunct}{\mcitedefaultseppunct}\relax
\EndOfBibitem
\bibitem[N\'u\~nez Fern\'andez \latin{et~al.}(2022)N\'u\~nez Fern\'andez,
  Jeannin, Dumitrescu, Kloss, Kaye, Parcollet, and
  Waintal]{Nunez_Learning_2022}
N\'u\~nez Fern\'andez,~Y.; Jeannin,~M.; Dumitrescu,~P.~T.; Kloss,~T.; Kaye,~J.;
  Parcollet,~O.; Waintal,~X. \emph{Phys. Rev. X} \textbf{2022}, \emph{12},
  041018\relax
\mciteBstWouldAddEndPuncttrue
\mciteSetBstMidEndSepPunct{\mcitedefaultmidpunct}
{\mcitedefaultendpunct}{\mcitedefaultseppunct}\relax
\EndOfBibitem
\bibitem[Erpenbeck \latin{et~al.}(2023)Erpenbeck, Lin, Blommel, Zhang, Iskakov,
  Bernheimer, N\'u\~nez Fern\'andez, Cohen, Parcollet, Waintal, and
  Gull]{Erpenbeck_Tensor_2023}
Erpenbeck,~A.; Lin,~W.-T.; Blommel,~T.; Zhang,~L.; Iskakov,~S.; Bernheimer,~L.;
  N\'u\~nez Fern\'andez,~Y.; Cohen,~G.; Parcollet,~O.; Waintal,~X.; Gull,~E.
  \emph{Phys. Rev. B} \textbf{2023}, \emph{107}, 245135\relax
\mciteBstWouldAddEndPuncttrue
\mciteSetBstMidEndSepPunct{\mcitedefaultmidpunct}
{\mcitedefaultendpunct}{\mcitedefaultseppunct}\relax
\EndOfBibitem
\bibitem[Cygorek \latin{et~al.}(2022)Cygorek, Cosacchi, Vagov, Axt, Lovett,
  Keeling, and Gauger]{cygorek_simulation_2022}
Cygorek,~M.; Cosacchi,~M.; Vagov,~A.; Axt,~V.~M.; Lovett,~B.~W.; Keeling,~J.;
  Gauger,~E.~M. \emph{Nat. Phys.} \textbf{2022}, \emph{18}, 662--668\relax
\mciteBstWouldAddEndPuncttrue
\mciteSetBstMidEndSepPunct{\mcitedefaultmidpunct}
{\mcitedefaultendpunct}{\mcitedefaultseppunct}\relax
\EndOfBibitem
\bibitem[Ng \latin{et~al.}(2023)Ng, Park, Millis, Chan, and
  Reichman]{ng_real-time_2023}
Ng,~N.; Park,~G.; Millis,~A.~J.; Chan,~G. K.-L.; Reichman,~D.~R. \emph{Phys.
  Rev. B} \textbf{2023}, \emph{107}, 125103\relax
\mciteBstWouldAddEndPuncttrue
\mciteSetBstMidEndSepPunct{\mcitedefaultmidpunct}
{\mcitedefaultendpunct}{\mcitedefaultseppunct}\relax
\EndOfBibitem
\bibitem[Thoenniss \latin{et~al.}(2023)Thoenniss, Sonner, Lerose, and
  Abanin]{thoenniss_efficient_2023}
Thoenniss,~J.; Sonner,~M.; Lerose,~A.; Abanin,~D.~A. \emph{Phys. Rev. B}
  \textbf{2023}, \emph{107}, L201115\relax
\mciteBstWouldAddEndPuncttrue
\mciteSetBstMidEndSepPunct{\mcitedefaultmidpunct}
{\mcitedefaultendpunct}{\mcitedefaultseppunct}\relax
\EndOfBibitem
\bibitem[Gull \latin{et~al.}(2011)Gull, Millis, Lichtenstein, Rubtsov, Troyer,
  and Werner]{Gull_Continuous-Time_2011}
Gull,~E.; Millis,~A.~J.; Lichtenstein,~A.~I.; Rubtsov,~A.~N.; Troyer,~M.;
  Werner,~P. \emph{Rev. Mod. Phys.} \textbf{2011}, \emph{83}, 349\relax
\mciteBstWouldAddEndPuncttrue
\mciteSetBstMidEndSepPunct{\mcitedefaultmidpunct}
{\mcitedefaultendpunct}{\mcitedefaultseppunct}\relax
\EndOfBibitem
\bibitem[Keiter and Kimball(1970)Keiter, and Kimball]{Keiter_Perturbation_1970}
Keiter,~H.; Kimball,~J.~C. \emph{Phys. Rev. Lett.} \textbf{1970}, \emph{25},
  672--675\relax
\mciteBstWouldAddEndPuncttrue
\mciteSetBstMidEndSepPunct{\mcitedefaultmidpunct}
{\mcitedefaultendpunct}{\mcitedefaultseppunct}\relax
\EndOfBibitem
\bibitem[Werner \latin{et~al.}(2006)Werner, Comanac, {de' Medici}, Troyer, and
  Millis]{Werner_Continuous-Time_2006}
Werner,~P.; Comanac,~A.; {de' Medici},~L.; Troyer,~M.; Millis,~A.~J.
  \emph{Phys. Rev. Lett.} \textbf{2006}, \emph{97}, 076405\relax
\mciteBstWouldAddEndPuncttrue
\mciteSetBstMidEndSepPunct{\mcitedefaultmidpunct}
{\mcitedefaultendpunct}{\mcitedefaultseppunct}\relax
\EndOfBibitem
\bibitem[Werner and Millis(2006)Werner, and Millis]{Werner_Hybridization_2006}
Werner,~P.; Millis,~A.~J. \emph{Phys. Rev. B} \textbf{2006}, \emph{74},
  155107\relax
\mciteBstWouldAddEndPuncttrue
\mciteSetBstMidEndSepPunct{\mcitedefaultmidpunct}
{\mcitedefaultendpunct}{\mcitedefaultseppunct}\relax
\EndOfBibitem
\bibitem[Haule(2007)]{Haule_Quantum_2007}
Haule,~K. \emph{Phys. Rev. B} \textbf{2007}, \emph{75}, 155113\relax
\mciteBstWouldAddEndPuncttrue
\mciteSetBstMidEndSepPunct{\mcitedefaultmidpunct}
{\mcitedefaultendpunct}{\mcitedefaultseppunct}\relax
\EndOfBibitem
\bibitem[M\"uhlbacher and Rabani(2008)M\"uhlbacher, and
  Rabani]{Muhlbacher_Real-Time_2008}
M\"uhlbacher,~L.; Rabani,~E. \emph{Phys. Rev. Lett.} \textbf{2008}, \emph{100},
  176403\relax
\mciteBstWouldAddEndPuncttrue
\mciteSetBstMidEndSepPunct{\mcitedefaultmidpunct}
{\mcitedefaultendpunct}{\mcitedefaultseppunct}\relax
\EndOfBibitem
\bibitem[Gull \latin{et~al.}(2010)Gull, Reichman, and
  Millis]{Gull_Bold-line_2010}
Gull,~E.; Reichman,~D.~R.; Millis,~A.~J. \emph{Phys. Rev. B} \textbf{2010},
  \emph{82}, 075109\relax
\mciteBstWouldAddEndPuncttrue
\mciteSetBstMidEndSepPunct{\mcitedefaultmidpunct}
{\mcitedefaultendpunct}{\mcitedefaultseppunct}\relax
\EndOfBibitem
\bibitem[Cohen \latin{et~al.}(2014)Cohen, Reichman, Millis, and
  Gull]{Cohen_Greens_2014}
Cohen,~G.; Reichman,~D.~R.; Millis,~A.~J.; Gull,~E. \emph{Phys. Rev. B}
  \textbf{2014}, \emph{89}, 115139\relax
\mciteBstWouldAddEndPuncttrue
\mciteSetBstMidEndSepPunct{\mcitedefaultmidpunct}
{\mcitedefaultendpunct}{\mcitedefaultseppunct}\relax
\EndOfBibitem
\bibitem[Cohen \latin{et~al.}(2014)Cohen, Gull, Reichman, and
  Millis]{Cohen_Greens_2014_1}
Cohen,~G.; Gull,~E.; Reichman,~D.~R.; Millis,~A.~J. \emph{Phys. Rev. Lett.}
  \textbf{2014}, \emph{112}, 146802\relax
\mciteBstWouldAddEndPuncttrue
\mciteSetBstMidEndSepPunct{\mcitedefaultmidpunct}
{\mcitedefaultendpunct}{\mcitedefaultseppunct}\relax
\EndOfBibitem
\bibitem[Antipov \latin{et~al.}(2017)Antipov, Dong, Kleinhenz, Cohen, and
  Gull]{Antipov_Currents_2017}
Antipov,~A.~E.; Dong,~Q.; Kleinhenz,~J.; Cohen,~G.; Gull,~E. \emph{Phys. Rev.
  B} \textbf{2017}, \emph{95}, 085144\relax
\mciteBstWouldAddEndPuncttrue
\mciteSetBstMidEndSepPunct{\mcitedefaultmidpunct}
{\mcitedefaultendpunct}{\mcitedefaultseppunct}\relax
\EndOfBibitem
\bibitem[Chen \latin{et~al.}(2017)Chen, Cohen, and
  Reichman]{Chen_Inchworm_2017}
Chen,~H.-T.; Cohen,~G.; Reichman,~D.~R. \emph{J. Chem. Phys.} \textbf{2017},
  \emph{146}, 054105\relax
\mciteBstWouldAddEndPuncttrue
\mciteSetBstMidEndSepPunct{\mcitedefaultmidpunct}
{\mcitedefaultendpunct}{\mcitedefaultseppunct}\relax
\EndOfBibitem
\bibitem[Chen \latin{et~al.}(2017)Chen, Cohen, and
  Reichman]{Chen_Inchworm_2017_2}
Chen,~H.-T.; Cohen,~G.; Reichman,~D.~R. \emph{J. Chem. Phys.} \textbf{2017},
  \emph{146}, 054106\relax
\mciteBstWouldAddEndPuncttrue
\mciteSetBstMidEndSepPunct{\mcitedefaultmidpunct}
{\mcitedefaultendpunct}{\mcitedefaultseppunct}\relax
\EndOfBibitem
\bibitem[Cai \latin{et~al.}(2020)Cai, Lu, and Yang]{Cai_Inchworm_2020}
Cai,~Z.; Lu,~J.; Yang,~S. \emph{Commun. Pure Appl. Math.} \textbf{2020},
  \emph{73}, 2430--2472\relax
\mciteBstWouldAddEndPuncttrue
\mciteSetBstMidEndSepPunct{\mcitedefaultmidpunct}
{\mcitedefaultendpunct}{\mcitedefaultseppunct}\relax
\EndOfBibitem
\bibitem[Cai \latin{et~al.}(2020)Cai, Lu, and Yang]{Cai_Numerical_2020}
Cai,~Z.; Lu,~J.; Yang,~S. Numerical analysis for inchworm Monte Carlo method:
  Sign problem and error growth. 2020;
  \url{https://arxiv.org/abs/2006.07654}\relax
\mciteBstWouldAddEndPuncttrue
\mciteSetBstMidEndSepPunct{\mcitedefaultmidpunct}
{\mcitedefaultendpunct}{\mcitedefaultseppunct}\relax
\EndOfBibitem
\bibitem[Cai \latin{et~al.}(2022)Cai, Lu, and Yang]{Cai_Fast_2022}
Cai,~Z.; Lu,~J.; Yang,~S. \emph{Comput. Phys. Commun.} \textbf{2022},
  \emph{278}, 108417\relax
\mciteBstWouldAddEndPuncttrue
\mciteSetBstMidEndSepPunct{\mcitedefaultmidpunct}
{\mcitedefaultendpunct}{\mcitedefaultseppunct}\relax
\EndOfBibitem
\bibitem[Boag \latin{et~al.}(2018)Boag, Gull, and
  Cohen]{Boag_Inclusion-Exclusion_2018}
Boag,~A.; Gull,~E.; Cohen,~G. \emph{Phys. Rev. B} \textbf{2018}, \emph{98},
  115152\relax
\mciteBstWouldAddEndPuncttrue
\mciteSetBstMidEndSepPunct{\mcitedefaultmidpunct}
{\mcitedefaultendpunct}{\mcitedefaultseppunct}\relax
\EndOfBibitem
\bibitem[Ridley \latin{et~al.}(2018)Ridley, Singh, Gull, and
  Cohen]{Ridley_Numerically_2018}
Ridley,~M.; Singh,~V.~N.; Gull,~E.; Cohen,~G. \emph{Phys. Rev. B}
  \textbf{2018}, \emph{97}, 115109\relax
\mciteBstWouldAddEndPuncttrue
\mciteSetBstMidEndSepPunct{\mcitedefaultmidpunct}
{\mcitedefaultendpunct}{\mcitedefaultseppunct}\relax
\EndOfBibitem
\bibitem[Ridley \latin{et~al.}(2019)Ridley, Gull, and Cohen]{Ridley_Lead_2019}
Ridley,~M.; Gull,~E.; Cohen,~G. \emph{J. Chem. Phys.} \textbf{2019},
  \emph{150}, 244107\relax
\mciteBstWouldAddEndPuncttrue
\mciteSetBstMidEndSepPunct{\mcitedefaultmidpunct}
{\mcitedefaultendpunct}{\mcitedefaultseppunct}\relax
\EndOfBibitem
\bibitem[Ridley \latin{et~al.}(2019)Ridley, Galperin, Gull, and
  Cohen]{Ridley_Numerically_2019}
Ridley,~M.; Galperin,~M.; Gull,~E.; Cohen,~G. \emph{Phys. Rev. B}
  \textbf{2019}, \emph{100}, 165127\relax
\mciteBstWouldAddEndPuncttrue
\mciteSetBstMidEndSepPunct{\mcitedefaultmidpunct}
{\mcitedefaultendpunct}{\mcitedefaultseppunct}\relax
\EndOfBibitem
\bibitem[Eidelstein \latin{et~al.}(2020)Eidelstein, Gull, and
  Cohen]{Eidelstein_Multiorbital_2020}
Eidelstein,~E.; Gull,~E.; Cohen,~G. \emph{Phys. Rev. Lett.} \textbf{2020},
  \emph{124}, 206405\relax
\mciteBstWouldAddEndPuncttrue
\mciteSetBstMidEndSepPunct{\mcitedefaultmidpunct}
{\mcitedefaultendpunct}{\mcitedefaultseppunct}\relax
\EndOfBibitem
\bibitem[Kim \latin{et~al.}(2022)Kim, Li, Eckstein, and
  Werner]{Kim_Pseudoparticle_2022}
Kim,~A.~J.; Li,~J.; Eckstein,~M.; Werner,~P. \emph{Phys. Rev. B} \textbf{2022},
  \emph{106}, 085124\relax
\mciteBstWouldAddEndPuncttrue
\mciteSetBstMidEndSepPunct{\mcitedefaultmidpunct}
{\mcitedefaultendpunct}{\mcitedefaultseppunct}\relax
\EndOfBibitem
\bibitem[Li \latin{et~al.}(2022)Li, Yu, Gull, and
  Cohen]{Li_Interaction-Expansion_2022}
Li,~J.; Yu,~Y.; Gull,~E.; Cohen,~G. \emph{Phys. Rev. B} \textbf{2022},
  \emph{105}, 165133\relax
\mciteBstWouldAddEndPuncttrue
\mciteSetBstMidEndSepPunct{\mcitedefaultmidpunct}
{\mcitedefaultendpunct}{\mcitedefaultseppunct}\relax
\EndOfBibitem
\bibitem[Pollock \latin{et~al.}(2022)Pollock, Gull, Modi, and
  Cohen]{Pollock_Reduced_2022}
Pollock,~F.; Gull,~E.; Modi,~K.; Cohen,~G. \emph{SciPost Phys.} \textbf{2022},
  \emph{13}, 027\relax
\mciteBstWouldAddEndPuncttrue
\mciteSetBstMidEndSepPunct{\mcitedefaultmidpunct}
{\mcitedefaultendpunct}{\mcitedefaultseppunct}\relax
\EndOfBibitem
\bibitem[Dong \latin{et~al.}(2017)Dong, Krivenko, Kleinhenz, Antipov, Cohen,
  and Gull]{Dong_Quantum_2017}
Dong,~Q.; Krivenko,~I.; Kleinhenz,~J.; Antipov,~A.~E.; Cohen,~G.; Gull,~E.
  \emph{Phys. Rev. B} \textbf{2017}, \emph{96}, 155126\relax
\mciteBstWouldAddEndPuncttrue
\mciteSetBstMidEndSepPunct{\mcitedefaultmidpunct}
{\mcitedefaultendpunct}{\mcitedefaultseppunct}\relax
\EndOfBibitem
\bibitem[Krivenko \latin{et~al.}(2019)Krivenko, Kleinhenz, Cohen, and
  Gull]{Krivenko_Dynamics_2019}
Krivenko,~I.; Kleinhenz,~J.; Cohen,~G.; Gull,~E. \emph{Phys. Rev. B}
  \textbf{2019}, \emph{100}, 201104\relax
\mciteBstWouldAddEndPuncttrue
\mciteSetBstMidEndSepPunct{\mcitedefaultmidpunct}
{\mcitedefaultendpunct}{\mcitedefaultseppunct}\relax
\EndOfBibitem
\bibitem[Kleinhenz \latin{et~al.}(2020)Kleinhenz, Krivenko, Cohen, and
  Gull]{Kleinhenz_Dynamic_2020}
Kleinhenz,~J.; Krivenko,~I.; Cohen,~G.; Gull,~E. \emph{Phys. Rev. B}
  \textbf{2020}, \emph{102}, 205138\relax
\mciteBstWouldAddEndPuncttrue
\mciteSetBstMidEndSepPunct{\mcitedefaultmidpunct}
{\mcitedefaultendpunct}{\mcitedefaultseppunct}\relax
\EndOfBibitem
\bibitem[Kleinhenz \latin{et~al.}(2022)Kleinhenz, Krivenko, Cohen, and
  Gull]{Kleinhenz_Kondo_2022}
Kleinhenz,~J.; Krivenko,~I.; Cohen,~G.; Gull,~E. \emph{Phys. Rev. B}
  \textbf{2022}, \emph{105}, 085126\relax
\mciteBstWouldAddEndPuncttrue
\mciteSetBstMidEndSepPunct{\mcitedefaultmidpunct}
{\mcitedefaultendpunct}{\mcitedefaultseppunct}\relax
\EndOfBibitem
\bibitem[Cresti \latin{et~al.}(2003)Cresti, Farchioni, Grosso, and
  Parravicini]{Cresti_Keldysh_2003}
Cresti,~A.; Farchioni,~R.; Grosso,~G.; Parravicini,~G.~P. \emph{Phys. Rev. B}
  \textbf{2003}, \emph{68}, 075306\relax
\mciteBstWouldAddEndPuncttrue
\mciteSetBstMidEndSepPunct{\mcitedefaultmidpunct}
{\mcitedefaultendpunct}{\mcitedefaultseppunct}\relax
\EndOfBibitem
\bibitem[Meir and Wingreen(1992)Meir, and Wingreen]{Meir_Landauer_1992}
Meir,~Y.; Wingreen,~N.~S. \emph{Phys. Rev. Lett.} \textbf{1992}, \emph{68},
  2512--2515\relax
\mciteBstWouldAddEndPuncttrue
\mciteSetBstMidEndSepPunct{\mcitedefaultmidpunct}
{\mcitedefaultendpunct}{\mcitedefaultseppunct}\relax
\EndOfBibitem
\bibitem[Borzenets \latin{et~al.}(2020)Borzenets, Shim, Chen, Ludwig, Wieck,
  Tarucha, Sim, and Yamamoto]{Borzenets_Observation_2020}
Borzenets,~I.~V.; Shim,~J.; Chen,~J.~C.; Ludwig,~A.; Wieck,~A.~D.; Tarucha,~S.;
  Sim,~H.-S.; Yamamoto,~M. \emph{Nature} \textbf{2020}, \emph{579},
  210--213\relax
\mciteBstWouldAddEndPuncttrue
\mciteSetBstMidEndSepPunct{\mcitedefaultmidpunct}
{\mcitedefaultendpunct}{\mcitedefaultseppunct}\relax
\EndOfBibitem
\bibitem[Pruschke \latin{et~al.}(1993)Pruschke, Cox, and
  Jarrell]{Pruschke_Hubbard_1993}
Pruschke,~T.; Cox,~D.~L.; Jarrell,~M. \emph{Phys. Rev. B} \textbf{1993},
  \emph{47}, 3553--3565\relax
\mciteBstWouldAddEndPuncttrue
\mciteSetBstMidEndSepPunct{\mcitedefaultmidpunct}
{\mcitedefaultendpunct}{\mcitedefaultseppunct}\relax
\EndOfBibitem
\bibitem[Eckstein and Werner(2010)Eckstein, and
  Werner]{eckstein_nonequilibrium_2010}
Eckstein,~M.; Werner,~P. \emph{Physical Review B} \textbf{2010}, \emph{82},
  115115\relax
\mciteBstWouldAddEndPuncttrue
\mciteSetBstMidEndSepPunct{\mcitedefaultmidpunct}
{\mcitedefaultendpunct}{\mcitedefaultseppunct}\relax
\EndOfBibitem
\end{mcitethebibliography}

\newpage
\begin{suppinfo}
 
\paragraph{Approximate impurity solvers gauged in the noninteracting limit:}
                
        Here, we discuss the accuracy of different impurity solvers based on the analytically solvable noninteracting case $U=\epsilon_0=0$. 
        In this case, the impurity site is indistinguishable from a lattice site, and any resistivity is an artifact of the impurity solver. 
        Note that, due to the conductive nature of the noninteracting system, this case poses significant challenges for hybridization expansion approaches.
        Achieving convergence in the presence of a Coulomb interaction is frequently more manageable (see below).

        \begin{figure}[b!]
            \centering
            \vspace*{-0.3cm}
            \hspace*{-9cm}
            a)\\             
            \includegraphics{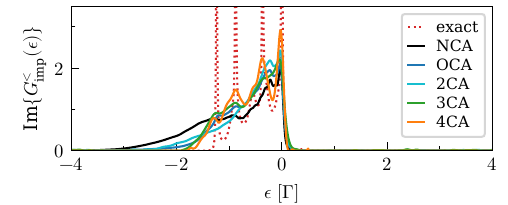}\\
            \vspace*{0.5cm}
            \hspace*{-1.4cm}
            b)  \hspace*{1cm} noninteracting system, NCA \\             
            \hspace*{1cm}
            \includegraphics[width=0.32\textwidth]{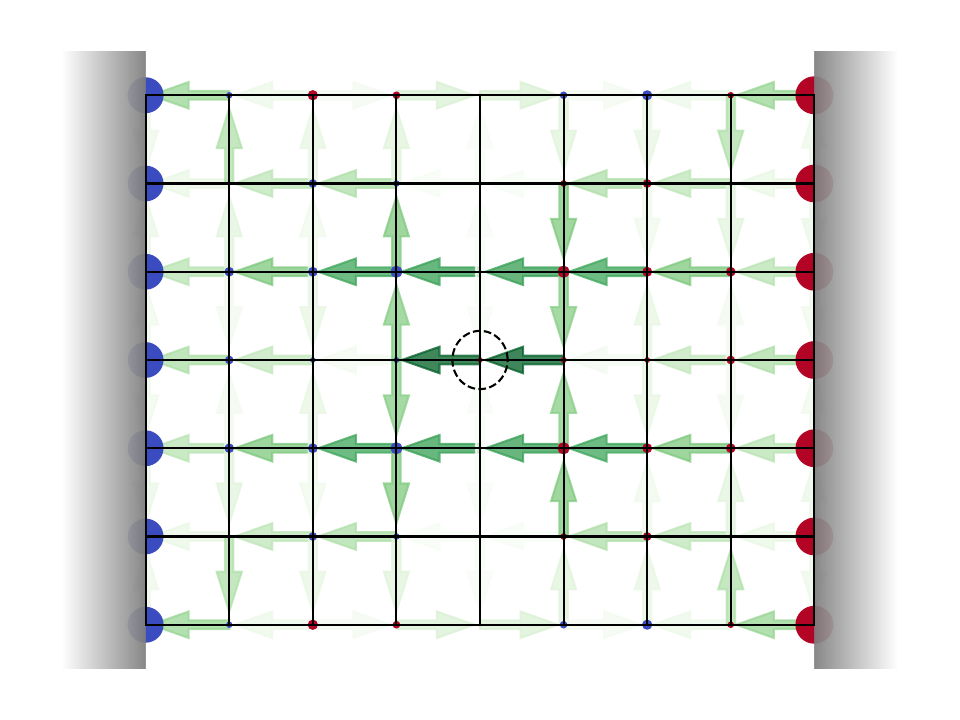}
            \hspace{-0.5cm}
            \includegraphics[width=0.0925\textwidth]{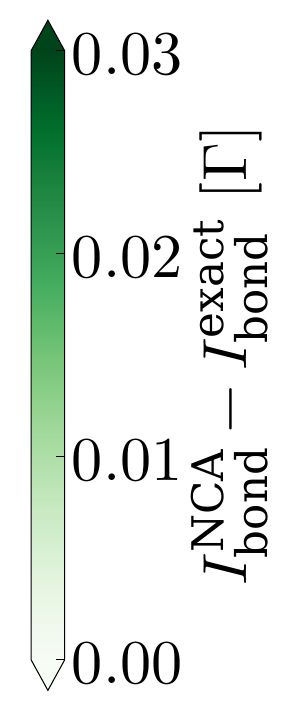}\\
            \caption{Transport behavior of the noninteracting system 1 (see main text for characterization).
                     a: lesser GF for $\Phi=0\Gamma$. The full lines correspond to the result calculated by different approximate methods of increasing order, the red dashed line is the exact result.
                     b: current profile for $\Phi=4.66\Gamma$ as calculated within the NCA. The size and color of the arrows is given by the corresponding bond current, the size of blue/red circles is proportional the charge accumulated/depleted at the respective site. The dashed circle indicates the location of the impurity.}
            \label{fig:nonint}
        \end{figure}
        Fig.~\ref{fig:nonint}a shows the lesser Green's function (GF) at the impurity site at zero bias voltage for different impurity solvers.
        Low-order approximate schemes, such as NCA and OCA \cite{Bickers_Review_1987, Pruschke_Anderson_1989, Pruschke_Hubbard_1993, Haule_Anderson_2001, eckstein_nonequilibrium_2010, Cohen_Greens_2014}, fail to accurately capture the sharp features in the lesser GF at the impurity site, while the four-crossing approximation (4CA, which corresponds to a resummation approach that includes any process that has up to four hybridization-line crossings in the associated Feynman diagram representation) shows improved results. This implies that larger hybridization expansion orders are generally required for an accurate description. While low-order schemes provide insights into the system's equilibrium physics and can be applied to systems out of equilibrium \cite{eckstein_nonequilibrium_2010, Cohen_Greens_2014, Erpenbeck_Revealing_2021, Erpenbeck_Resolving_2021}, an artifact associated to the NCA is that it underestimates the total current by approximately $2\%$, which is attributed to an artificial resistivity caused by imperfect description of the interface between the impurity and the sheet, as exemplified by the reconstructed current profile from NCA in Fig.~\ref{fig:nonint}b.
        The emergence of a spurious scattering center is inherent to all approximate hybridization expansion schemes.
        
        To overcome these artifacts, the steady-state inchworm scheme is employed \cite{Erpenbeck_Quantum_2023}, offering systematic convergence to the correct results without the need for hybridization function representation. This scheme directly provides results in the steady state, which is crucial for the system with sharp resonances indicating long-lived oscillations that govern transient dynamics.

\paragraph{Convergence analysis:}
    \begin{figure}[htb!]
        \centering
        \vspace*{-0.3cm}
        \hspace*{-9cm}
        a)\\             
        \includegraphics{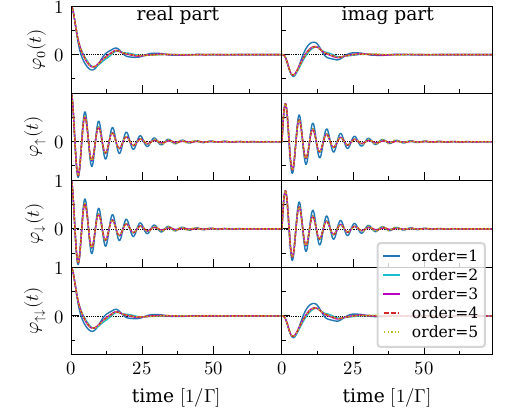}\\
        \centering
        \vspace*{-0.3cm}
        \hspace*{-9cm}
        b)\\             
        \includegraphics{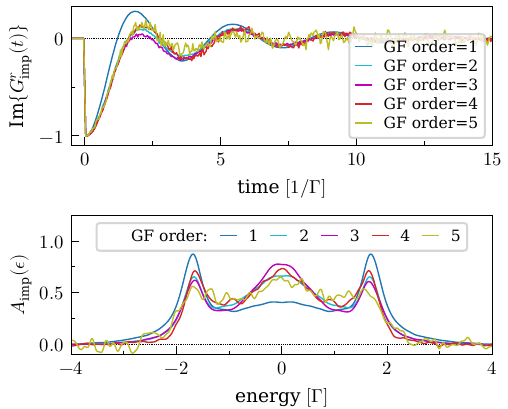}\\
        \vspace*{-4.05cm}
        \hspace*{-9cm}
        c)\\             
        \vspace*{3.25cm}
        \caption{
                Convergence analysis for system 1 at $\Phi=0$ and $T=\Gamma/3$.
                a: propagator $\varphi_\alpha$ for different hybridization orders.
                b: retarded GF for different orders in the hybridization expansion for the GF. The order for the propagators used here is $5$.
                c: spectral function $A_{\mathrm{imp}}(\epsilon)=-\mathrm{Im}\lbrace G^r_{\mathrm{imp}}(\epsilon) \rbrace$ calculated from the data in panel b, that is for different orders in the hybridization expansion for the GF.
                Note that we use propagator order $5$, and GF order $4$ for the data presented in the main text, which provides reasonable accuracy while remaining numerically feasible.
                }
        \label{fig:convergence}
    \end{figure}
    
    We present a convergence analysis for a selected bias and temperature value, $\Phi=0$ and $T=\Gamma/3$ for system 1, aiming to demonstrate the accuracy of our findings. This is a challenging parameter set to obtain converged results; converged results for higher temperatures and higher bias voltages can be obtained at smaller hybridization orders.
    
    For the scope of this work, we employ the steady-state inchworm Quantum Monte Carlo method (iQMC) method \cite{Erpenbeck_Quantum_2023}.
    The iQMC is based on the calculation of restricted propagators, 
    \begin{eqnarray}
            \varphi_\alpha(t)	&=&	
                                \mathrm{Tr}_\mathrm{B} 
                                \left\lbrace \rho_\mathrm{B}
                                \bra{\alpha} e^{iH t} \ket{\alpha}                                
                                \right\rbrace ,
    \end{eqnarray}
    where $\alpha$ is a state of the impurity subspace, $\rho_\mathrm{B}$ is the density matrix of the environment, and $\mathrm{Tr}_\mathrm{B}$ is the trace over the environment's degrees of freedom.
    These propagators are calculated by applying the inchworm scheme based on the hybridization expansion in the coupling between the impurity and the environment. The steady-state GFs are obtained from the restricted propagators \cite{Cohen_Greens_2014, Antipov_Currents_2017, Erpenbeck_Quantum_2023}.

    Fig.~\ref{fig:convergence}a illustrates the restricted propagator of the four impurity states $0, \uparrow, \downarrow, \uparrow\downarrow$ for different hybridization orders. 
    Notably, we observe that the propagator effectively converges at order 3, with the data from orders 3 to 5 overlapping each other in the plot. 
    This significant convergence at a relatively low hybridization order is made possible by the inchworm approach, which serves as an efficient resummation scheme \cite{Cohen_Taming_2015, Chen_Inchworm_2017, Cai_Numerical_2020}. 
    For higher temperatures and bias voltages, we anticipate achieving convergence at even lower hybridization orders. 
    For the purposes of this study, all data presented is based on restricted propagators at hybridization order 5. 
    Based on the convergence analysis depicted in Fig.~\ref{fig:convergence}a, we are confident that we are using converged restricted propagators throughout this work.
    
    Fig.~\ref{fig:convergence}b displays the retarded GFs calculated at different hybridization orders, while the corresponding spectral functions are presented in Fig.~\ref{fig:convergence}c. 
    It is worth noting that calculating the GF from the propagators becomes numerically challenging at higher orders. 
    Consequently, the GF and spectral function at order 5 exhibit considerable Monte Carlo noise, making it impractical to go beyond this order. 
    Nevertheless, we find that GFs calculated at orders 3-5 exhibit qualitatively similar features. 
    This consistency gives us confidence in the correctness of the extracted properties.
    However, determining the exact value of spectral features may prove to be numerically unfeasible. 
    For this study, we present the GF calculated at hybridization order 4.

    \begin{figure}[htb!]
        \centering
        \vspace*{-0.3cm}
        \hspace*{-9cm}
        a)\\           
        \includegraphics{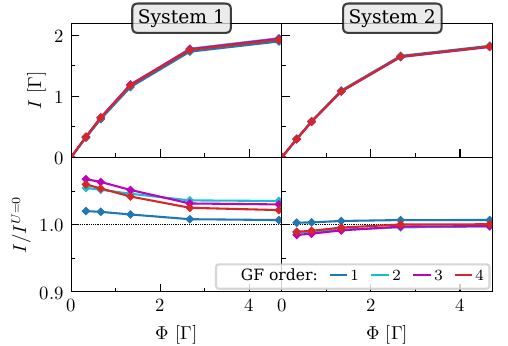}\\
        \centering
        \vspace*{-0.3cm}
        \hspace*{-9cm}
        b)\\           
        \includegraphics{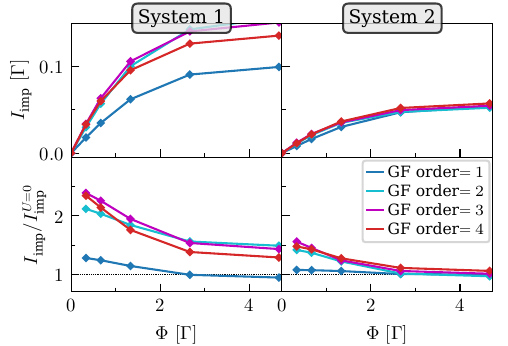}\\
        \centering
        \vspace*{-0.3cm}
        \hspace*{-9cm}
        c)\\           
        \includegraphics{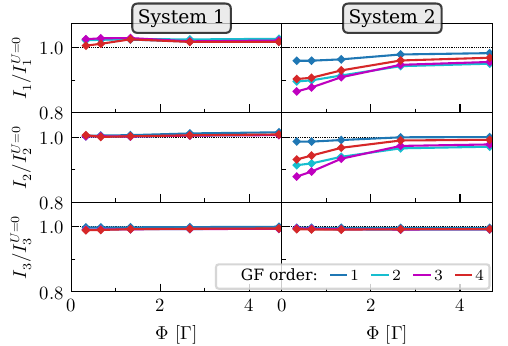}
        \caption{
            Convergence analysis of the total current of the system and the bond currents for $T=\Gamma/3$ for different hybridization orders in the GF. 
            The hybridization order for the propagators is set to $5$.
            This convergence analysis pertains Fig.~3 of the main text.
            a: convergence of the full current.
            b: convergence of the current flowing through the impurity. 
            c: convergence of the current flowing parallel to the impurity.
        }
        \label{fig:convergence_current}
    \end{figure}
    Finally, we present the convergence analysis for the current, which is the primary focus of this study. 
    Specifically, Fig.~\ref{fig:convergence_current} provides the convergence analysis for the data presented in Fig.~3 of the main text, focusing on the temperature $T=\Gamma/3$. 
    We focus on only varying the order of the GF, which represents the challenging aspect of this calculation, while ensuring that the restricted propagators are converged at order 5.
    Establishing convergence for the GF is a numerically challenging task, and this is directly reflected variations observed in the currents. 
    However, while order 1 displays evident deviations from the actual results, orders 2-5, although exhibiting slight differences in absolute values, consistently depict the same behavior. 
    This consistency provides us with confidence regarding the reliability of the results discussed in the main text. 
    We note that comparing ratios of currents at low bias voltages is particularly challenging, as it involves the comparison of two small numbers.

\paragraph{Absolute currents:}
    \begin{figure*}[tb!]
            \raggedright \hspace*{0.65cm} System 1, $\Phi=0.33\Gamma$ \hspace*{1.07cm} System 1, $\Phi=1.33\Gamma$ \hspace*{1.07cm} System 1, $\Phi=2.66\Gamma$
            \vspace{-0.25cm} \\
            \centering
            \includegraphics[width=0.3\textwidth]{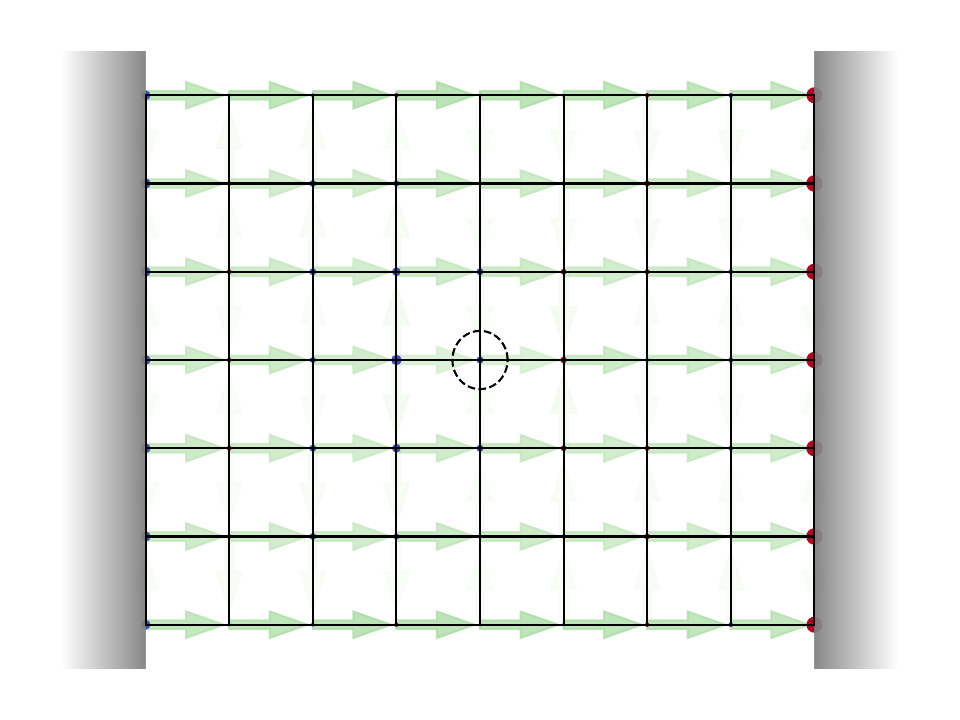}
            \includegraphics[width=0.3\textwidth]{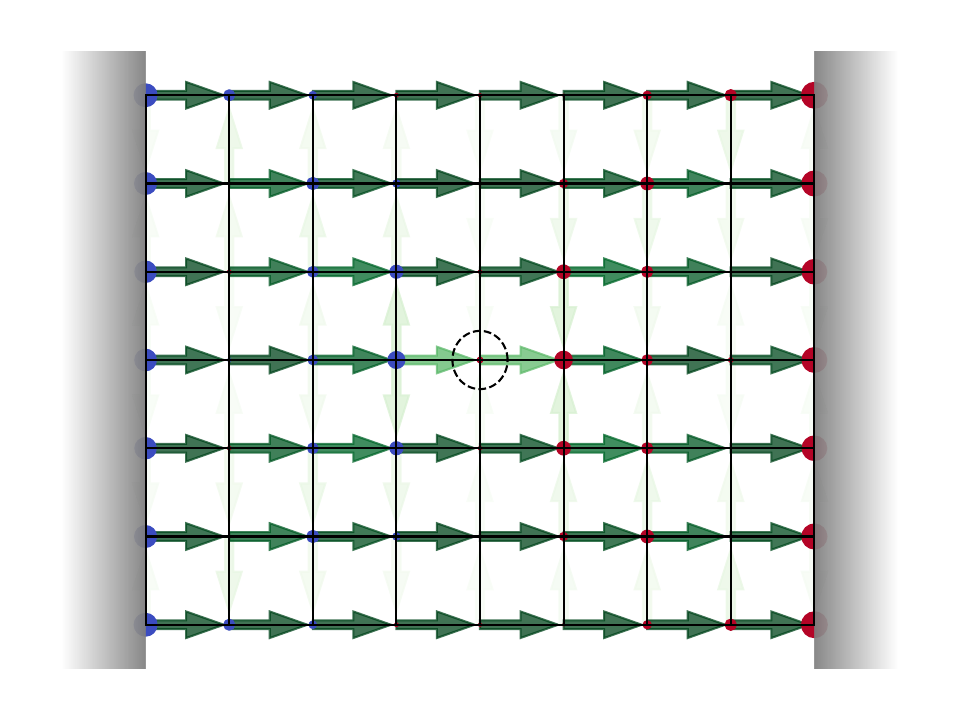}
            \includegraphics[width=0.3\textwidth]{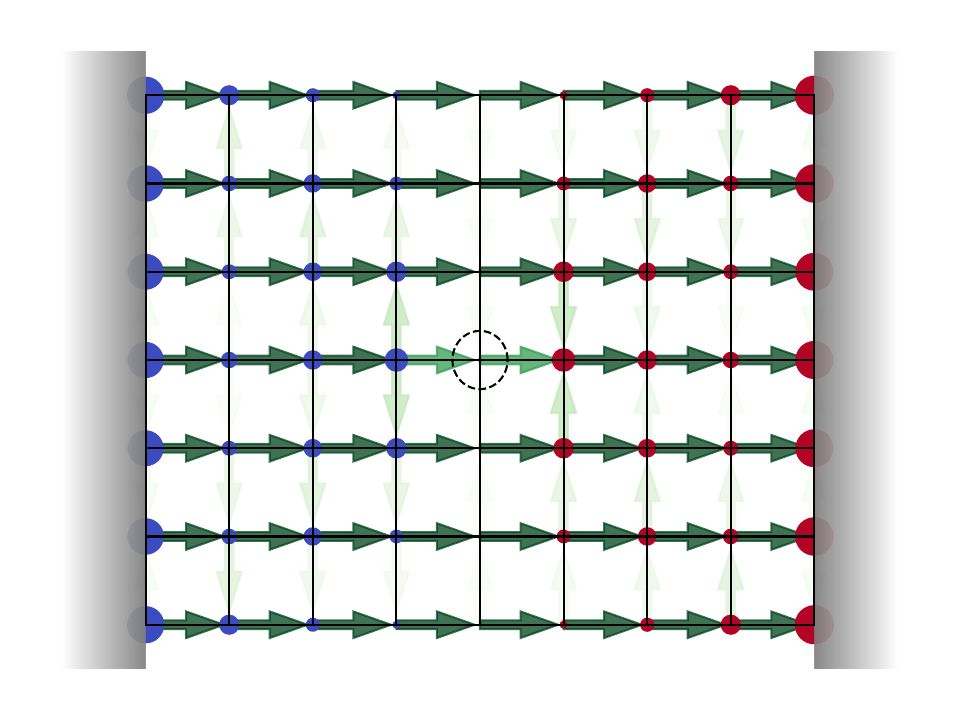}
            \hspace{-0.5cm}
            \includegraphics[width=0.075\textwidth]{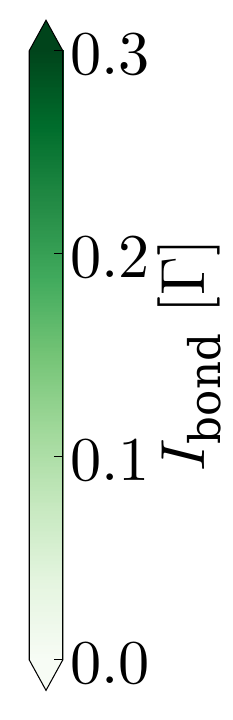}\\
            \vspace*{0.5cm}
            \raggedright \hspace*{0.65cm} System 1, $\Phi=0.33\Gamma$ \hspace*{1.07cm} System 1, $\Phi=1.33\Gamma$ \hspace*{1.07cm} System 1, $\Phi=2.66\Gamma$
            \vspace{-0.25cm} \\
            \centering
            \includegraphics[width=0.3\textwidth]{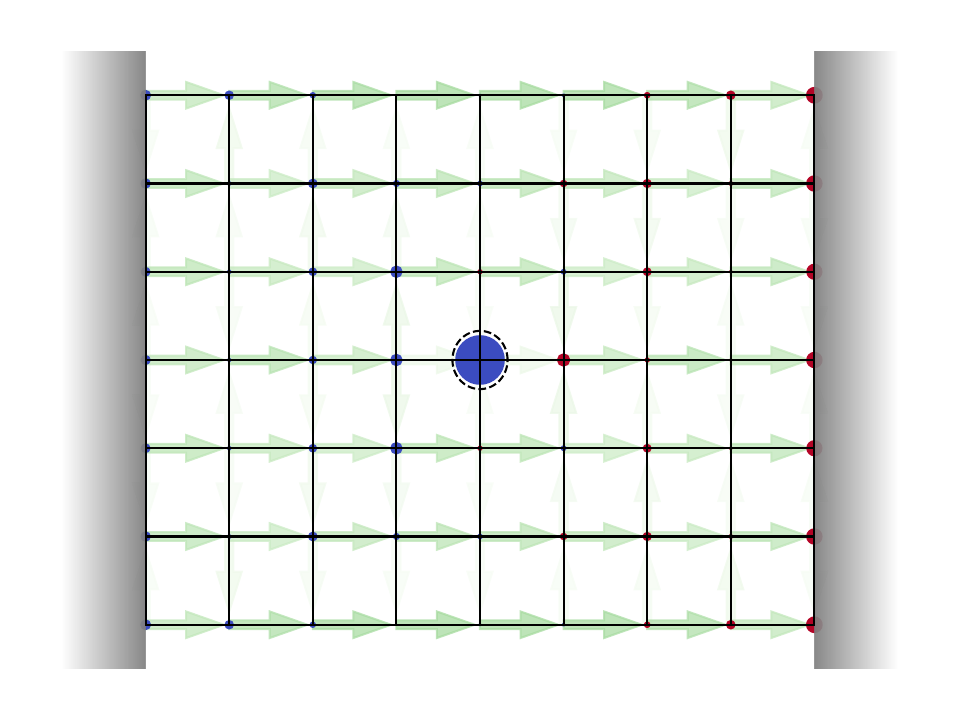}
            \includegraphics[width=0.3\textwidth]{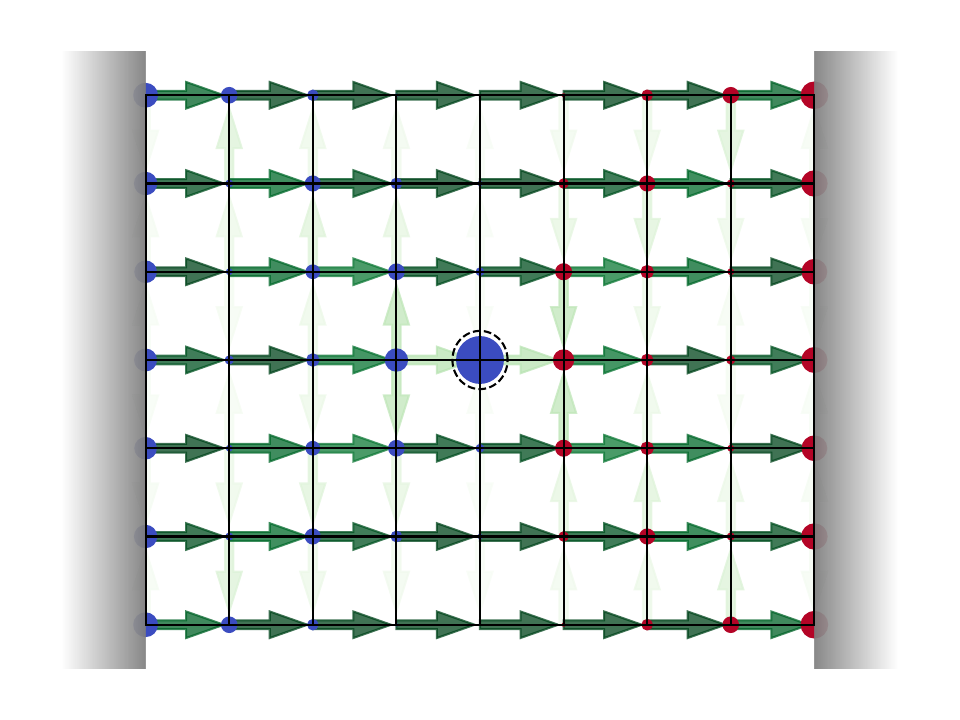}
            \includegraphics[width=0.3\textwidth]{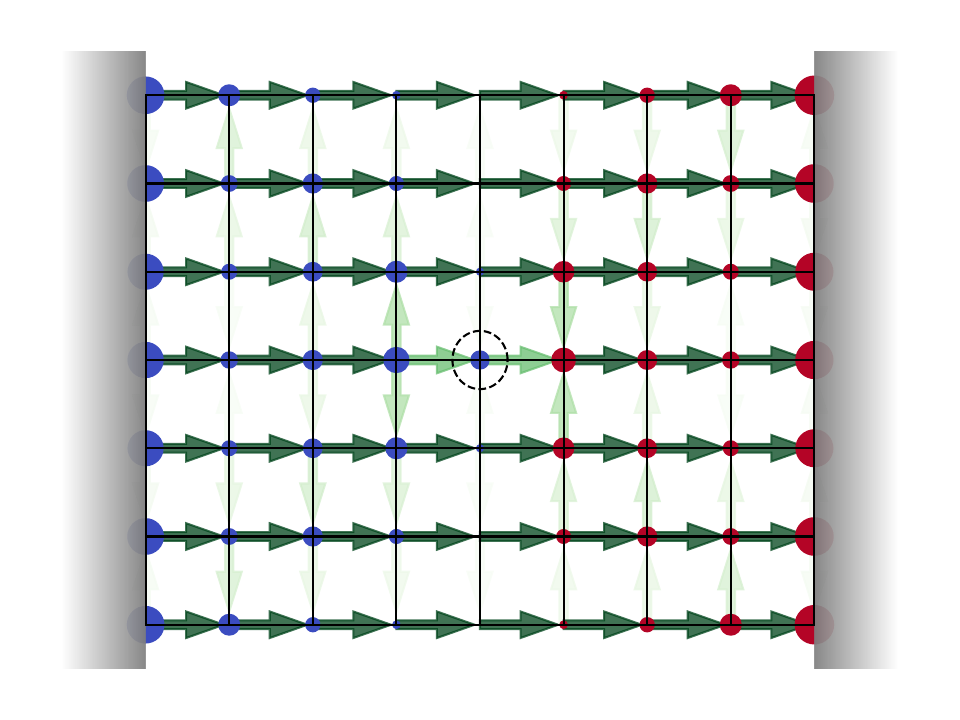}
            \hspace{-0.5cm}
            \includegraphics[width=0.075\textwidth]{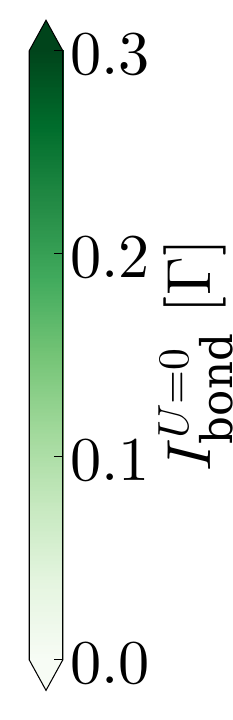}\\
            \caption{
                    Current and charge distribution for system 1 for three representative bias voltages (left to right) at temperature $T=0.167\Gamma$.
                    Top: interacting case. Bottom: noninteracting case $U=0$.
                    Green arrows: bond currents. 
                    Red/blue circles: charge depleted/accumulated, with the size of the circle indicating the amount of charge. 
                    Dashed circle: position of the impurity.
                }
        \label{fig:transport_map_sys1}
     \end{figure*}
     \begin{figure*}[htb!]
            \raggedright \hspace*{0.65cm} System 2, $\Phi=0.33\Gamma$ \hspace*{1.07cm} System 2, $\Phi=1.33\Gamma$ \hspace*{1.07cm} System 2, $\Phi=2.66\Gamma$
            \vspace{-0.25cm} \\
            \centering
            \includegraphics[width=0.3\textwidth]{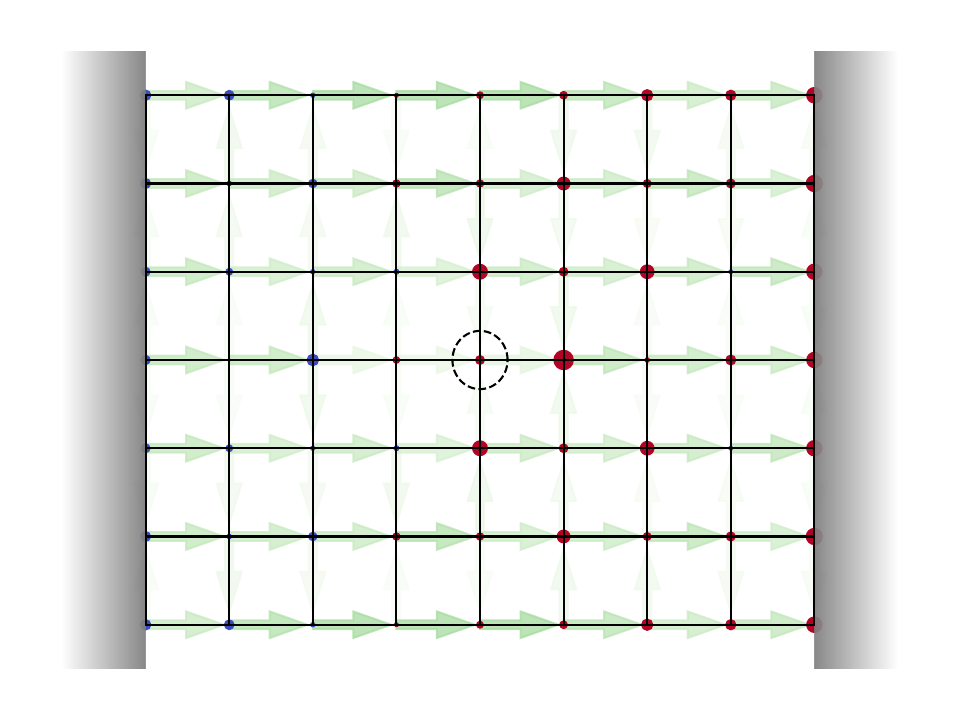}
            \includegraphics[width=0.3\textwidth]{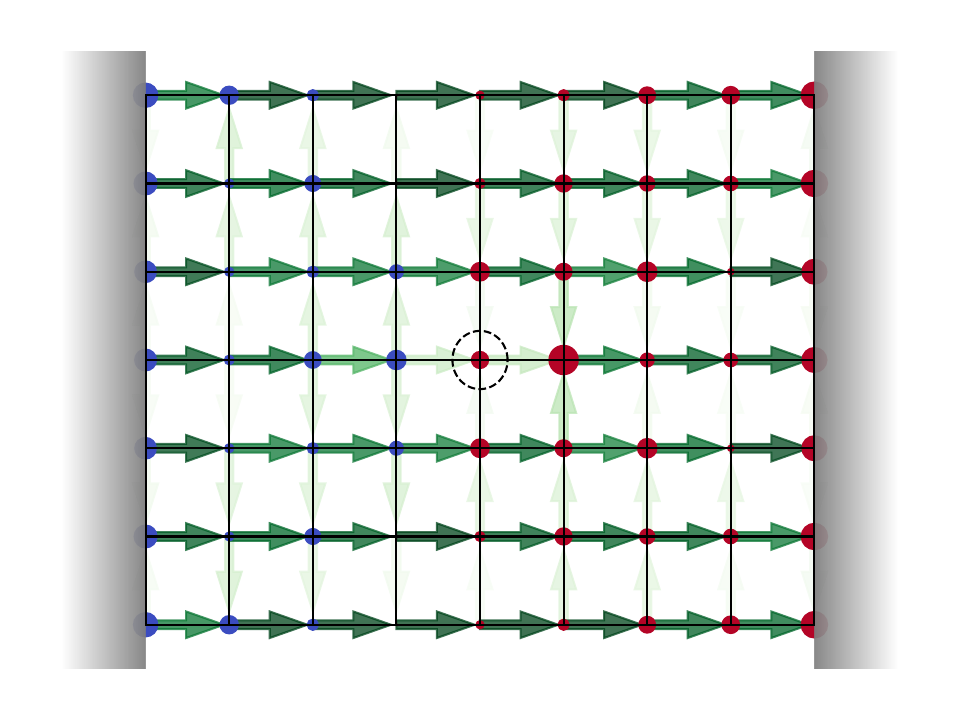}
            \includegraphics[width=0.3\textwidth]{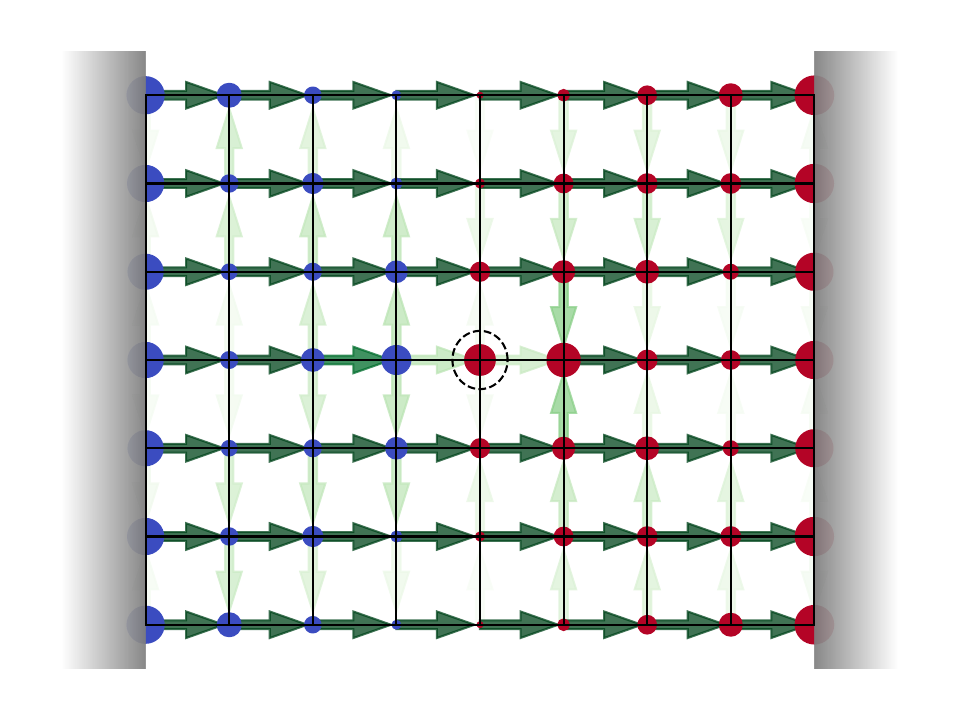}
            \hspace{-0.5cm}
            \includegraphics[width=0.075\textwidth]{{colorbar_int}.pdf}\\
            \vspace*{0.5cm}
            \raggedright \hspace*{0.65cm} System 2, $\Phi=0.33\Gamma$ \hspace*{1.07cm} System 2, $\Phi=1.33\Gamma$ \hspace*{1.07cm} System 2, $\Phi=2.66\Gamma$
            \vspace{-0.25cm} \\
            \centering
            \includegraphics[width=0.3\textwidth]{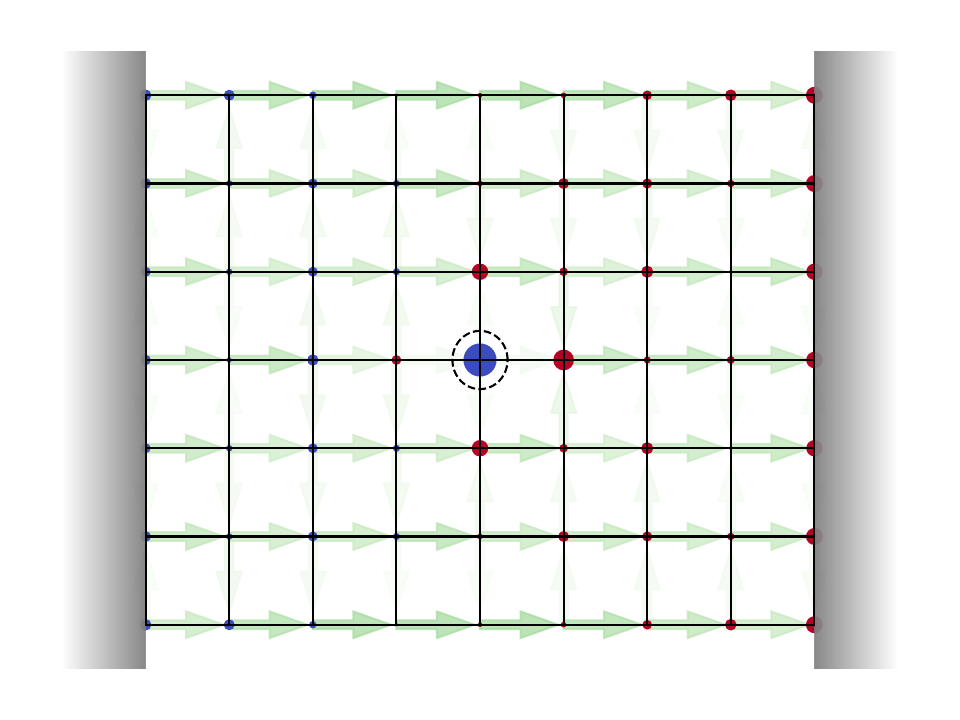}
            \includegraphics[width=0.3\textwidth]{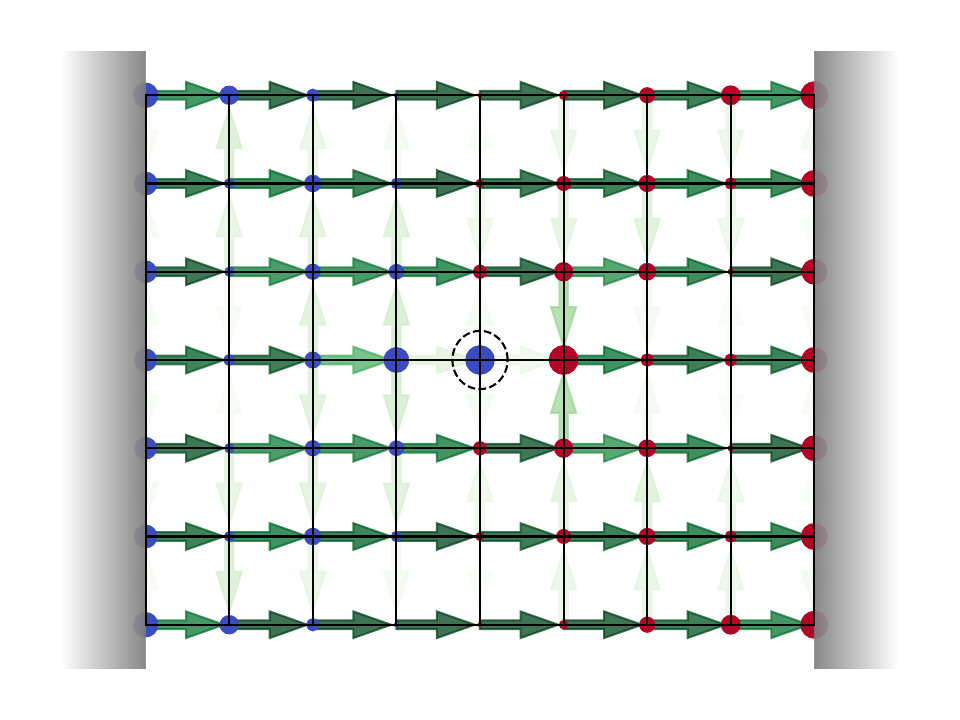}
            \includegraphics[width=0.3\textwidth]{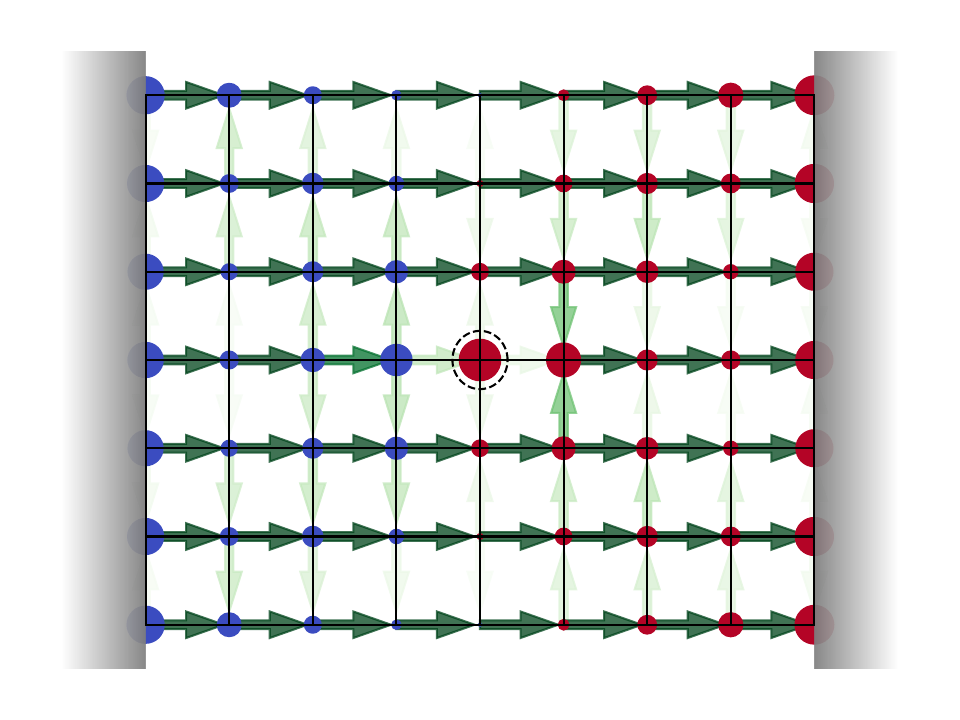}
            \hspace{-0.5cm}
            \includegraphics[width=0.075\textwidth]{{colorbar_nonint}.pdf}\\
            \caption{
                    Current and charge distribution for system 2 for three representative bias voltages (left to right) at temperature $T=0.167\Gamma$.
                    Top: interacting case. Bottom: noninteracting case $U=0$.
                    Green arrows: bond currents. 
                    Red/blue circles: charge depleted/accumulated, with the size of the circle indicating the amount of charge. 
                    Dashed circle: position of the impurity.
                    }
        \label{fig:transport_map_sys2}
    \end{figure*}
    In Fig.~2 of the main text, we show the difference in current between the interacting system and the noninteracting system. 
    For completeness, we present the raw data, i.e.\ the full current for the interacting and the noninteracting case in Figs.~\ref{fig:transport_map_sys1} and \ref{fig:transport_map_sys2} for system 1 and system 2, respectively.
    
    The data presented in Figs.~\ref{fig:transport_map_sys1} and \ref{fig:transport_map_sys2} is dominated by the overall current flow through the system as well as by finite size effects. Nevertheless, we find that the effect of the impurity, which is to introduce a resistivity into the system, is pronounced for both systems. Moreover, we observe a clear change of charge accumulated at the impurity site attributed to $\epsilon_0 \neq 0$,

\paragraph{Spectral function and Kondo-(anti)-resonance:}
    While the primary focus of this study revolves around the analysis of currents flowing through systems 1 and 2, some readers may find it valuable to examine the spectral functions at specific positions within the system. 
    In Fig.~\ref{fig:spectrum}, we present the spectrum $A_i(\epsilon)=-\mathrm{Im}\lbrace G^r_{ii}(\epsilon) \rbrace$,
    with $i$ representing both, the impurity site, as well as lattice sites positioned one, two, and three sites away from the impurity. 
    The notation for the lattice sites aligns with the notation used for the highlighted bond currents in Fig.~1 of the main text. The red dashed line corresponds to the noninteracting case, while the colored lines correspond to the spectra at different temperatures at zero bias voltage.
    \begin{figure}[htb]
        \centering
        \includegraphics{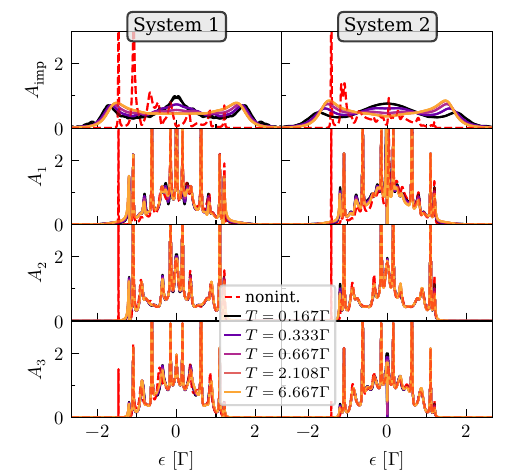}
        \caption{
                Spectral functions for system 1 (left panels) and system 2 (right panels)
                at for different temperatures and for bias voltage $\Phi=0$.
                Top panels: spectral function of the impurity. 
                Subsequent panels: the three sites located directly above the impurity. 
                The labeling for these lattice sites is consistent with the notation used for the highlighted bond currents in Fig.~1 of the main text.
                }
        \label{fig:spectrum}
    \end{figure}
    
    Regarding the impurity, particularly in system 1, we observe the emergence of a broadened peak at $\epsilon=0$, which shrinks with increasing temperature.
    This observation aligns with the interpretation that the parameter regime considered in this work is at the boundary of the Kondo regime. 
    In the case of system 2, we identify distinct sharp anti-resonances at $\epsilon=0$ for the site right above the impurity as well as two lattice sites above the impurity. This corroborates the interpretation of a current suppression, which aligns with previous findings in side-coupled quantum dots \cite{kang_anti-kondo_2001, Aligia_Kondo_2002, Sato_Observation_2005, Feng_Anti_2005, Kiss_Numerical_2011, Huo_Fano_2015, Wang_Unified_2022, Lara_Kondo_2023}.

\paragraph{Influence of the reference system for Identifying correlation effects at the atomic level:}
        \begin{figure}[htb!]
            \centering
            \vspace*{-0.3cm}
            \hspace*{-9cm}
            a)\\ 
            \includegraphics{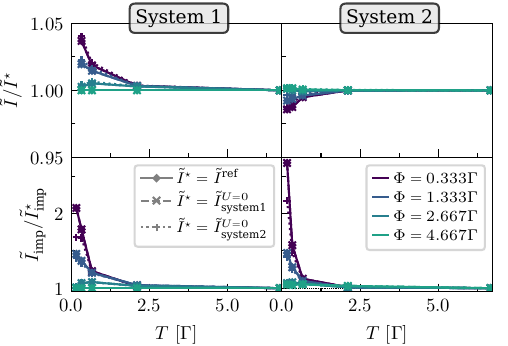}\\
            \centering
            \vspace*{-0.3cm}
            \hspace*{-9cm}
            b)\\ 
            \includegraphics{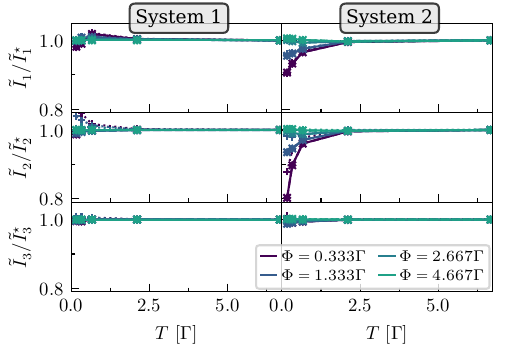}
            \caption{
                    Temperature dependence of the different currents of the interacting system $\tilde I$, divided by their noninteracting counterparts of the reference system $\tilde I^{\mathrm{ref}}$.
                    Left panels: system 1. Right panels: system 2.
                    a: ratio for the total current (top) and the current through the impurity (bottom).
                    b: ratios for currents flowing parallel to the impurity along the bonds $I_1$, $I_2$, and $I_3$, as indicated in Fig.~1 of the main text.
                    Colors correspond to different bias voltages, line styles signify the usage of different reference systems. 
                    }
            \label{fig:temperature}
        \end{figure}
        
    In the last section of the main text, we present a scheme to identify correlation effects in the current. To this end, we employ a reference system and study the temperature dependence of the current. Here, we demonstrate that this scheme is relatively robust against the choice of reference system upon redoing the same analysis as in the main text using three different reference systems that differ in their microscopic details.
    
    Fig.~\ref{fig:temperature} pertains Fig.~4 of the main text.
    Fig.~\ref{fig:temperature}a displays the data for the total current and the current flowing through the impurity, Fig.~\ref{fig:temperature}b provides the data for the currents flowing parallel to the impurity as defined in Fig.~1 of the main text.
    The usage of the three different reference systems is signified by different line styles.
    As reference systems, we will use system ``ref'' defined by $\epsilon_0=U=0$ and $t_{i0}=t\ \forall i\in N_i$, which was also used in the main text. Additionally, we also use the noninteracting systems 1 and 2 with $U=0$ and $\epsilon_0 = -\Gamma$.
    System ``ref'' differs from systems 1 and 2 in the value of $\epsilon_0$, while system ``ref'' and 1 differ from system 2 in their coupling of the impurity to its left site and its overall symmetry.
    
    Comparing the results obtained by using three different reference systems in Fig.~\ref{fig:temperature}, we find that all three model systems leads to qualitatively similar results. While the absolute values differ to some degree as expected, all reference systems can be used to identify an increase in current through the impurity for system 1 and 2, as well as a suppression of current flowing parallel to the impurity for system 2. 
    This is remarkable as the microscopic details -- especially the symmetry of the system which ultimately determines the current flow through the system -- differs for the three reference systems used here.

\end{suppinfo}

\end{document}